\documentclass[usenatbib,useAMS,espfig]{mn2e}
\usepackage[dvips]{graphicx}
\usepackage{times}
\usepackage{mathptm}
\usepackage{longtable,afterpage}
\usepackage{caption}
\usepackage{subcaption}
\usepackage{url}
\usepackage{multirow}
\usepackage{amssymb}
\usepackage{lscape}
\usepackage[reqno]{amsmath}
\usepackage{color}
\usepackage{nameref}
\usepackage[bookmarks=true,pdftitle={},pdfauthor={Roberto Raddi},pdfsubject={},pdfcreator={Roberto Raddi}, colorlinks=true, linkcolor=black, citecolor=black]{hyperref}
\usepackage[figure, figure*]{hypcap}
\usepackage{bbding}

\def\aap{A\&A}
\def\aapr{A\&ARv}
\def\aaps{A\&AS}
\def\aj{AJ}
\def\apj{ApJ}
\def\apjs{ApJS}
\def\apjl{ApJ}
\def\apss{APSS}

\def\mnras{MNRAS}
\def\pasp{PASP}

\defcitealias{Raddi13}{paper~I}
\defcitealias{RF09}{RF09}
\defcitealias{SFD98}{SFD98}
\defcitealias{KW99}{KW99}

\def\Halpha{{\rm H}\alpha}

\DeclareGraphicsExtensions{.pdf}

\title[A deep catalogue of classical Be stars]{A deep catalogue of classical Be stars in the direction of the Perseus Arm: spectral types and interstellar reddenings.}
\author[Raddi et al.]{R. Raddi$^{1,2}$\thanks{E-mail:
    r.raddi@warwick.ac.uk}, J. E. Drew$^{1}$, D. Steeghs$^{2}$, N. J. Wright$^{1}$, J. J. Drake$^3$, G. Barentsen$^1$, 
\newauthor  J. Fabregat$^4$, S. E. Sale$^{5}$\\
$^{1}$Centre for Astrophysics Research, STRI, University of Hertfordshire, College Lane Campus, Hatfield, AL10 9AB, U.K.\\
$^{2}$Department of Physics, University of Warwick, Gibbet Hill Road, Coventry, CV4 7AL, U.K.\\
$^{3}$Smithsonian Astrophysical Observatory, 60 Garden Street, Cambridge, MA 02138, USA\\
$^{4}$Observatorio Astron\'{o}mico, Universidad de Valencia, 46100 Burjassot, Spain\\ 
$^{5}$Rudolf Peierls Centre for Theoretical Physics, Keble Road, Oxford OX1 3NP, UK\\}

\begin{document}

\date{Accepted 2014 October 05. Received 2014 October 05; in original form 2014 June 30}

\pagerange{\pageref{firstpage}--\pageref{lastpage}} \pubyear{}

\maketitle

\label{firstpage}

\begin{abstract}
We present a catalogue of 247 photometrically and spectroscopically
confirmed fainter classical Be stars ($13 \lesssim r \lesssim 16$) 
in the direction of the Perseus Arm of the Milky Way ($-1^{\circ} < b 
< +4^{\circ}$, $120^{\circ} < \ell < 140^{\circ}$).  The catalogue
consists of 181 IPHAS-selected new classical Be stars, in addition to 
66 objects that were studied by \citet{Raddi13} more closely, and 3 
stars identified as classical Be stars in earlier work.  This study 
more than doubles the number known in the region.  
Photometry spanning 0.6 to 5\,$\mu$m, spectral types, and interstellar
reddenings are given for each object. The spectral types were
determined from low-resolution spectra
($\lambda$ / $\Delta$ $\lambda \approx 800$--2000), to a precision of
1--3 subtypes. The interstellar reddenings are derived from 
the $(r - i)$ colour, using a method that corrects for circumstellar disc 
emission.   The colour excesses obtained range from $E(B-V) = 0.3$ up 
to 1.6 -- a distribution that modestly extends the range reported in 
the literature for Perseus-Arm open clusters. For around half
 the sample, the reddenings obtained are compatible 
with measures of the total sightline Galactic extinction.  Many of
these are likely to lie well beyond the Perseus Arm.

\end{abstract}


\section{Introduction}
\label{intro}
The first classical Be star to be classified was $\gamma$\,Cas, identified by Father Angelo Secchi almost 150 years ago. Secchi described it
as showing a particularly curious bright line -- in that case, H$\beta$ -- as opposed to the dark lines observed in the spectra
of stars with similar colours \citep{Secchi66}.
Presently, the family of classical Be stars numbers over 2000 known members in the all-sky Be Star Spectra database 
\citep[BeSS\footnote{\url{http://basebe.obspm.fr}},][]{Neiner11}. 
\begin{figure*}
\centering
\includegraphics[width=\textwidth]{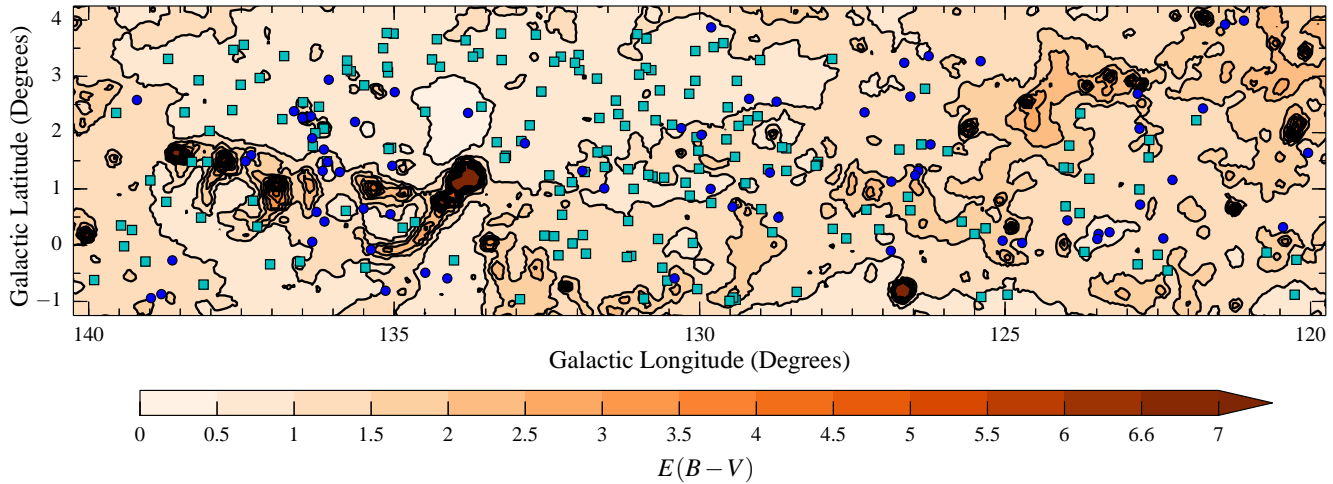}
\caption{The spatial distribution on the sky of the studied classical Be stars is plotted on the map of the interstellar dust extinction
computed by \citetalias{SFD98}. We applied the correction ($0.86 \times E(B-V)$) suggested by \citet{Schlafly10} to the \citetalias{SFD98} maps, 
colour coding the levels of extinction that are binned in 0.5\,mag steps and cut at $E(B-V) = 7$ -- with the lightest colour
mapping the 0--0.5\,mag range. The cyan squares symbols denote the 181 classical Be stars
observed with FLWO-1.5m/FAST, while the blue circles represent the 67
classical Be stars observed in La Palma.  We note that the minimum \citetalias{SFD98} is measured above the W4 H~{\sc ii} bubble ($\ell \approx 134$,
$b \approx 2.5$), where $E(B-V) \sim 0.3$, corresponding to the first reddening bin of the colour-scale. The most obscured regions with
$E(B-V)>7$ correspond to the densest areas of the Perseus arm, where ongoing star formation is observed (specially W3/W4/W5 and Cas\,OB\,7).}
\label{f:map}
\end{figure*}

Following the original definition given by \citet{Collins87}, classical Be
stars are non-supergiant early-type stars whose spectra have, or have
had at some time, one or 
more Balmer lines in emission \citep{Porter03}. They do not present
forbidden emission lines in their spectra. They are also characterised
by excess continuum emission at ultraviolet (UV), optical, and infrared
(IR) wavelengths \citep*[e.g.\ see][\citealt{Kaiser89, ZB91, Dougherty94}]{Dachs88}. 
The presence of dense and hot ($n_{e} \approx 10^{12}$\,cm$^{-3}$, 
$T_{e} \approx 10000$\,K) circumstellar decretion discs is the cause of
the observed free-free and free-bound optically-thin continuum emission,
and also the line emission \citep*[][\citealt{Dachs88, Carciofi06}]{Gehrz74}. The 
continuum excess emission has been carefully characterised in the past, in 
order to understand its origins and to separate out its contribution to
observed optical and IR fluxes \citep[][]{Dachs88, ZB91, Dougherty94}.
Fast rotation is specially brought up as one of the main factors
behind the formation of the circumstellar discs \citep{Townsend04, Cranmer05}.
\citet{Carciofi12} have provided evidence that the circumstellar disc of classical Be stars
 approximates a viscous decretion disc as described by
\citet*{Lee91}.  Evidence for Keplerian rotation of the disc matter
around the B star has been obtained by \citet{Meilland07} and 
\citet{Oudmaijer11}.  More recently still, the requirement of a 
fast-rotating B-type stars has been added to the definition of classical Be star
\citep*{Rivinius13}.

In the past, there has been some debate regarding the evolutionary
status of classical Be stars \citep{Mermilliod82, Slettebak85,
  Fabregat00}. More recently the controversy has started to settle.
\citet{Martayan07} found, through studying classical Be stars both in the Milky
Way and in the Magellanic Clouds, that the evolution of the rotation speeds with
age is mass and metallicity dependent. This work concluded that the Be 
phenomenon, in our Galaxy, appears earlier in the main sequence (MS)
life at higher stellar masses ($\sim 12$\,M$_{\odot}$) and earlier
spectral types, while it is delayed in lower mass stars ($\sim
5$\,M$_{\odot}$), or equivalently, in later B-types, confirming
previous observations of classical Be stars in open clusters
\citep{Fabregat00}.  In general, classical Be stars can be viewed as
relatively (but not very)
young stars that are typically observed in open clusters up to 100\,Myr old, 
with highest incidence in clusters with ages in the range of
13--25\,Myr, which are typical of the turn-off age of B2 stars \citep{Fabregat00}.
On the other hand, observations of Galactic-field bright classical Be stars
show a flat distribution across the B subtypes \citep{ZB97}, which
could be explained by the relatively smaller numbers of known old open 
clusters \citep[e.g. in][]{Dias02} and high "infant-mortality''
\citep[see e.g.][]{Goodwin06}.

In \citet[hereafter paper~I]{Raddi13}, we studied a sample of 67 classical Be
stars, located within a section of the Galactic plane between 
$120^{\circ} \leq \ell \leq 140^{\circ}$ and $-1^{\circ} \leq b \leq
+4^{\circ}$ for which intermediate-dispersion optical spectra were
available.  Our stars were picked out from a larger set of candidate 
emission line stars selected photometrically from the Isaac Newton
Telescope (INT) Photometric $\Halpha$ Survey of the Northern Galactic
Plane \citep[IPHAS,][]{Drew05}.  In \citetalias{Raddi13}, we investigated whether
the dereddened sample could provide any hints of spiral arm structure
in this section of the thin disc, but concluded that the errors in the
spectroscopic parallaxes wash out any distinction between a smooth or
spiral-arm dominated stellar density profile.  Nevertheless, the study 
represented a step change to a sample of classical Be stars that is appreciably 
fainter, more distant, and more reddened than the previously known classical Be 
stars in the area. We will return to this point later on comparing the
expanded sample presented here with the properties of the classical Be stars in 
the BeSS database and the early-type emission line stars in the
catalogue of \citet[KW99 from now on]{KW99}. 

In this work, we add a further 181 classical Be stars to the sample analysed in 
\citetalias{Raddi13}.  These stars have also been
spectroscopically-confirmed, but with lower-resolution spectra. Our
aim here is to present a homogeneous catalogue for all 248 stars, that
incorporates photometry from WISE, alongside 2MASS data, and 
provides a sound estimate of the interstellar extinction towards each
of them.  The available range of data for the stars in the catalogue
is presented in Section\,\ref{chap2}.   The spectral typing for the new 
classical Be stars is described in Section\,\ref{chap3}, where we compare the 
results with those obtained in \citetalias{Raddi13}. In
Section\,\ref{chap4}, we will discuss the measurement of interstellar
reddening and we will describe the procedure adopted for removing the 
circumstellar colour-excess from the observed $(r-i)$ colour.
Finally, the discussion in Section\,\ref{chap5} focuses on the
comparison between the interstellar reddenings measured for the classical Be
stars and the total integrated values for their Galactic sightlines, 
by \citet*[hereafter SFD98]{SFD98} and \citet[hereafter RF09]{RF09},
 and we confront our results with the most recent
3D extinction maps of the northern Galactic plane by \citet{Sale14}. 

\section{The sample}
\label{chap2}
The classical Be stars in the catalogue are found in the strip of the Galactic
plane, contained within $120^{\circ} \leq \ell
\leq 140^{\circ}$ and $-1^{\circ} \leq b \leq +4^{\circ}$. Their
spatial distribution is shown in Fig.\,\ref{f:map}, overplotted on the 
SFD98 extinction map of the area. As we reported in \citetalias{Raddi13},
the follow-up spectroscopy targeted $\Halpha$ emitters, identified in
IPHAS mainly by \citet{Witham08}. The great majority of our sample
falls in the magnitude range $12.5 < r < 16.5$ (noting that the IPHAS
survey is calibrated in the Vega system).  In terms of the 
optical spectroscopy we have available, the sample falls into two groups:
\begin{figure*}
\centering
\begin{subfigure}{\textwidth}
\includegraphics[width=\textwidth]{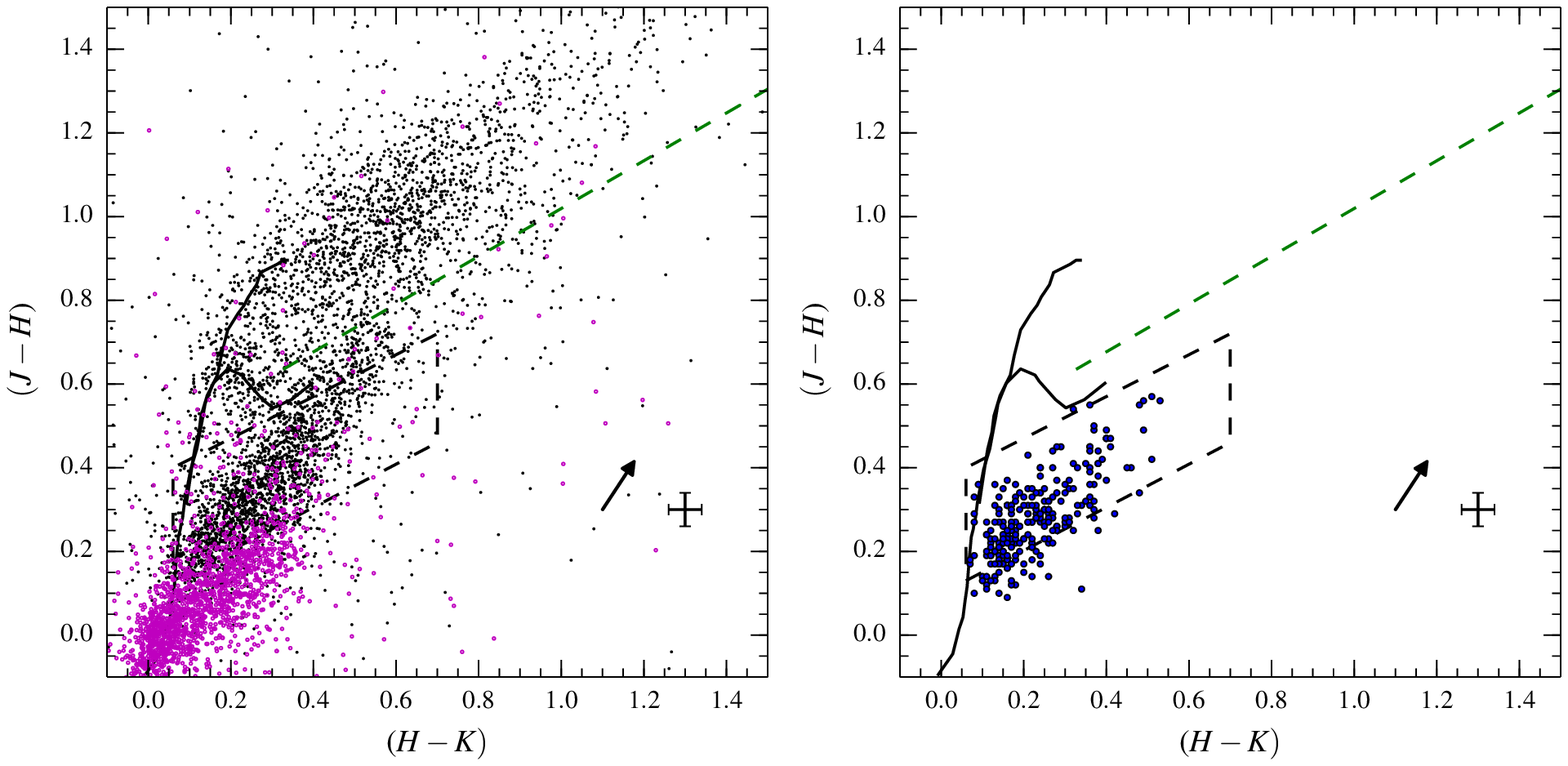}
\end{subfigure}
\begin{subfigure}{\textwidth}
\includegraphics[width=\textwidth]{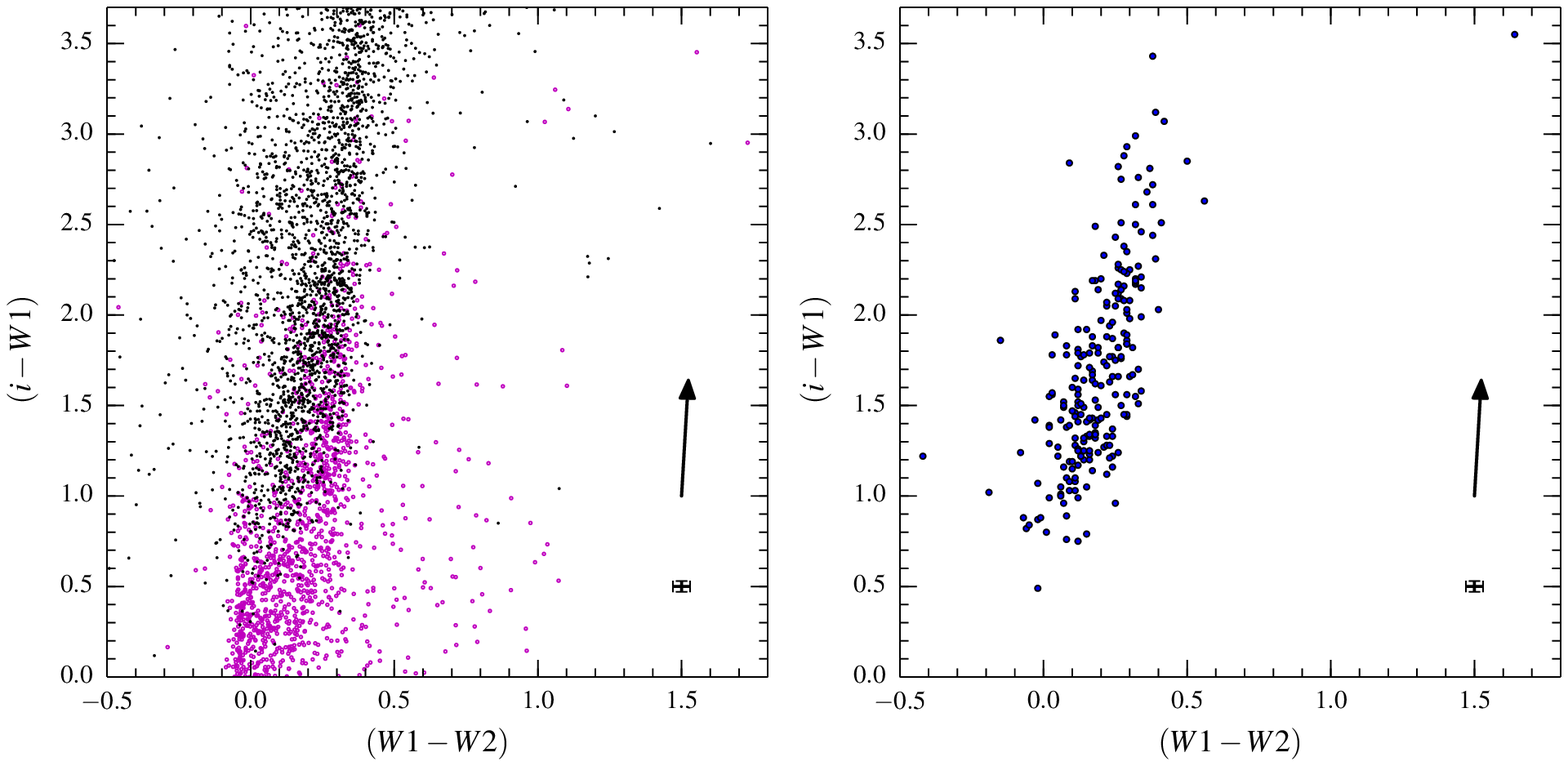}
\end{subfigure}
\caption{The top panels present the 2MASS near-IR
    colour-colour diagrams.   The studied classical Be stars are plotted as
    blue circles in the right-hand panel, while on the left 
    the \citet{Witham08} catalogue of $\Halpha$ emitters is shown in
    black (from which most of the classical Be stars sample is drawn) and the
classical Be stars from the catalogue of \citet{Neiner11} in lilac. The
black/solid curves sketched in the top panels are the MS and giant-star
loci from \citet{Rieke85}, while the green-dashed line is the
classical T~Tauri star locus \citep[][]{Meyer97} and the black-dashed
box is the generously-defined zone in which classical Be stars with 
$A_V = 4$ would fall \citep{Corradi08}. The lower panels present
the optical-WISE colour-colour counterparts to those shown in the top
panels.  The average photometric uncertainties are shown in the
bottom-right corner of each plot. The black arrows are the reddening 
vectors for $A_{V} = 1$. }
\label{f:ccd_ir}
\end{figure*}
\begin{itemize}
\item 181 objects were identified in the initial follow-up optical 
spectroscopy, presented in Section\,2 of \citetalias{Raddi13} and taken
no further. Observations were taken in queue mode between 2005 and
2011 at the 1.5\,m Fred Laurence Whipple Observatory (FLWO) using the
FAst Spectrograph for the Tillinghast Telescope (FAST)
\citep[][]{Fabricant98} and processed at the Telescope Data Centre of
the Smithsonian Astrophysical Observatory.
The final spectral resolution is $\Delta \lambda \approx 6$\,\AA\, and
the spectral coverage is 3500-7500\,\AA.   Since the principal aim of
this initial follow-up was the characterisation of $(r - \Halpha)$ excess 
objects identified by IPHAS at $r \lesssim 17$, the standards needed
for nightly flux calibration were not obtained.   

\item 67 objects are the sample of confirmed classical Be stars, observed
using the INT and Nordic Optical Telescope in La Palma, that were
studied in depth in paper I. For 60 of them, FLWO-1.5m/FAST spectra
are also available.  Details regarding the selection and observation
of this sub-sample are given in Sections\,2 and 3 of that paper, while
their spectral typing is described there in Section\,4.  The spectral
resolution of the La Palma data is 2--3 times higher than that
delivered by the FAST instrument, and sufficient flux standards were
observed to permit relative spectrophotometry.  These stars can now 
provide some
checks on the inferences drawn from the larger FLWO-1.5m/FAST-only sample.
\end{itemize}

The criteria satisfied by all stars in the spectroscopic sample we
present are: (i) the spectrum has a B-type spectral appearance with the low-order Balmer lines in emission
-- $\Halpha$-emission is necessary for identification -- and no 
clearly visible forbidden emission lines \citep{Porter03,Rivinius13}; (ii) the near-IR colours are
consistent with those of classical Be stars; 
(iii) $(J-H) \leq 0.6$ \citep[see e.g.][and top right panel of Fig.\,\ref{f:ccd_ir}]{Corradi08}. 
\begin{table*}
\centering
\caption{Classical Be stars with matching entries within a 5-arcsec search radius in \citet{Koenig08}.
Ours and their ID number (recno) are given for each, along with IPHAS
$r$ and Spitzer magnitudes and YSO class according to \citet{Koenig08}.}
\begin{tabular}{@{}lrrrrrrrlr@{}}
\hline
N & recno & IPHAS2 $r$ &3.6\,$\mu$m &   4.5\,$\mu$m & 5.8\,$\mu$m & 8.0\,$\mu$m &  24\,$\mu$m &  class & Separation \label{t:koenig08}\\
  &       &   (mag)    &   (mag)      &   (mag)    &   (mag)    &   (mag)    &   (mag)   &        &  (arcsec) \\
\hline
  212  &  2833   & 13.37  &  $9.78   \pm0.01$  &    $9.55   \pm0.01$  &    $9.25   \pm0.01$  &    $8.95   \pm0.01$  &  $7.78  \pm0.04$   &  II &0.10\\    
  222  &  10258  & 16.17  &  $12.83  \pm0.01$  &    $12.69  \pm0.01$  &    $12.53  \pm0.05$  &    $12.06  \pm0.12$   &               & III &   0.04  \\
  223  &  10573  & 14.52  &  $11.93  \pm0.01$  &    $11.57  \pm0.01$  &    $11.24  \pm0.01$  &    $10.76  \pm0.03$   &  $9.10   \pm0.08$   &  II &0.70 \\    
  225  &  12190  & 15.01  &  $12.07  \pm0.01$  &    $11.81  \pm0.01$  &    $11.54  \pm0.02$  &    $11.21  \pm0.02$   &               &  II &   0.21 \\
  230  &  15241  & 13.88  &  $11.80  \pm0.01$  &    $11.59  \pm0.0$1  &    $11.40  \pm0.01$  &    $11.11  \pm0.03$   &               &  III &    0.09 \\
  234  &  16947  & 14.38  &  $12.14  \pm0.01$  &    $11.95  \pm0.01$  &    $11.73  \pm0.01$  &    $11.51  \pm0.04$   &               &  III &   0.08 \\
\hline
\end{tabular}
\end{table*}

To better confirm the classical Be nature of the stars in our sample, we now 
complement the IPHAS and near-IR photometry from the Two-Micron All-Sky Survey
\citep[2MASS,][]{Skrutskie06}, with $W1$ ($3.4\,\mu$m) and $W2$
($4.6\,\mu$m) data from the Wide-field IR Survey Explorer 
\citep[WISE,][]{Wright10}. The combination of
optical and near-IR colours, plotted against $(W1-W2)$ colours from
WISE, can reveal the presence of a warm dust shell that would imply
the object is more likely to be a young Herbig Be star.
The circumstellar contribution to the near- and mid-IR spectral energy 
distribution (SED) of a classical Be star is very small in comparison to that seen
in dust-enshrouded Herbig and other more exotic Be stars
\citep[][]{Gehrz74, Dachs88, Waters91,Lada92}.   One would expect the
stars in our sample to be stretched over a relatively limited portion of the
colour-colour diagrams of Fig.\,\ref{f:ccd_ir}, chiefly due to the
effect of interstellar reddening.  That is, their locus should roughly
parallel the reddening vector -- which they do, in both panels of Fig.\,\ref{f:ccd_ir}.   
The expectation regarding the SEDs of
classical Be stars is that they do {\em not} present the double-peaked
appearance or the flat/rising profile through the 1--5\,$\mu$m range, that is
typical of young stellar objects (YSOs). 
\begin{figure}
\includegraphics[width=\linewidth]{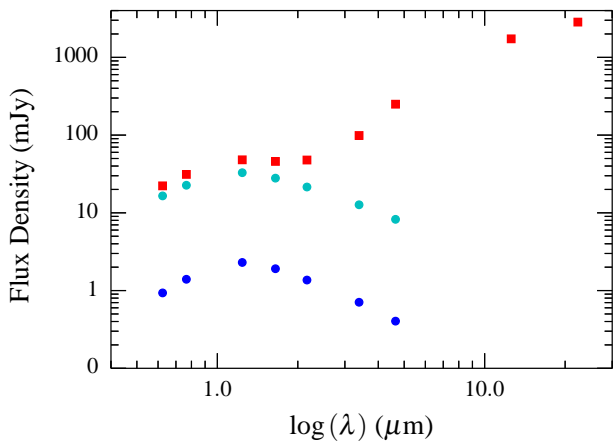}
\caption{The 0.6 -- 5 $\mu$m SEDs of three objects in the catalogue. 
Red squares are used for \#\,174 that we rule out as classical Be star, because its SED suggests
the presence of warm dust. For this object we plot the WISE W3 and W4 fluxes, which were also available.
The other two SEDs shown are for two typical classical Be stars in our sample,
i.e.\ \#\,44, and \#\,184.}
\label{f:sed}
\end{figure}

It turns out that the added WISE photometry mainly confirms the
results of the classification via a combination of visual inspection
of optical spectra and 2MASS photometric criteria employed in \citetalias{Raddi13}.
Just one object, star \#\,174 \citepalias[\#\,45 in][]{Raddi13} is
excluded on the basis of its $W1$ and $W2$ magnitudes. This object is
more likely to be a Herbig~Be star approaching the zero-age MS:
its $(W1 - W2)$ colour is very extreme, placing it in the top right
corner of the upper-right panel of Fig\,\ref{f:ccd_ir}. For comparison, we plot in
Fig.\,\ref{f:sed} the SED of star \#\,174 together with the SEDs of
two more typical classical Be stars in our catalogue. While the SEDs of the 
two classical Be stars do not present a marked circumstellar colour-excess out
to $\sim 5\,\mu$m, the SED of \#\,174 rises strongly through the WISE
bands indicating the presence of warm dust. 
In addition, a few objects are found to have $(W1-W2)<0$, which does not
make physical sense even for early-type stars. Although in most cases 
their colours are consistent with zero (to within 3 times the quoted
errors), the WISE team note a tendency to overestimate the background 
subtraction and,
as a result, underestimate the magnitudes when $W1>14$ and $W2>13.5$.
This is plausibly the source of difficulty for objects \# 127, 135,
and 229 that show the largest negative values.

Finally, we also identify 6 objects in common with \citet{Koenig08} 
(\#\,212, 222, 223, 225, 230, and 234 in Table\,\ref{t:koenig08}). The authors used Spitzer near- and mid-IR photometry
to classify the YSOs belonging to the star forming region W5. Their classification
is determined from the slope of the SEDs in the 1.25--24\,$\mu$m range, after correcting for foreground extinction via the use
 of 2MASS-generated extinction maps \citep[][and references therein]{Gutermuth08}. Sources \#\,222, 230, and 234 are labelled as class III, 
i.e.\ objects that are dominated even at IR wavelengths by photospheric
emission. This is consistent with the classical Be-star definition. On the other hand,
\#\,212, 223, and 225 are labelled as class II YSOs, which means the emission from the circumstellar disc in the IR dominates
over the photospheric emission, as seen in T~Tauri stars or Herbig~Ae/Be stars with optically thick discs. 
We question this classification, noting that the colours of these
objects in the Spitzer colour-colour diagrams of fig.\,5 and 6 in
\citet{Koenig08} overlap with objects that the authors identify as
transition-disc and class III.  They do not stand out in Fig.\,\ref{f:ccd_ir}.   
Our sample also includes one star (\#\,217) that falls within the area surveyed by \citet{Koenig08} that does not find a match with their
sources. This object is reported to vary in the WISE source catalogue
and, indeed, the listed $W1$ and $W2$ fluxes are at odds with both
2MASS and IPHAS photometric data, in being much brighter than
extrapolation would suggest. Neither 2MASS nor WISE images show evidence for nearby blended sources.

In Table\,\ref{t:photometry}, we report the photometry for all the 248
stars in the sample. Star \#\,174 is included in this list, but it will
be excluded from further analysis. Unless otherwise noted, the $r$, $i$,
and $\Halpha$ magnitudes given here take advantage of the 
most recent IPHAS release \citep[DR2,][]{Barentsen14}. We also 
indicate whether a star in the catalogue has a La Palma spectrum, a 
FLWO-1.5m/FAST spectrum, or both.
\subsection{Completeness and a comparison with previous catalogues}
\begin{figure}
\centering
\includegraphics[width=\linewidth]{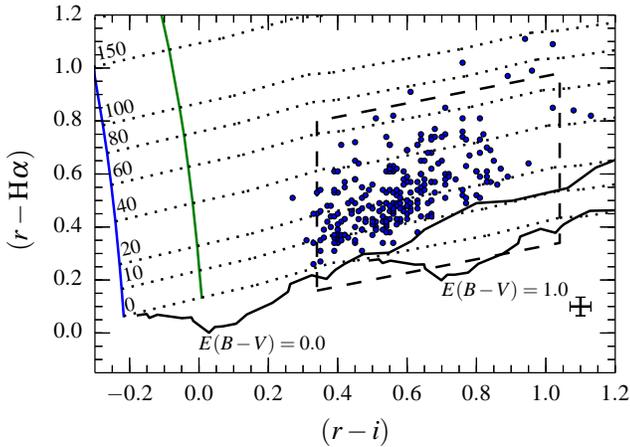}
\caption{IPHAS colour-colour diagram of the 248 objects studied here
  (blue symbols). The black thin-curves are synthetic main sequence (MS) loci \citep[see table\,2 in][]{Drew05}. 
The lower dashed curve is the early-A reddening curve, while the dot-dashed curves are lines of constant $\Halpha$ emission 
(corresponding equivalent widths are indicated on the left-hand-side of the plot). The vertical curves are the
curves of growth of $\Halpha$ emission for the Rayleigh-Jeans case (blue, $F_{\lambda} \sim \lambda^{-4}$) and an A0V SED (green, $F_{\lambda} \sim \lambda^{-3}$).
The thick black line, for an equivalent width of $EW(\Halpha) = -15$\,\AA\, represents the completeness limit for $E(B-V)=1$. 
On the bottom right side of the figure, the mean colour uncertainties
are indicate. For reference, we also overplot the expected domain for classical Be stars with $A_V = 4$ \citep{Corradi08}, as a black-dashed box.}
\label{f:ccd_iphas}
\end{figure}

The selection and identification of classical Be stars from IPHAS photometry,
is limited mainly by the combination of interstellar reddening and the
strength of $\Halpha$ emission. In other words the likelihood of
inclusion in the sample can be defined with reference to $(r-i)$
colour (dominantly a measure of extinction for B stars), and 
$(r - \Halpha)$ excess
(which is proportional to emission equivalent width $EW(\Halpha)$).
From the IPHAS colour-colour plane plotted in  Fig.\,\ref{f:ccd_iphas},
it can be estimated that the sample is 
reasonably complete down to $EW(\Halpha) \lesssim -15$\,\AA\, over the 
colour excess range, $0.5 \lesssim E(B-V) \lesssim 1.5$.  Note,
however, that objects with weaker line emission can be selected in the 
lower half of the reddening range -- as long as they 
remain fainter than $r \sim 12$ (the saturation limit of IPHAS).  Indeed
the bright limit potentially cuts out some modestly reddened B0-B1 classical Be 
stars.   In principle, classical Be stars more reddened than $E(B-V) \sim 1.5$ 
can be picked out if they present with correspondingly stronger
H$\alpha$ emission, lifting $(r - H\alpha)$ above the main locus of 
late-type MS stars.  There are no such candidates in the present 
magnitude-limited sample.
\begin{table}
\centering
\caption{Classical Be stars with matching entries within 5-arcsec in the \citetalias{KW99} and BeSS catalogues.
Their visual magnitudes and spectral types (from \citetalias{KW99}) and angular separation are supplied along 
with the ID name and coordinates.  The IPHAS2 $r$ magnitudes,
calibrated in the Vega system, are
brought forward from the full photometry list, Table\,\ref{t:photometry}.  }
\begin{tabular}{@{}rllllll@{}}
\hline
N &\citepalias{KW99}& BeSS &  $V$ & SpT & IPHAS2 $r$ & Separation \label{t:matches}\\
  &      &       & (mag) &  & (mag) & (arcsec) \\
\hline
  8  & 3-27   &  &  15.1 &  & 14.85 & 4.0\\
  37  & 5- 3  &  &  14.6 & A & 12.64 & 2.9\\
  73  & 7-25  &  &  12.4 & B & 12.46 & 1.0\\
  74  & 7-31   &  SAN 28 &  13.4 & B &
  13.76 & 0.3\\
  76  &        & BG 82 &   &  & 13.37 & 0.1\\
  115 & 9- 7  &  &  13.9 &  & 13.40 & 1.8\\
  124 & 9-29  &  &  12.9 &  & 12.58 & 2.3\\
  144 & 9-55  &  &   &  & 12.87 & 3.2\\
  174 & 12-38 &  &  12.5 &  & 12.88 & 4.2\\
  186 & 12-58 &  &  13.6 &  & 13.26 & 0.2\\
  205 & 13-31 &  &   &  & 13.43 & 0.5\\
  208 & 13-36 &  &  12.5 &  & 12.91 & 0.5\\
  244 & 14-40 &  &  15.1 &  & 14.46 & 0.9\\
\hline
\end{tabular}
\end{table}
\begin{figure}
\centering
\includegraphics[width=\linewidth]{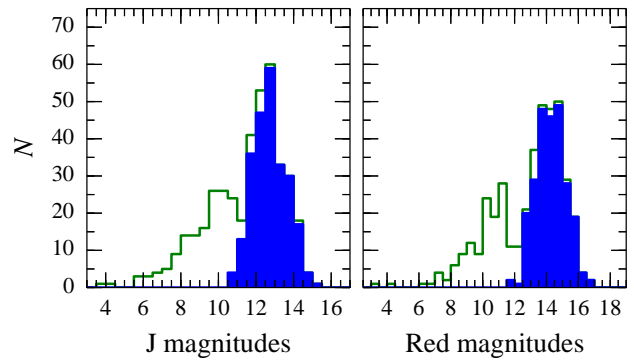}
\caption{
2MASS $J$ and red magnitudes distributions, for the 248 classical Be stars in the sample (solid blue) and for the previously known classical Be stars in the area
(step histogram). Not all the previously known classical Be stars have 2MASS photometry so that only 152 are plotted in the left figure. Because fewer of the previously known
classical Be stars in the area do not possess IPHAS $r$ photometry either, we used USNO\,B1.0 photometry \citep{Monet03} as a close proxy.}
\label{f:red_magnitudes}
\end{figure}

Comparing our sample with previously known classical Be stars
gives some insight into the photometric selection and how the 
bright and faint magnitude limits ($r \approx 17$) affect the 
outcome. In the area there are 64 known classical Be stars listed in the BeSS 
database, plus 148 $\Halpha$ emitters in the \citetalias{KW99} catalogue,
with spectral types and near-IR colours that match our classical Be star definition.  
These two databases have in common 45 objects, while the overlap with 
our sample is only 13 objects cross-matching within 5-arcsec (in 
Table\,\ref{t:matches}).  The matches generally have $V$ magnitudes
within a few tenths of the IPHAS2 $r$ magnitudes, as should occur if
they refer to the same object.  Also these stars are all sufficiently 
isolated in the IPHAS images, relative to other similarly bright
stars, that we can have confidence in the cross-identification.

This modest overlap mainly reflects the magnitude ranges sampled, as is
evident from the distributions plotted in Fig.\ref{f:red_magnitudes}. 
The stars in our new catalogue are on average 2-3 magnitudes fainter 
than in either the BeSS database or KW99.
We notice that the merged magnitude distributions seem to
peak at 12-13\,mags in J and 14-15\,mags in the red, but 
present a deficit of stars at respectively $J \approx 11$ and 
$R \approx 12$. This feature may very well be due to the combined 
incompleteness of \citetalias{KW99} and \citet{Witham08} lists, in
particular, at the intersection of their faint and bright limits. 
Furthermore, the median magnitudes of classical Be stars in the BeSS catalogue 
in this same region are also relatively bright at $R \sim 10$ and $J \sim 9$.
It can be inferred that, if the magnitude distributions are smooth, in reality,
there remains a deficit of 30-40 classical Be stars at $r \sim 12$
awaiting discovery.   
\section{Spectral typing of the FLWO-1.5\lowercase{m}/FAST sample}
\label{chap3}
Compared to the classification of the La Palma spectra reported in
\citetalias{Raddi13}, classification based only on FLWO-1.5m/FAST spectra
is more challenging, due to the generally lower S/N ratio
 -- with a median of S/N = 28, at $\lambda 4500$\,\AA\, -- and lower
 spectral resolution. 
However, for the purpose of determining 
interstellar reddenings from IPHAS $(r-i)$ colours, even a coarser
spectral type assignment, as opposed to the precision of 1-2 sub-types reached in \citetalias{Raddi13}, is
still acceptable, due to the weak dependence of intrinsic $(r-i)$
colours on spectral type among B-type stars (see also Section\,\ref{chap4}).

Here, we will estimate spectral types relying on the most easily detected blue
spectral features -- the He~{\sc i} ($\lambda 4471$\,\AA) and Mg~{\sc ii} 
($\lambda 4481$\,\AA) lines -- alongside the higher Balmer series. 

We stressed in \citetalias{Raddi13} the value of measuring the ratio of equivalent 
widths, $\log{(W_{\lambda 4471}/W_{\lambda 4481})}$, as it is well
established as a guide to effective temperature.  
We also showed before that the reliability of the measure is strongly 
dependent on the S/N ratio measured in the proximity of the He~{\sc i} and
Mg~{\sc ii} transitions, finding in particular that for spectra with S/N 
lower than $\sim 25$ there is a marked increase in the spectral
typing error, and that at $\rm{S/N} \sim 10$ almost no constraint is possible.
This evaluation was based on measuring line ratios from model
atmospheres \citep[][]{Munari05} many times, to which controlled amounts of 
random noise had been added. For S/N~$\geq 25$ the distribution of measured
line ratios from model atmospheres was found to be sufficiently narrow 
to allow a spectral-type determination within 1-2 subtypes.
\begin{figure}
\centering
\includegraphics[width=\linewidth]{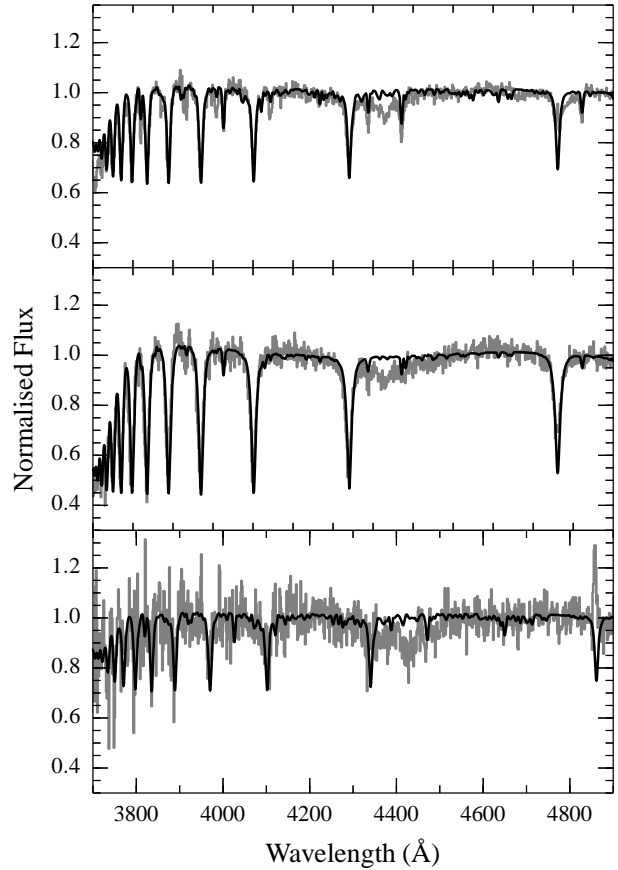}
\caption{Three example of spectra observed with FLWO-1.5m/FAST are shown. From top to bottom: \#\,121 is a B3 star with S/N = 62; 
\#\,183 is a B7 star, with S/N = 45; \#\,115 is an Early-B type star with S/N = 20. The observed spectra are plotted in grey, 
while the appropriate model atmospheres from \citet{Munari05} are in black.}
\label{f:example_spectra}
\end{figure}

Our approach to typing here has to be effective on less optimal data.
Where we can, we assign a trial spectral type from measurement of the 
$\log{(W_{\lambda 4471}/W_{\lambda 4481})}$ ratio, and otherwise appraise
types by means of comparisons with model spectra
\citep[][]{Munari05}. The spectral classification of the classical Be stars
arising from both FLWO-1.5m/FAST spectra and the
La~Palma data presented in \citetalias{Raddi13} have been compared, to 
check for overall consistency of outcome: this confirms that typing to
within 1-2 subtypes starts failing at $\rm{S/N} \sim 25$ (we return to this 
at the end of this section).  Hence we have split the FLWO-1.5m/FAST 
sample into two, according to $\rm{S/N}$ ratio:  (i) for spectra with $\rm{S/N} \geq 25$,
we use $\log{(W_{\lambda 4471}/W_{\lambda 4481})}$, as an effective
temperature proxy, and then check via comparison with model
atmospheres; (ii) for the spectra with S/N~$< 25$ and some 
better-exposed cases where method (i) gives uncertain solutions, we
are restricted to assessing the best spectral type via a simple $\chi^{2}$ 
minimisation, using resolution-degraded \citet{Munari05} model 
atmospheres.  To facilitate (ii), both the observed spectrum and the 
models are normalised to unity over the spectral range
$\lambda\lambda\,3800-5000$\,\AA. In Fig.\,\ref{f:example_spectra}, three
examples with decreasing $\rm{S/N}$ ratio at $\lambda\, 4500$\AA\ are
given. The agreement is convincing for the spectra shown in the top
two panels while the classification of the least well-exposed example
is clearly much more uncertain.  

We have found that spectral types assigned with method (i) rarely
needed a sub-type shift of more than a 1-2 sub-types, in order to
better satisfy a visual comparison with model spectra.
Where a star has been observed at higher resolution from La Palma, 
we continue to use the La Palma spectral typing for the purpose of 
estimating reddenings in Section\,\ref{chap4}. We also
note that 3 stars (\#\,73, 74, 76) of the 181 stars with
FLWO-1.5m/FAST spectra were already classified by \citet{Mathew11} as 
B2V, B5-7V, and B5V, to be compared with our B2, B4, and B3 spectral 
types (these three stars are also found in 
\citetalias{KW99} and in the BeSS catalogue, e.g. cf Table\,\ref{t:matches}).

On the other hand, for the spectra with S/N~$< 25$ that were
classified with method (ii), the grid of \citet{Munari05} models that
we used is limited to $\log{g} = 4$, rotational velocity of 250\,km/s 
\citep[typically observed in Be stars,][]{Chauville01}, and 
$T_{\rm{eff}} = 9000$--$30000$\,K, which we map onto spectral types using
the \citet{KH95} scale. In the minimisation procedure, we  
masked out the H$\beta$, and H$\gamma$ spectral regions that 
can be more strongly affected by emission/infilling, along with the 
spectral region between $\lambda\lambda$ 4400--4500\,\AA\, that includes
the diffuse interstellar band at $\lambda\,4428$\,\AA\ (often strong in 
our spectra, whilst absent from the model atmospheres). Including the 
Balmer jump helps the spectral-type 
determination, since the jump itself carries a $T_{\rm{eff}}$
dependence. Since the final output of the $\chi^{2}$ minimisation does
not allow a typing better than 3-subtypes on average, we classified
the stars put through method (ii) as one of Early-B (B0 to B3), Mid-B (B4 to
B6), or Late-B (B7 to A0), according to the range in which the measured 
minimum $\chi^{2}$ falls.  In cases where the FLWO~1.5m/FAST spectra
are relatively underexposed (S/N $\sim 10$), they are classified even more loosely as ``continuum+Balmer line emission''

The spectral types assigned to the FLWO-1.5m/FAST spectra and the S/N ratio at $\lambda\, 4500$\,\AA\, are given in Table\,\ref{t:parameters}
along with the spectral types and S/N ratio measured for the La Palma spectra in \citetalias{Raddi13}. 
\begin{figure}
\centering
\includegraphics[width=\linewidth]{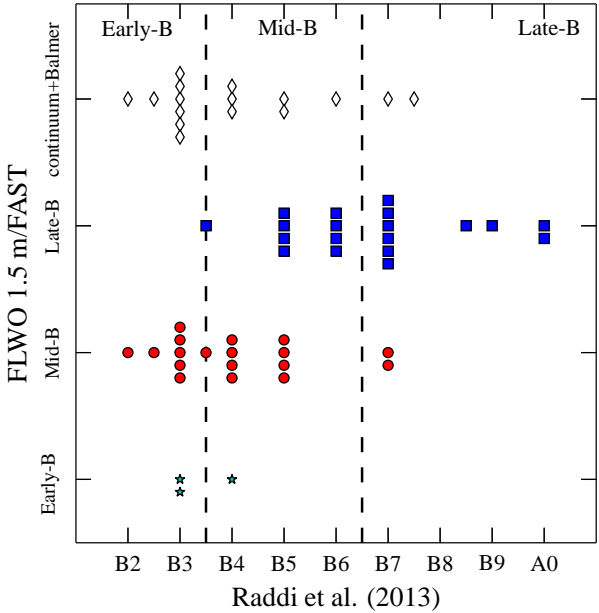}
\caption{Comparison between spectral types determined from La Palma 
spectra and from FLWO-1.5m/FAST observations, where $\rm{S/N} < 25$ (using
method ii). The latter are grouped as explained in the text.}
\label{f:spt_comparison}
\end{figure}
\begin{figure}
\centering
\includegraphics[width=\linewidth]{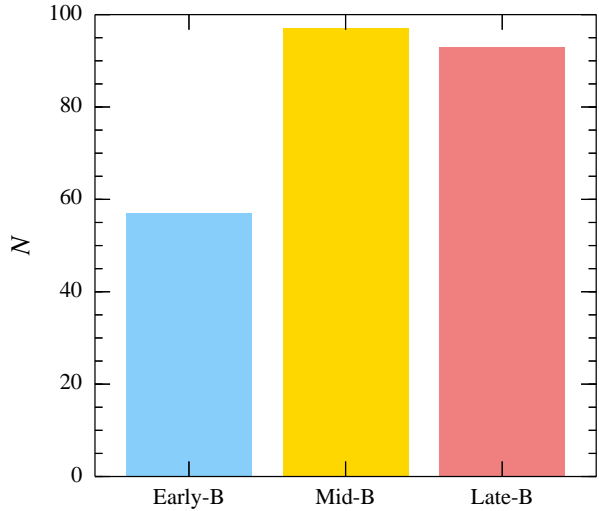}
\caption{Grouping of spectral types of the 247 classical Be stars in the catalogue. 
Early-B: B1--B3, Mid-B: B4--B6, Late-B: B7--A0.}
\label{f:hist-numbers}
\end{figure}

The comparison between the spectral types
that are determined with method (ii) and more precisely from 
\citetalias{Raddi13} 
La~Palma spectra is shown in Fig.\,\ref{f:spt_comparison}.  This offers
a useful insight into the dataset. The FLWO-1.5m/FAST classifications are broadly
consistent, with the expected large scatter and perhaps a slight
tendency to shift to later type. Interestingly, the spectra classified as 
``continuum+Balmer line emission'' are revealed by the superior La Palma data to be 
overwhelmingly early and mid B sub-type -- a finding that may fit with 
their being more distant, more highly-reddened (tougher to observe) 
objects.

To conclude, we display the spectral type distribution of the 247 classical Be
stars in Fig.\,\ref{f:hist-numbers}, arranging them as Early-B, Mid-B, and 
Late-B types. The distribution is fairly even between types,
with the Early-B group being less populous than the other
two, accounting for around a quarter of the sample. This modest 
favouring of later types stands in interesting contrast with the 
distribution by type present in the BeSS catalogue in which around
half of the objects are B3 or earlier \citep{Neiner11}.  This may
originate in small part from the spectral-typing bias we find
affecting the FLWO-1.5m/FAST spectra 
(Fig.\,\ref{f:spt_comparison}).  However it is also true that previous 
studies on Be stars have focused on brighter, magnitude-limited
samples, compared to our search that reaches to fainter magnitudes and
takes the surveyed volume more closely to the limits of the Galactic disc (see 
Section\,\ref{chap5}).  Furthermore, it is worth noting that the impact of 
interstellar extinction on our method of selection should be to favour 
earlier-type classical Be stars with stronger H$\alpha$ emission.   
The fact that now the earlier-type classical Be stars are a minority group hints that few
B0-B3 stars remain to be discovered at fainter magnitudes.

\section{Interstellar reddening}
\label{chap4}

In \citetalias{Raddi13}, we adopted a spectroscopic method for
measuring the $(B-V)$ colour excesses of 67 classical Be stars, from well
flux-calibrated spectra.  As part of the confirmation that these stars 
are classical Be stars, we also determined the $(B-V)$ colour excesses indirectly
by using IPHAS $r,i,\Halpha$ photometry.  The expectation that 
colour excesses inferred from IPHAS photometry were larger than the
direct blue/visual SED-fitting was confirmed -- indicating the
SED-reddening presence of circumstellar emission, usually interpreted
as evidence for discs around these stars. Paper~I
adapted existing methodologies for the correction of the measurements
for the effects of the characteristic superposed disc emission.  
Here we compute colour excesses for the entire FLWO-1.5m/FAST sample
using the photometric approach, because the spectra do not offer a reliable
flux calibration. 

The excess in the $(r-i)$ colour can be written as:
\begin{equation}
E(r-i) = (r-i) - (r-i)_{\circ}
\label{eq:1}
\end{equation}
The intrinsic-colour scale we adopt is the same we used in
\citetalias{Raddi13}. Because the red-optical SED of B-type stars
tends towards the Rayleigh-Jeans limit of the Planck function,
the intrinsic colour $(r-i)_{\circ}$ is not very sensitive to
effective temperature: across the complete B0 to A0 range, it changes
from $-0.17$ to 0.0.   This implies that even if spectral types
determined from FLWO-1.5m/FAST spectra are more uncertain than those
derived from better-resolved La Palma spectra, the intrinsic colour we 
assign to any one star will typically carry uncertainty of 
$\pm 0.05$.

The measured colour excess, $E(r-i)$, of Eq.\,\ref{eq:1}, includes a 
circumstellar contribution, $E^{cs}(r-i)$, along with the dominant
interstellar component, $E^{is}(r-i)$: 

\begin{equation}
E(r-i) = E^{is}(r-i) + E^{cs}(r-i),
\label{eq:2}
\end{equation}
as is true also of the $(B-V)$ colour excess \citep[e.g.][]{Dachs88, Kaiser89}.
The disc emission at red wavelengths produces a contribution to the
colour excess that is larger than the disc emission in the $(B-V)$
colour we corrected for in \citetalias{Raddi13}. In order to estimate  
the $E^{cs}(r-i)$, we use the grid of recombination continua that was 
modelled in \citetalias{Raddi13}, to represent the disc emission.
The grid of models has been generated for 10 different spectral types,
with the electron temperatures in the circumstellar disc scaling
to the stellar effective temperature according to 
$T_{e} = 0.6 \times T_{\rm{eff}}$ \citep[as is typically observed in both
hot-star winds and classical Be star discs,][]{Drew89, Carciofi06}, and for
finely spaced (0.01\,dex) disc fractions $f_D$.  The latter quantity
specifies the ratio between the disc flux and the total flux emitted
by the star and disc together at $\lambda\,5500$\,\AA. 
The circumstellar continuum excess, in $(r-i)$, is defined as:

\begin{equation}
E^{cs}(r-i) = (r-i)_{\rm{star+disc}} - (r-i)_{\rm{star}}.
\end{equation} 
The theoretical values for different combinations of $T_{e}$ and
$f_{D}$ are determined by convolving the IPHAS filters' profiles
with both the star+disc and star-only models. To model the stellar 
photospheres we chose the most appropriate models from the \citet{Munari05}
library. A sample of theoretical estimates is given in
Table\,\ref{t:corrections} as a function of the spectral type, or 
$T_{\rm{eff}}$, and $f_{D}$.
\begin{table}
\caption[]{Circumstellar colour excess $E^{is}(r-i)$, estimated for a 
given spectral type and disc contribution to the total flux.}
\centering
\begin{tabular}{@{}l l c c c c  @{}}
\hline
 SpT & $T_{e} (K)$  &   $f_{\rm{D}}$ = 0.05  &  $f_{\rm{D}}$ = 0.10 &
       $f_{\rm{D}}$ = 0.20 &  $f_{\rm{D}}$ = 0.30 \\
\hline
B1   & 18000 &  0.055 &0.106 &0.194 &0.269 \\ 
B3   & 13200 &  0.063 &0.119 &0.216 &0.297 \\  
B5   &  9300 &  0.071 &0.134 &0.240 &0.328 \\  
B7   &  7800 &  0.082 &0.153 &0.272 &0.367 \\  
A0   &  5700 &  0.112 &0.205 &0.352 &0.464 \label{t:corrections}\\  
\hline
\end{tabular}
\end{table}

As in \citetalias{Raddi13}, we estimate the appropriate circumstellar colour excesses for each star by linking the disc fraction to the $EW(\Halpha)$, 
by means of the empirical equation:
\begin{equation}
f_{D} = 0.1 \times \frac{EW(\Halpha)}{-30\, \rm{\AA}},
\label{eq:dachs}
\end{equation}
first proposed by \citet{Dachs88}. The scatter associated with
Eq.\,\ref{eq:dachs} is about $\sim 0.02$\,dex.  To maintain
contemporaneity between the measures of $E(r-i)$ and $f_{D}$, we estimated 
$EW(\Halpha)$ by interpolating the IPHAS colours between the theoretical
curves of growth that were plotted in Fig.\,\ref{f:ccd_iphas}
and computed in \citet{Drew05}. In Table\,\ref{t:parameters}, we give
all the derived and measured quantities used in obtaining the
estimates of $E^{cs}(r-i)$ and $E^{is}(r-i)$ via Eq.\,\ref{eq:1}--\ref{eq:dachs}. 
Since interstellar reddenings are generally expressed as the colour
excess in the $(B-V)$ colour, we compute it using
the following transformation:
\begin{equation}
E^{is}(B-V) = \frac{E^{is}(r-i)}{0.65},
\end{equation}
where 0.65 is the monochromatic conversion-factor\footnote{The conversion factor is 5 per cent smaller than the 
band-averaged value of 0.69 that we used in \citetalias{Raddi13}.}, computed from the
\citet{Fitzpatrick99} $R_V = 3.1$ reddening law. 
The errors on $E^{is}(B-V)$ are typically
in the range of 0.05--0.1\,mag. They combine the IPHAS photometric
uncertainty, an uncertainty on spectral type, the astrophysical spread
of intrinsic colours for MS stars \citep{Houk97}, and the uncertainty
arising from the estimate of disc fraction.
\begin{figure}
\centering
\includegraphics[width=\linewidth]{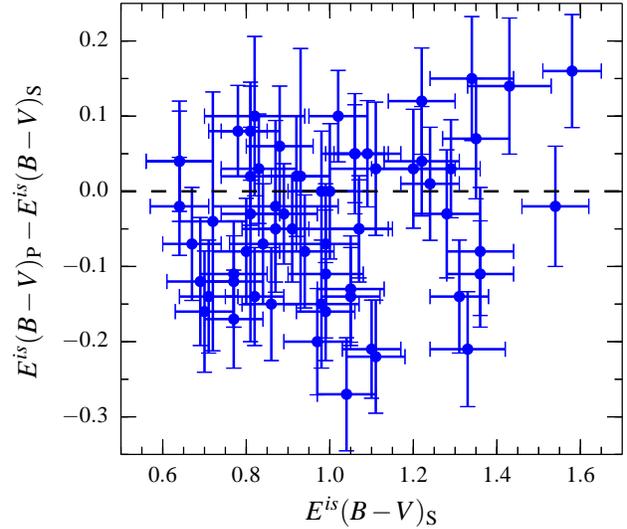}
\caption{The difference between the two estimates of interstellar reddenings, for the 63 classical Be stars with La Palma spectra, are plotted against
the $E^{is}(B-V)_{\rm{S}}$ determined via SED fitting to the blue spectral region \citepalias{Raddi13}. The equality line is also shown. The mean difference is found
to be $\sim -0.04 \pm 0.01$. }
\label{f:corrected_difference}
\end{figure}

For 63 of the 67 classical Be stars that were observed in La Palma, the
interstellar reddenings derived in the first place via the SED fitting
procedure of \citetalias{Raddi13} are compared to the photometric
procedure described here, in Fig.\,\ref{f:corrected_difference}. From
the comparison, we notice a systematic negative difference: the mean 
offset being $\Delta E^{is}(B-V) = -0.04 \pm 0.01$, with a sample
standard deviation of 0.1\,mag.  This is just significant at the $3\sigma$
level, and it may connect to the impression noted in Section 2 that
the spectral type estimates based on FLWO-1.5m/FAST data are inclined to be 
later than those based on La Palma spectra.  Nevertheless the bias is 
not large and we will recall it in Section 5, when appropriate.  
The level of scatter apparent is a little high relative to the
individual errors on $E(B-V)$ and may in part be due to the application
of a single reddening law.
\begin{figure}
\centering
\includegraphics[width=\linewidth]{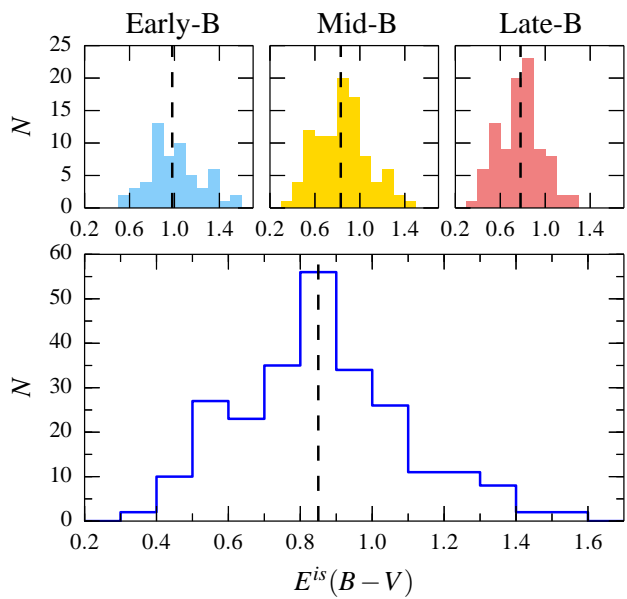}
\caption{Histogram distribution of the photometric colour excess. 
In the top panels, objects are sorted by spectral-type groups, while
the bottom panel shows the combined distribution. The overall mean is found at $E^{is}(B-V) = 0.86 \pm 0.23$,
and the dashed lines mark the median of each distribution.}
\label{f:reddenings_histograms}
\end{figure}
\begin{figure*}
\centering
\includegraphics[width=\linewidth]{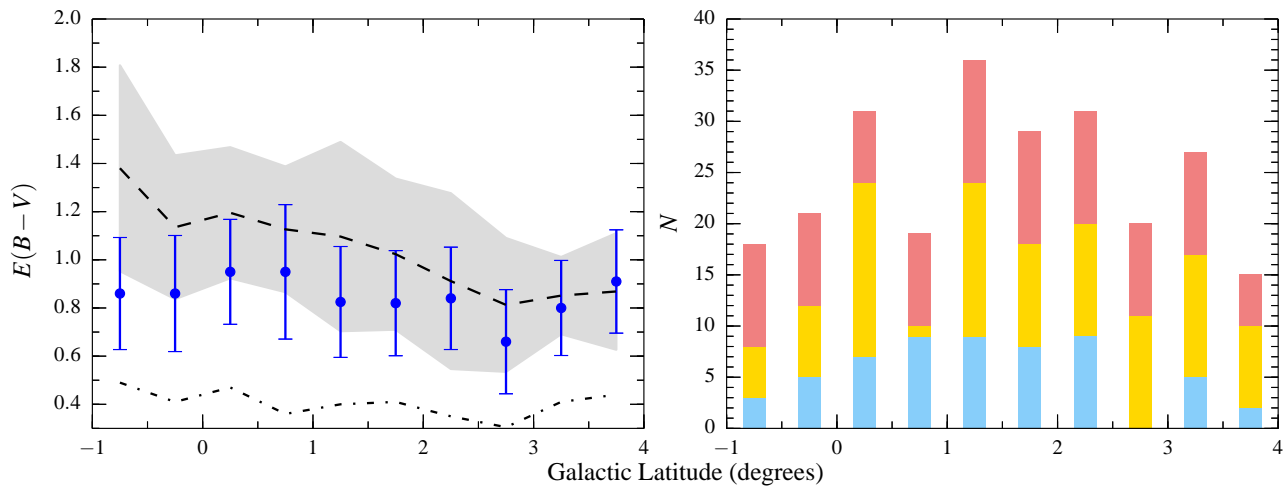}
\caption{Left panel: mean interstellar reddenings derived from the 247 classical Be stars in the catalogue on
  binning up into 0.5\,deg strips parallel to the Galactic plane
  (marginalising over longitude). The error bars represent the 1-$\sigma$ dispersion around the mean values.
The upper dashed-curve and the shaded area trace the mean \citetalias{SFD98} values in each strip and the relative 1-$\sigma$ dispersion, while the lower dot-dashed curve follows the
mean \citetalias{RF09} values in each strip. Right panel: histogram bars representing the number of classical Be stars in each
latitude bin, colour coded accordingly to the spectral classification (light blue: Early-B; gold: Mid-B; Late-B: pink).}
\label{f:reddenings_latitude}
\end{figure*}

The distribution of derived $E^{is}(B-V)$ is shown in 
Fig.\,\ref{f:reddenings_histograms}. The overall mean 
colour excess is 0.86\,mag, with a sample dispersion of 
0.23\,mag. We first note that the median $(B-V)$ colour excess, measured for 
the group of Early-B stars, is found at 0.98\,mags that is larger than the median values
of $E^{is}(B-V) = 0.83,\ 0.78$ obtained for the Mid-B and Late-B groups.
This difference is not surprising, as the intrinsically brighter group of stars can be
detected behind a larger column of dust. We also note that the stars observed in La Palma
are on average more reddened and therefore fainter (median $E^{is}(B-V) = 1$, and $r = 14.7$) 
than the FLWO-1.5m/FAST objects, for which the 
distribution medians are $E^{is}(B-V) = 0.81$ and $r = 14$. 
This difference arising from the larger aperture offered by the INT with 
respect to the FLWO-1.5m. The
median contribution to the $(r-i)$ colour excess due to the
circumstellar disc is 0.11 mag, reflecting a median disc fraction,
$f_D$, of 0.07.

\section{Inferences on the spatial distribution of the classical Be stars sample from interstellar extinction}
\label{chap5}
Armed with approximate spectral types, we have derived interstellar
colour excesses across the sample to a median precision of 
$\Delta E(B-V) = 0.08$, after making the necessary correction for 
circumstellar emission.
This typical uncertainty is in the region of 10 per cent of the typical 
reddening towards these stars.  It allows more than enough precision to
make a broad comparison with existing measures of integrated
extinction in the area in order to gain a first impression of how the
population of classical Be stars is distributed.  To achieve this, we turn to
comparisons with three comprehensive studies of the pattern of 
Galactic-Plane extinction -- namely, \citetalias{SFD98},
\citetalias{RF09}, and \citet{Sale14}.

To begin with, we recall some constraints from studies of open clusters.
The lowest classical Be star reddenings obtained here are compatible with less obscured
parts of the Perseus Arm, such as the sightline to the open cluster NGC\,637 
with $E(B-V) \sim 0.6$ \citep[][]{MM07}.  The average colour excesses
resemble those measured for more prominent clusters
in the Perseus Arm, such as $E(B-V) \simeq 0.8$\,mag for both IC\,1805
\citep*[][]{Massey95} and NGC\,663 \citep[][]{Pandey05}.
If we consider all the open clusters in the \citet{Dias02}
catalogue seen in this part of the Galactic Plane, we find that clusters
in the range of distances that is regarded as typical of the Perseus Arm
\citep*[i.e.\ 2--3\,kpc][]{Russeil07, Vallee14} are also reddened on
average by 0.8\,mags, although the full range runs from 0.3 up to 1.2.
The high-end extinctions tend to be associated with IR-identified
clusters.  Clusters listed by \citet{Dias02} at greater distances have
colour excesses only a little larger, spanning 0.5 up to 1.3 mag.  In
view of this it is reasonable to anticipate that the great majority of classical Be
stars identified here are at least as distant as the Perseus Arm, and
some may be appreciably more distant still. 

The left panel of Fig.\,\ref{f:reddenings_latitude} illustrates the binned distribution of
reddenings per Galactic latitude range. The stars have been
grouped into strips 0.5\,deg wide in Galactic latitude and the median and 
standard deviation of the reddenings obtained within each is shown. We 
also plot the median values within these same strips of the total
Galactic reddening obtained from \citetalias{SFD98} and that inferred
by \citetalias{RF09} from median 2MASS near-IR colour excesses, along the
same sets of sightlines.   The \citetalias{SFD98} results have been 
adjusted using the finding of \citet{Schlafly10} that they should be 
scaled down by multiplying by a factor of 0.86.
In the left panel of Fig.\,\ref{f:reddenings_latitude} 
it can be seen that there is only a 
$\sim 0.25$\,mag difference between the maximum classical Be star median reddening, 
obtained between $b = 0^{\circ}$--$0.5^{\circ}$, and the minimum measured in
the $b = 2.5^{\circ}$--$3^{\circ}$ strip.  It was noted above that
the overall sample mean is $E^{is}(B-V) = 0.86$.  The variation is within
the 1-$\sigma$ dispersion present in each latitude strip, indicating no
strong overall trend. 

It is striking, however, that the colour excesses obtained from
\citetalias{RF09} are systematically $\sim$0.5\,mag lower than the ones we
measure for the classical Be stars. Even at the level of comparison between the
individual objects in our catalogue and the \citetalias{RF09} data,
there are only 3 for which the latter supplies a higher reddening.
The discrepancy calculated on an object-by-object basis is $0.45 \pm
0.01$ mag with a dispersion of 0.2 mag. In contrast, the distribution
of corrected \citetalias{SFD98} colour excesses overlaps or exceeds the
general run of our classical Be-star measures. The latter result is expected
for the reason that the \citetalias{SFD98} maps measure the asymptotic 
reddening along a given sightline -- the maximum colour excess
expected for a star behind the Galactic dust column.

Given that $E(B-V)$ to the least obscured Perseus Arm open clusters
is only infrequently as low as 0.4\,mag \citep{Dias02}, it
would seem that the extinction measured with the method of \citetalias{RF09} is
limited to sampling the column of dust between the Sun and the Perseus
Arm -- a half or less of the likely total Galactic column.
This is in part attributable to the reliance on median near-IR extinction,
and may also be a consequence of the relatively bright magnitude
limits of the 2MASS survey.  Unfortunately the UKIDSS surveys do not
reach to this part of the Galactic plane, so it cannot be established 
whether the \citetalias{RF09} method would be sensitive to more of the 
column if it were applied to deeper near-IR data. The main point for now
is that, despite its excellent angular resolution (2--3\,arcmin), the 
map due to \citetalias{RF09} does not provide a useful comparison with
our catalogue of classical Be stars.
\begin{figure*}
\centering
\includegraphics[width=\linewidth]{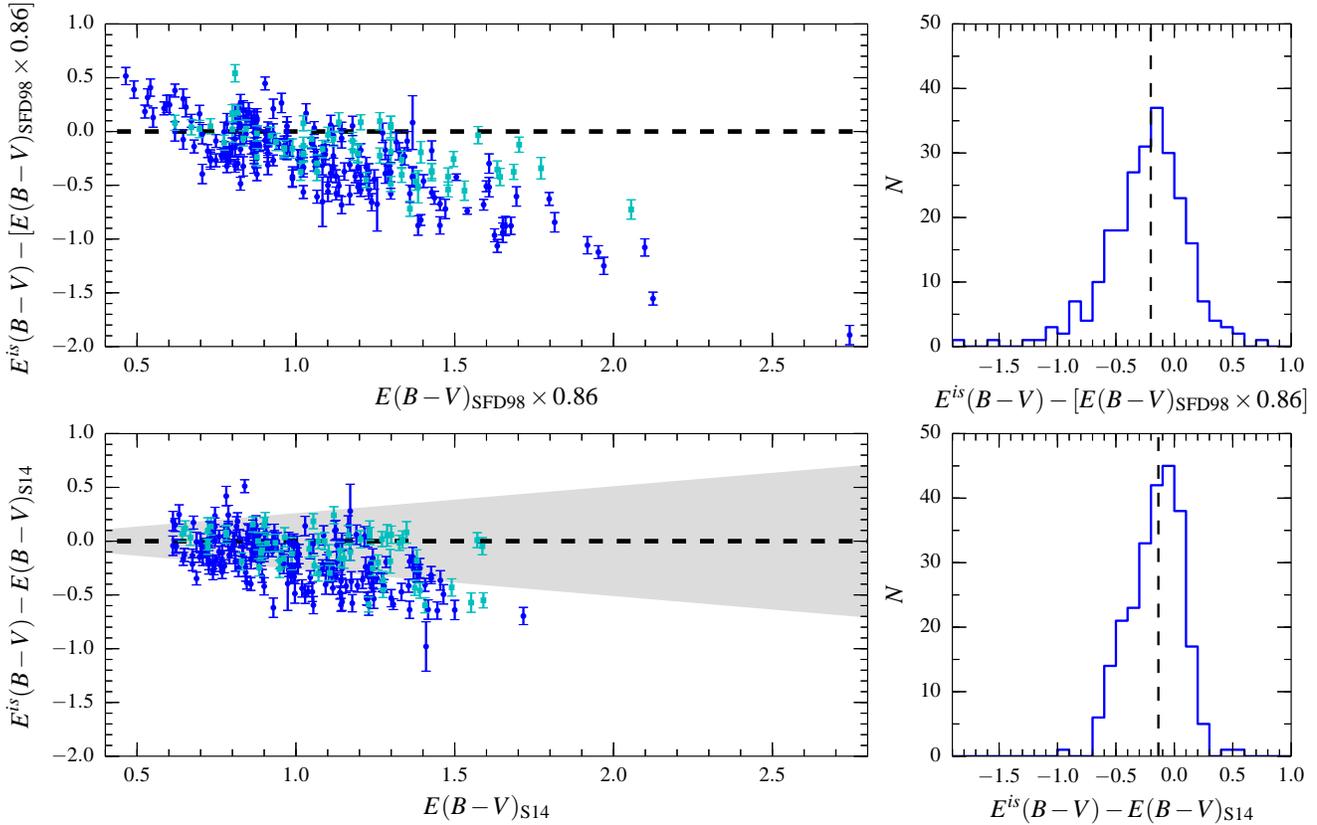}
\caption{Upper panels: on the left, comparison between the colour excesses we
  measure for the classical Be
  stars and the values from \citetalias{SFD98} for the
  same sightlines (after applying the \citet{Schlafly10} correction).  
  Squares with error bars are used for the stars in
  \citetalias{Raddi13}, while circles are plotted for the new classical Be
  stars presented here. The dashed line represents the equality line.
 On the right, is an histogram-plot displaying the collapsed difference between
\citetalias{SFD98} and the colour excesses we measure for the classical Be stars. The dashed vertical line
marks the median of the distribution at $-0.2$\,mag.
  Lower panel: on the left, the same comparison with respect to the colour excess measured at the 90th percentile of the distance
distribution from the 3D extinction-distance map of \citet{Sale14}.  The shaded
  area roughly traces the scatter caused by differential extinction 
  \citep[cf. fig.\,2 and 6 in][]{Sale14}. On the right, the difference between
the x- and y-axis is collapsed into histogram bars, with the median of the distribution at $-0.14$\,mag
as traced by the vertical dashed line.}
\label{f:extin_comparison}
\end{figure*}

In the right panel of Fig.\,\ref{f:reddenings_latitude}
we plot the number distribution of classical Be stars per latitude bin, distinguishing
the stars belonging to different spectral type groups. First, we
notice that the stars in the catalogue are more concentrated between 
$b = 0^{\circ}$--$3^{\circ}$.   Each bin includes a range of spectral
types that in turn will be characterised by different extinction and distance
ranges (given the common magnitude bounds used in selection, see
figure 10). The overall latitude distribution crudely follows a broad
Gaussian peaking near $b \simeq +1.5^{\circ}$, such that there is a 
difference of 21 stars between the most and the least populous bins.
In detail, smaller number counts are observed between $b = 0.5^{\circ}$--$1^{\circ}$
and between $b = 1.5^{\circ}$--$2^{\circ}$, due to the masking effect of
the larger dust columns located near W4/W5 and Cas\,OB\,7.  The Early-B
type classical Be stars seem to favour the $b = 0^{\circ}$--$2.5^{\circ}$ range,
as one would expect for an intrinsically-brighter population
able to trace the warp: Freudenreich et al 2004 showed this to lie in
the $1 \lesssim b \lesssim 2$ latitude range at these longitudes (in
HI and dust emission).  The less-luminous later spectral 
types, at shorter distances, show a flatter distribution.

To permit a more detailed comparison between our measured reddenings and the
corrected \citetalias{SFD98} values, we plot the difference between
  them against the SFD98 values on a star-by-star basis in the
upper-left panel of Fig.\,\ref{f:extin_comparison}.  
The outcome is not qualitatively different from the results for the
smaller sample presented by \citetalias{Raddi13}, although there are 
now more stars that are observed with reddenings significantly 
exceeding $[E(B-V)_{\rm{SFD98}} \times 0.86]$. 
Paper~I found two stars (\#\.164 and 167 here) located
well above the equality line -- here, these are joined by 10 more stars
(i.e.\ \# 161, 162, 166, 182, 184, 221, 225, and 232). These stars are
the most extreme with $E^{is}(B-V) - E(B-V)_{\rm{SFD98}}\times 0.86 >
  +0.3$\,mag in the upper-right histogram in Fig.\,\ref{f:extin_comparison}.
The most likely origin of this problem is the coarser angular
resolution of the dust-temperature data ($\sim 1$\,deg) used by 
\citetalias{SFD98} to infer the dust column from far-infrared
emissivity data ($\sim 6$\,arcmin angular resolution): in the vicinity
of dust warmed by strong H~{\sc ii} region emission in the Galactic Plane, 
this results in underestimation of the dust column \citep[see][ and
  \citet{Sale14}]{Schlafly10}. If we count the number of stars
with reddenings that are the same or larger than the \citetalias{SFD98}
measure, less our typical 0.1\,mag uncertainty, 
$\sim 37$ per cent (91 stars) seem to be lying at or beyond the
  limits of the dust disc.  We measure the median of the distribution
of differences to be $-0.2$\,mag, with a standard deviation of 0.35\,mags. 

We now turn to the IPHAS-based extinction map of Sale et al (2014),
based on $r$, $i$ and H$\alpha$ data on stars of spectral type earlier than K, to
see if this particular limitation of \citetalias{SFD98} at low
galactic latitudes can be overcome.  In the lower-left panel of 
Fig.\,\ref{f:extin_comparison}, the measured reddening of each star is 
plotted against the reddening obtained at the 90th percentile of the 
distance distribution derived from the nearest-neighbour
extinction-distance relationships of \citet[][see fig.\,3]{Sale14}.  
The result is tidier in that the lowest estimated total colour 
excesses have disappeared and fewer classical Be star reddenings lie far above the 
equality line.  The angular resolution of this map in our sky area is 
mostly 10 or 15\,arcmin, rising to 30 arcmin in the vicinity of the 
dark clouds, W3 and W5 (where there are not so many selected stars). 

The fractal nature of the ISM imposes a differential
extinction \citep[cf. fig.\,6 in][]{Sale14} of up to 25 per cent in the 
most distant bins reached by the map (expressed as a standard
deviation, and shown as the shaded area in the bottom left panel of 
Fig.\,\ref{f:extin_comparison}). 
From inspection, we can identify 166 objects (i.e.\ 67 per cent of the
sample) with reddenings that fall within the shaded area, that may be
regarded as consistent with the reddenings at the 90th percentile of
the distance in the extinction-distance relationships
of \citet{Sale14}.  This sets a generous upper limit on the fraction
of the classical Be stars sample likely to lie beyond most of the dust column -- to be 
compared with 37 per cent, deduced from the \citetalias{SFD98} comparison.
We also notice that the binned distribution of differences between the 
reddenings we measure and the colour excess derived from the 3D
extinction map peaks at $-0.14$\,mag (lower-right panel in 
Fig.\,\ref{f:extin_comparison}) and its standard deviation is just 0.23\,mags. 
We recall that our photometric method is inclined toward slight under- 
rather than over-estimation of classical Be-star interstellar colour excess
(see Fig.\,\ref{f:corrected_difference}).  

However, there is a modest systematic offset between the maximum extinction 
estimates of \citetalias{SFD98} and \citet{Sale14}.
For the sightlines relevant here, the median difference between the
two measures is 7 percent, in the sense that the scaled \citetalias{SFD98}
results are 1.07 times the \citet{Sale14} values.  The most obvious 
expression of this difference in Fig.\,\ref{f:extin_comparison} is the 
longer tail of objects toward higher colour excess in the upper
left panel presenting the \citetalias{SFD98} comparison that is absent
from the \citet{Sale14} comparison.  If we attempt
a simple correction for this offset by multiplying the \citet{Sale14} 
maximum reddenings by 1.07, before considering again how many classical Be stars have
reddenings compatible with the Galactic maximum, we find the fraction
of the classical Be stars sample that may be located beyond the Galactic disc dust column drops 
from 67 per cent to 58 per cent.  

Of the three estimates (37, 58, and 67 per cent) of the fraction of the sample
of classical Be stars lying behind most or all the dust column, we favour the first two.
In rough terms there is a case to argue that about half of the sample fits this description.

The seeming downward trend in the data in the left panels of 
Fig.\,\ref{f:extin_comparison} has its origins in the magnitude limits 
imposed on the present sample of classical Be stars.  It can be seen that at small
asymptotic $E(B-V)$ the observed points cluster more tightly to the
equality line than at larger asymptotic $E(B-V)$: only classical Be stars
brighter than $r \sim 13^{{\rm th}}$ {\em could} be less reddened.  
In contrast, along the most reddened sightlines, nearer classical Be stars are 
picked up that present colour excesses falling progressively
further below the estimate of the asymptotic limit.  This is 
the faint magnitude limit, at $\sim16^{{\rm th}}$ magnitude,
expressing itself.  But in both distributions, it is noteworthy 
that up to a $(B-V)$ colour excess of 1.4\,mags, there are classical Be stars with reddenings clearly compatible with 
these measures of maximum reddening.  We deduce from this that, along 
sightlines where the total Galactic colour excess is less than 1.4\,mags, 
the classical Be stars sample is either weakly or not at all extinction-limited.

In principle, we can use the \citet{Sale14} 3D extinction-distance 
relations to provide a guide to where the dust layer runs out.  The
reader is referred to the data available on the IPHAS survey website
(\url{www.iphas.org}) to investigate any sightline of interest across this 
region.  Nevertheless, it is not recommended to use these data to estimate 
distances to individual classical Be stars -- the uncertainties presently 
involved are too large to make this worthwhile.  Instead, a very
simple example of spectroscopic parallax illustrates the point clearly
that typical distances to them will be large, reaching beyond the
Perseus Arm at 2--3\,kpc: specifically, a Mid-B star \citepalias[$M_r <-0.5$,][]{Raddi13} 
of median brightness in the sample ($r \simeq 14.5$), experiencing a median extinction of 
$A_r \sim 2.2$, cannot lie
at a shorter distance than 3.6\,kpc. The faintest most reddened stars
of the sample will lie at distances exceeding 6\,kpc.  
If there is the coherent structure at these longer distances, such as
the Outer Arm, then it is possible that a significant portion of
the newly-identified classical Be stars, lying behind all or most of the 
Galactic dust column, are associated with it. 

A similar conclusion was reached in 
\citetalias{Raddi13} based on the smaller sample benefitting from
higher quality spectroscopy.  Indeed there it was argued that 
10--15 stars in the smaller sample of 66 objects may even lie outside 
the Galactocentric radius of 13\,kpc (or heliocentric distances larger 
than 7\,kpc), where the Galactic stellar density gradient is expected
to be steepening \citep[][]{Ruphy96, Sale10}.  Without question, this catalogue of classical Be stars  
adds many candidates to the handful of O stars proposed by other authors as 
belonging to the Outer Arm \citep[e.g.][]{Negueruela03}.

\begin{figure*}
\centering
\includegraphics[width=\linewidth]{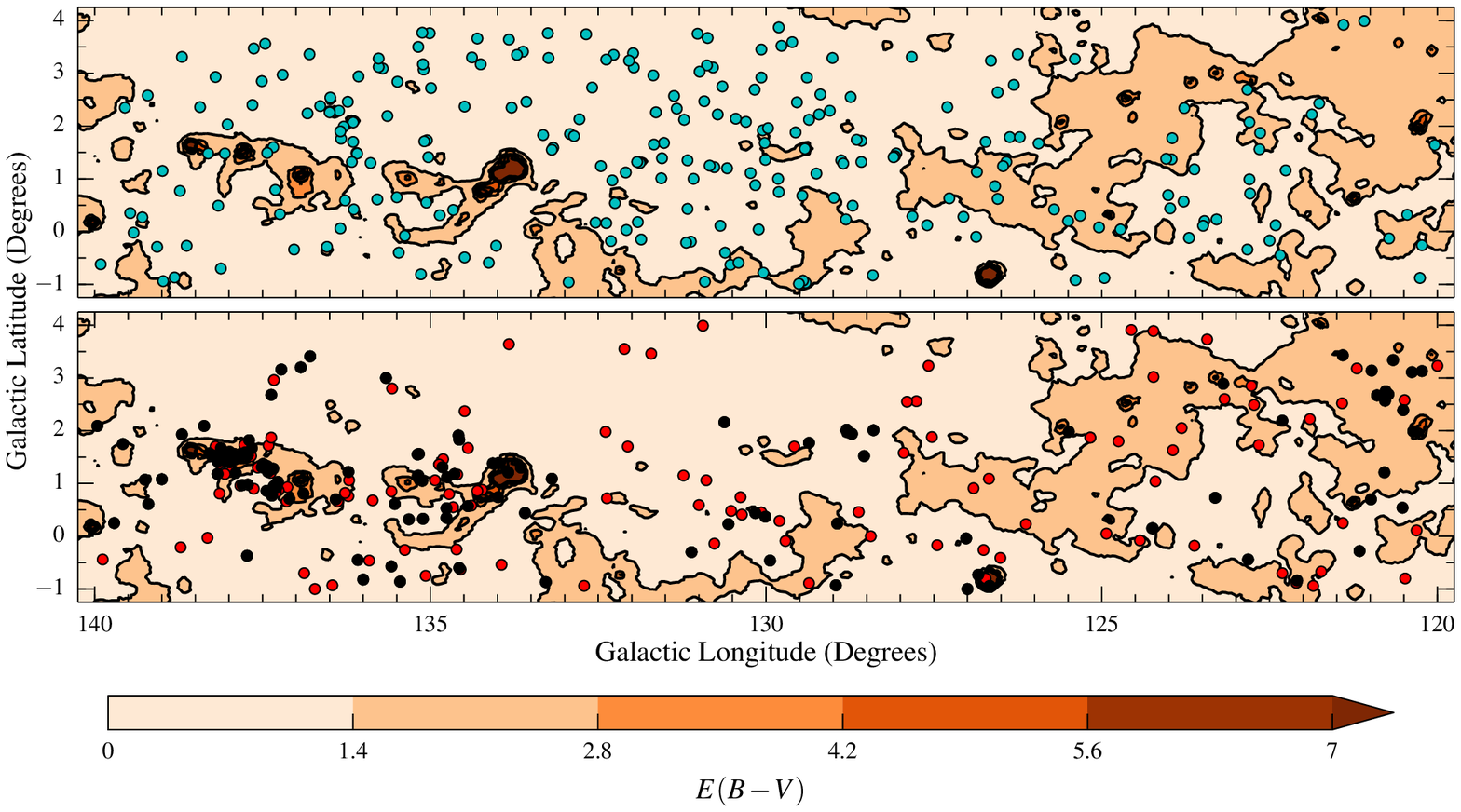}
\caption{The spatial distributions of the 247 classical Be stars studied here (upper panel)
and the 250 $\Halpha$ emitters from \citet{Witham08} (lower panel) with $16.5 < r < 18$ (red dots) and $r \geq 18$ (black dots) are plotted on the map of 
 the interstellar dust extinction computed by \citetalias{SFD98}, corrected accordingly to \citet{Schlafly10}. 
The extinction map is binned in 1.4\,mag steps, colour coded, and cut at $E(B-V) = 7$.
Comparing the two panels, we notice the higher order of clustering towards areas of higher extinction
for the faintest stars from \citet{Witham08}, while fewer remain
unexplored and unclassified in areas where the maximum colour excess is
below 1.4\,mags. In fact, we have found that our sample of classical Be
stars seem not to be significantly limited by total colour excesses of
up to this limit. The remaining scattered candidate emission line
stars down to $r = 18$ in these locations may prove to be later-type
classical Be stars.}
\label{f:faint_map}
\end{figure*}
\section{Summary and concluding remarks}
\label{chap6}

We have presented an expanded catalogue of 247 classical Be stars, drawn from
a $5 \times 20$\,deg$^2$ section of the Galactic Plane
including the Perseus Arm and its much-studied W3--5 star-forming
clouds.  Close to three quarters of the included stars are newly 
identified, with most of the rest taken from the more detailed study 
by \citetalias{Raddi13}. Our
list also includes: 3 stars (\#\,73, 74, and 76) that were already 
recognised via spectroscopy as classical Be stars in the open cluster NGC\,663 
\citep[see e.g.][]{Mathew11}; 11 objects that had already been
designated as $\Halpha$ emitters from objective prism spectroscopy 
\citep[][]{KW99}; 6 stars that were classified as either class II or 
class III SEDs by \citet{Koenig08}, according to their near- and
mid-IR colours. This work moves the magnitude limit down to 
$r \simeq 16$ from 12--13 in the BeSS catalogue \citep{Neiner11}. 

We can already conclude that the number of classical Be stars still awaiting
discovery at fainter optical magnitudes in this part of the Galactic
Plane is not large.  This is shown by Fig.\,\ref{f:faint_map} in 
which the spatial distributions of candidate emission line stars
brighter and fainter than $r = 16.5$ (the latter being drawn 
from \citet{Witham08}) are compared. The
character of the faint distribution is very different from that of 
the bright distribution: at brighter magnitudes, the confirmed classical Be stars are
distributed in a loose unclustered fashion, while the faintest 
candidate emission line stars $(r > 18)$ are evidently much more clustered  
around the star-forming regions associated with high extinction.  
These objects will mainly be T~Tauri
and Herbig Ae/Be stars -- leaving just a few tens of objects scattered
across the general field as the more probable more mature classical Be
stars in the magnitude range $16.5 < r < 18$.  The job is largely done, and our enlarged list of classical Be stars -- 
set out in the Appendices\footnote{Also available for download at \url{www.iphas.org}.} -- represents a 
significant step
towards a full census of the fainter classical Be stars in this part of the 
Galactic Plane.  It is likely that the undiscovered objects will be
preferentially later type classical Be stars.

We have determined spectral types with an uncertainty ranging between
1 to 3 subtypes, depending on the S/N of the spectra available.
In contrast to brighter samples, we find that later-type classical Be stars
are at least as well represented as the earlier sub-types. 
Even with only approximate spectral sub-types, we compute reddenings 
that are on average good to 10 per cent.  The photometric method
needed to compute the corrected colour excesses, $E^{is}(B-V)$, from 
$r$, $i$ and $\Halpha$ IPHAS data only, has been shown to produce
satisfactory results when compared to alternate determinations using 
blue data, that suffer less contamination by circumstellar
disc emission. 

We estimate that perhaps half of the sample, as shown by
  comparison with the corrected map of \citet{SFD98} and the new map
  due to \citet{Sale14}, is compatible with heliocentric distances of
4--5\,kpc or more.  

The major remaining challenge is to pin down the distances to these 
usefully luminous objects.  For this we look to the future, and the eventual
release of {\em Gaia} parallaxes that will enable our catalogue of classical Be stars
to be mapped out across the outer Galactic Plane.  From the pre-launch
performance data presented by \citet{deBruijne12}, the end-of-mission
parallax precision for $14 < r < 16$ is expected to be in the region of
20--30 $\mu$as, or a relative error of $\sim 10$ per cent at
distances of $\sim$5\,kpc.  This will be sufficient to grasp at last
how the outer Galactic disc is organised.

\section*{Acknowledgements}
This publication make use of spectroscopic data that were obtained
at the FLWO-1.5m with FAST, which is operated by
Harvard-Smithsonian Centre for Astrophysics. 
In particular, we would like to thank Perry Berlind and Mike Calkins for 
their role in obtaining most of the FLWO-1.5m/FAST data. 

This paper makes use of data obtained as part of the INT Photometric $\Halpha$ Survey of the Northern Galactic Plane 
(IPHAS, \url{www.iphas.org}) carried out at the Isaac Newton Telescope (INT). The INT is operated on the island of 
La Palma by the Isaac Newton Group in the Spanish Observatorio del Roque de los Muchachos of the Instituto de Astrofisica de Canarias. 
All IPHAS data are processed by the Cambridge Astronomical Survey Unit, at the Institute of Astronomy in Cambridge. The catalogue presented in this work
was assembled at the Centre for Astrophysics Research, University of Hertfordshire, supported by a grant from the Science \& Technology Facilities Council of the UK (STFC, ref
ST/J001335/1).

This publication makes use of data products from the Two Micron All Sky Survey, 
which is a joint project of the University of Massachusetts and 
the Infrared Processing and Analysis Center/California Institute of Technology, 
funded by the National Aeronautics and Space Administration and the National Science Foundation.
This publication makes use of data products from the Wide-field Infrared Survey Explorer, 
which is a joint project of the University of California, Los Angeles, and the 
Jet Propulsion Laboratory/California Institute of Technology, funded by the National Aeronautics and Space Administration.

The research leading to these results has received funding from the
European Research Council under the European Union's Seventh Framework
Programme (FP/2007-2013) / ERC Grant Agreement n.~320964 (WDTracer).
JED, GB, and SES acknowledge support from
the STFC of the United
Kingdom (JED and GB ST/J001333/1, SES
ST/K00106X/1). NJW is in receipt of a Fellowship funded by the Royal Astronomical
Society of the United Kingdom. JF is supported by the Spanish Plan Nacional de
I+D+i and FEDER under contract AYA2010-18352. 

\bibliographystyle{mn2e}


\appendix
\section{Photometric data}
\begin{table*}
\tiny
\begin{minipage}{180mm}
\caption[]{Optical, near-IR and mid-IR photometry of the classical Be stars in the catalogue. The columns list: the star ID number; 
name based on the IPHAS coordinates; Galactic coordinates ($\ell$, $b$);
IPHAS $r$ magnitudes and $(r-i)$ and $(r-\Halpha)$ colours; 2MASS $J$ magnitudes and $(J-H)$ and $(H-K)$ colours; 
WISE $W1$ magnitudes and $W1-W2$, $i-W1$, and $J-W2$ colours. The tick symbols in
the last two columns indicate if the object has a La Palma spectrum and/or a FLWO-1.5m/FAST spectrum. 
Typical uncertainties in IPHAS magnitudes and colours are 0.02 and 0.03\,mag respectively.}
\centering
\begin{tabular}{@{}llrrrrrrrrrrrrcc@{}}
\hline
\# & Name & $\ell$ & $b$ & $r$ & $(r-i)$ & $(r-\Halpha)$ & $J$ & $(J-H)$ & $(H-K)$ & $W1$ & $(W1-W2)$ & $(i-W1)$ & $(J-W1)$ & LP & FF \label{t:photometry} \\
 & Jhhmmss.ss+ddmmss.s & (deg) & (deg) & (mag) & (mag) & (mag) & (mag) & (mag) & (mag) & (mag) & (mag) & (mag) & (mag) &  &   \\
\hline
1 & J002441.72+642137.9 & 120.04 & $1.64$ & 14.75 & 0.83 & 0.61 & $12.51 \pm 0.02$ & $0.30 \pm 0.04$ & $0.26 \pm 0.04$ & $11.79 \pm 0.02$ & $0.11 \pm 0.03$ & $2.13 \pm 0.03$ & $0.72 \pm 0.03$ & \checkmark &  \\
2 & J002758.97+622906.1 & 120.23 & $-0.26$ & 15.03 & 0.59 & 0.46 & $13.30 \pm 0.02$ & $0.34 \pm 0.04$ & $0.24 \pm 0.05$ & $12.32 \pm 0.02$ & $0.26 \pm 0.03$ & $2.12 \pm 0.03$ & $0.99 \pm 0.03$ &  & \checkmark \\
3 & J002843.24+615216.2 & 120.26 & $-0.88$ & 14.38 & 0.48 & 0.42 & $13.16 \pm 0.02$ & $0.17 \pm 0.04$ & $0.20 \pm 0.05$ & $12.58 \pm 0.03$ & $0.18 \pm 0.04$ & $1.32 \pm 0.03$ & $0.58 \pm 0.03$ &  & \checkmark \\
4 & J002926.93+630450.2 & 120.45 & $0.32$ & 14.07 & 0.40 & 0.36 & $13.11 \pm 0.02$ & $0.12 \pm 0.04$ & $0.18 \pm 0.04$ & $12.87 \pm 0.03$ & $0.01 \pm 0.04$ & $0.80 \pm 0.03$ & $0.24 \pm 0.03$ & \checkmark & \checkmark \\
5 & J003210.31+623929.2 & 120.72 & $-0.13$ & 13.39 & 0.47 & 0.51 & $12.33 \pm 0.02$ & $0.22 \pm 0.04$ & $0.15 \pm 0.04$ & $11.37 \pm 0.02$ & $0.32 \pm 0.03$ & $1.55 \pm 0.03$ & $0.96 \pm 0.03$ &  & \checkmark \\
6 & J003248.02+664759.6 & 121.09 & $3.99$ & 14.41 & 1.02 & 0.85 & $12.11 \pm 0.02$ & $0.55 \pm 0.03$ & $0.36 \pm 0.03$ & $10.45 \pm 0.02$ & $0.29 \pm 0.03$ & $2.93 \pm 0.03$ & $1.66 \pm 0.03$ & \checkmark & \checkmark \\
7 & J003559.27+664503.3 & 121.40 & $3.92$ & 15.85 & 0.71 & 0.60 & $14.11 \pm 0.04$ & $0.40 \pm 0.06$ & $0.46 \pm 0.06$ & $12.30 \pm 0.03$ & $0.50 \pm 0.03$ & $2.85 \pm 0.03$ & $1.82 \pm 0.05$ & \checkmark & \checkmark \\
8 & J004014.89+651644.0 & 121.76 & $2.43$ & 14.82 & 0.68 & 0.77 & $12.95 \pm 0.02$ & $0.29 \pm 0.04$ & $0.26 \pm 0.05$ & $12.09 \pm 0.02$ & $0.22 \pm 0.03$ & $2.05 \pm 0.03$ & $0.86 \pm 0.03$ & \checkmark &  \\
9 & J004121.36+650413.5 & 121.87 & $2.22$ & 13.80 & 0.67 & 0.50 & $12.21 \pm 0.02$ & $0.38 \pm 0.04$ & $0.38 \pm 0.04$ & $10.41 \pm 0.02$ & $0.38 \pm 0.03$ & $2.72 \pm 0.03$ & $1.80 \pm 0.03$ &  & \checkmark \\
10 & J004517.05+640124.4 & 122.26 & $1.16$ & 15.61 & 0.95 & 0.64 & $13.37 \pm 0.02$ & $0.41 \pm 0.04$ & $0.38 \pm 0.04$ & $11.92 \pm 0.02$ & $0.27 \pm 0.03$ & $2.75 \pm 0.03$ & $1.45 \pm 0.03$ & \checkmark & \checkmark \\
11 & J004620.80+622503.9 & 122.34 & $-0.45$ & 13.18 & 0.63 & 0.54 & $11.98 \pm 0.02$ & $0.26 \pm 0.04$ & $0.28 \pm 0.04$ & $10.79 \pm 0.02$ & $0.24 \pm 0.03$ & $1.77 \pm 0.03$ & $1.19 \pm 0.03$ &  & \checkmark \\
12 & J004651.68+625914.7 & 122.41 & $0.12$ & 14.88 & 0.60 & 0.45 & $13.32 \pm 0.02$ & $0.22 \pm 0.04$ & $0.26 \pm 0.05$ & $12.79 \pm 0.03$ & $0.14 \pm 0.04$ & $1.49 \pm 0.03$ & $0.53 \pm 0.04$ & \checkmark & \checkmark \\
13 & J004741.54+624203.3 & 122.50 & $-0.17$ & 14.15 & 0.46 & 0.43 & $13.01 \pm 0.02$ & $0.20 \pm 0.04$ & $0.13 \pm 0.04$ & $12.50 \pm 0.02$ & $0.14 \pm 0.04$ & $1.20 \pm 0.03$ & $0.51 \pm 0.03$ &  & \checkmark \\
14 & J004842.93+644411.1 & 122.64 & $1.87$ & 14.74 & 0.58 & 0.42 & $13.33 \pm 0.03$ & $0.36 \pm 0.04$ & $0.09 \pm 0.04$ & $12.72 \pm 0.03$ & $0.13 \pm 0.04$ & $1.45 \pm 0.03$ & $0.61 \pm 0.04$ &  & \checkmark \\
15 & J004850.12+642533.7 & 122.65 & $1.56$ & 15.64 & 0.86 & 0.56 & $13.92 \pm 0.03$ & $0.35 \pm 0.05$ & $0.15 \pm 0.06$ & $12.95 \pm 0.03$ & $0.29 \pm 0.04$ & $1.84 \pm 0.03$ & $0.98 \pm 0.04$ &  & \checkmark \\
16 & J005011.87+635129.9 & 122.80 & $0.99$ & 13.75 & 0.40 & 0.53 & $12.75 \pm 0.03$ & $0.19 \pm 0.04$ & $0.11 \pm 0.04$ & $12.28 \pm 0.02$ & $0.11 \pm 0.04$ & $1.08 \pm 0.03$ & $0.47 \pm 0.04$ &  & \checkmark \\
17 & J005011.87+633526.1 & 122.79 & $0.72$ & 15.38 & 0.64 & 0.60 & $13.69 \pm 0.02$ & $0.28 \pm 0.04$ & $0.37 \pm 0.05$ & $12.86 \pm 0.03$ & $0.24 \pm 0.04$ & $1.87 \pm 0.03$ & $0.83 \pm 0.03$ & \checkmark & \checkmark \\
18 & J005012.69+645621.6 & 122.80 & $2.07$ & 14.14 & 0.56 & 0.49 & $12.65 \pm 0.03$ & $0.30 \pm 0.05$ & $0.14 \pm 0.05$ & $11.90 \pm 0.02$ & $0.17 \pm 0.04$ & $1.69 \pm 0.03$ & $0.76 \pm 0.04$ & \checkmark & \checkmark \\
19 & J005029.22+653331.1 & 122.83 & $2.69$ & 14.63 & 0.70 & 0.65 & $12.97 \pm 0.03$ & $0.26 \pm 0.04$ & $0.30 \pm 0.04$ & $11.99 \pm 0.02$ & $0.23 \pm 0.03$ & $1.94 \pm 0.03$ & $0.98 \pm 0.03$ & \checkmark & \checkmark \\
20 & J005032.31+623155.5 & 122.83 & $-0.34$ & 15.43 & 0.59 & 0.47 & $13.95 \pm 0.05$ & $0.31 \pm 0.08$ & $0.21 \pm 0.08$ & $12.92 \pm 0.03$ & $0.12 \pm 0.04$ & $1.92 \pm 0.03$ & $1.03 \pm 0.06$ &  & \checkmark \\
21 & J005436.84+630549.8 & 123.29 & $0.23$ & 14.91 & 0.64 & 0.75 & $13.30 \pm 0.03$ & $0.33 \pm 0.05$ & $0.36 \pm 0.05$ & $12.11 \pm 0.02$ & $0.32 \pm 0.03$ & $2.17 \pm 0.03$ & $1.19 \pm 0.03$ & \checkmark & \checkmark \\
22 & J005611.62+630350.5 & 123.47 & $0.20$ & 14.40 & 0.55 & 0.46 & $12.95 \pm 0.02$ & $0.18 \pm 0.03$ & $0.18 \pm 0.03$ & $12.47 \pm 0.02$ & $0.08 \pm 0.03$ & $1.38 \pm 0.03$ & $0.47 \pm 0.03$ & \checkmark & \checkmark \\
23 & J005619.50+625824.0 & 123.49 & $0.11$ & 14.63 & 0.42 & 0.37 & $13.46 \pm 0.02$ & $0.23 \pm 0.03$ & $0.20 \pm 0.04$ & $13.02 \pm 0.03$ & $0.16 \pm 0.04$ & $1.20 \pm 0.03$ & $0.44 \pm 0.03$ & \checkmark & \checkmark \\
24 & J005743.72+640235.6 & 123.62 & $1.18$ & 14.18 & 0.45 & 0.40 & $13.07 \pm 0.02$ & $0.20 \pm 0.04$ & $0.13 \pm 0.04$ & $12.74 \pm 0.03$ & $0.12 \pm 0.04$ & $0.99 \pm 0.03$ & $0.34 \pm 0.03$ &  & \checkmark \\
25 & J005809.86+624412.9 & 123.70 & $-0.12$ & 14.67 & 0.50 & 0.56 & $13.32 \pm 0.02$ & $0.27 \pm 0.04$ & $0.15 \pm 0.03$ & $12.73 \pm 0.03$ & $0.17 \pm 0.04$ & $1.43 \pm 0.03$ & $0.59 \pm 0.03$ &  & \checkmark \\
26 & J005859.24+632603.0 & 123.78 & $0.57$ & 13.18 & 0.45 & 0.72 & $12.06 \pm 0.02$ & $0.30 \pm 0.04$ & $0.25 \pm 0.03$ & $11.22 \pm 0.02$ & $0.27 \pm 0.03$ & $1.50 \pm 0.03$ & $0.84 \pm 0.03$ &  & \checkmark \\
27 & J005926.64+651157.0 & 123.77 & $2.34$ & 13.45 & 0.49 & 0.54 & $12.06 \pm 0.02$ & $0.21 \pm 0.04$ & $0.14 \pm 0.04$ & $11.55 \pm 0.02$ & $0.12 \pm 0.03$ & $1.41 \pm 0.03$ & $0.52 \pm 0.03$ &  & \checkmark \\
28 & J010045.61+631740.4 & 123.98 & $0.44$ & 15.43 & 0.77 & 0.63 & $13.53 \pm 0.02$ & $0.32 \pm 0.04$ & $0.29 \pm 0.05$ & $12.57 \pm 0.03$ & $0.27 \pm 0.04$ & $2.09 \pm 0.03$ & $0.96 \pm 0.03$ & \checkmark & \checkmark \\
29 & J010051.26+641327.3 & 123.96 & $1.37$ & 13.61 & 0.53 & 0.42 & $12.37 \pm 0.02$ & $0.19 \pm 0.03$ & $0.14 \pm 0.04$ & $11.86 \pm 0.02$ & $0.13 \pm 0.03$ & $1.22 \pm 0.03$ & $0.51 \pm 0.03$ &  & \checkmark \\
30 & J010054.58+643729.6 & 123.95 & $1.77$ & 13.02 & 0.61 & 0.57 & $11.30 \pm 0.03$ & $0.34 \pm 0.04$ & $0.19 \pm 0.04$ & $10.35 \pm 0.02$ & $0.25 \pm 0.03$ & $2.05 \pm 0.03$ & $0.95 \pm 0.04$ &  & \checkmark \\
31 & J010107.85+633227.0 & 124.01 & $0.69$ & 13.84 & 0.57 & 0.49 & $12.47 \pm 0.02$ & $0.22 \pm 0.03$ & $0.18 \pm 0.04$ & $11.89 \pm 0.02$ & $0.18 \pm 0.03$ & $1.39 \pm 0.03$ & $0.58 \pm 0.03$ &  & \checkmark \\
32 & J010138.04+641349.9 & 124.04 & $1.38$ & 13.32 & 0.63 & 0.71 & $11.74 \pm 0.02$ & $0.34 \pm 0.03$ & $0.24 \pm 0.03$ & $10.80 \pm 0.02$ & $0.28 \pm 0.03$ & $1.90 \pm 0.03$ & $0.94 \pm 0.03$ &  & \checkmark \\
33 & J010707.68+625117.0 & 124.72 & $0.04$ & 14.57 & 0.74 & 0.58 & $12.65 \pm 0.02$ & $0.38 \pm 0.03$ & $0.24 \pm 0.03$ & $11.71 \pm 0.02$ & $0.25 \pm 0.03$ & $2.12 \pm 0.03$ & $0.95 \pm 0.03$ & \checkmark & \checkmark \\
34 & J010841.17+615511.8 & 124.96 & $-0.88$ & 13.72 & 0.38 & 0.44 & $12.84 \pm 0.02$ & $0.10 \pm 0.03$ & $0.14 \pm 0.04$ & $12.60 \pm 0.02$ & $0.12 \pm 0.04$ & $0.75 \pm 0.03$ & $0.24 \pm 0.03$ &  & \checkmark \\
35 & J010958.80+625229.3 & 125.04 & $0.08$ & 14.00 & 0.85 & 0.65 & $12.44 \pm 0.02$ & $0.47 \pm 0.04$ & $0.40 \pm 0.04$ & $10.95 \pm 0.02$ & $0.32 \pm 0.03$ & $2.20 \pm 0.03$ & $1.49 \pm 0.03$ & \checkmark & \checkmark \\
36 & J011216.30+615051.2 & 125.39 & $-0.92$ & 13.56 & 0.35 & 0.27 & $12.65 \pm 0.02$ & $0.14 \pm 0.03$ & $0.10 \pm 0.03$ & $12.35 \pm 0.02$ & $-0.02 \pm 0.04$ & $0.87 \pm 0.03$ & $0.30 \pm 0.03$ &  & \checkmark \\
37 & J011234.21+630432.5 & 125.32 & $0.30$ & 12.64 & 0.67 & 0.62 & $10.94 \pm 0.02$ & $0.27 \pm 0.03$ & $0.24 \pm 0.03$ & $10.34 \pm 0.02$ & $0.23 \pm 0.03$ & $1.63 \pm 0.03$ & $0.60 \pm 0.03$ &  & \checkmark \\
38 & J011402.43+625735.3 & 125.50 & $0.20$ & 12.91 & 0.56 & 0.34 & $11.41 \pm 0.02$ & $0.27 \pm 0.03$ & $0.25 \pm 0.03$ & $$ & $$ & $$ & $$ &  & \checkmark \\
39 & J011543.99+660116.2 & 125.40 & $3.27$ & 14.21 & 0.94 & 1.11 & $11.95 \pm 0.02$ & $0.49 \pm 0.04$ & $0.40 \pm 0.04$ & $10.46 \pm 0.02$ & $0.37 \pm 0.03$ & $2.81 \pm 0.03$ & $1.49 \pm 0.03$ & \checkmark & \checkmark \\
40 & J011604.41+630926.7 & 125.71 & $0.42$ & 14.83 & 0.76 & 0.54 & $12.93 \pm 0.02$ & $0.31 \pm 0.03$ & $0.16 \pm 0.03$ & $11.91 \pm 0.02$ & $0.28 \pm 0.03$ & $2.16 \pm 0.03$ & $1.02 \pm 0.03$ &  & \checkmark \\
41 & J011918.18+642233.8 & 125.94 & $1.67$ & 13.44 & 0.55 & 0.71 & $12.22 \pm 0.02$ & $0.29 \pm 0.03$ & $0.23 \pm 0.03$ & $11.23 \pm 0.02$ & $0.30 \pm 0.03$ & $1.66 \pm 0.03$ & $0.99 \pm 0.03$ &  & \checkmark \\
42 & J012158.75+642813.1 & 126.22 & $1.79$ & 14.35 & 0.67 & 0.74 & $12.74 \pm 0.03$ & $0.29 \pm 0.05$ & $0.33 \pm 0.05$ & $11.66 \pm 0.02$ & $0.29 \pm 0.03$ & $2.01 \pm 0.03$ & $1.07 \pm 0.04$ & \checkmark & \checkmark \\
43 & J012320.11+635830.9 & 126.42 & $1.32$ & 14.08 & 0.89 & 0.97 & $11.89 \pm 0.02$ & $0.47 \pm 0.03$ & $0.40 \pm 0.03$ & $10.51 \pm 0.02$ & $0.36 \pm 0.03$ & $2.68 \pm 0.03$ & $1.38 \pm 0.03$ & \checkmark & \checkmark \\
44 & J012325.80+642638.7 & 126.37 & $1.79$ & 16.32 & 0.67 & 0.57 & $14.60 \pm 0.03$ & $0.27 \pm 0.05$ & $0.11 \pm 0.08$ & $14.10 \pm 0.03$ & $0.03 \pm 0.06$ & $1.56 \pm 0.04$ & $0.51 \pm 0.04$ &  & \checkmark \\
45 & J012339.47+631544.2 & 126.55 & $0.62$ & 15.04 & 0.80 & 0.61 & $12.97 \pm 0.02$ & $0.45 \pm 0.03$ & $0.29 \pm 0.03$ & $11.90 \pm 0.02$ & $0.21 \pm 0.03$ & $2.33 \pm 0.03$ & $1.07 \pm 0.03$ &  & \checkmark \\
46 & J012339.76+635313.0 & 126.47 & $1.24$ & 15.07 & 0.82 & 0.75 & $12.96 \pm 0.02$ & $0.41 \pm 0.03$ & $0.35 \pm 0.03$ & $11.99 \pm 0.02$ & $0.26 \pm 0.03$ & $2.26 \pm 0.03$ & $0.96 \pm 0.03$ & \checkmark & \checkmark \\
47 & J012358.07+652615.4 & 126.31 & $2.78$ & 13.72 & 0.47 & 0.47 & $12.51 \pm 0.02$ & $0.18 \pm 0.03$ & $0.14 \pm 0.03$ & $12.07 \pm 0.02$ & $0.12 \pm 0.03$ & $1.17 \pm 0.03$ & $0.43 \pm 0.03$ &  & \checkmark \\
48 & J012405.46+660100.0 & 126.25 & $3.36$ & 15.09 & 0.63 & 0.57 & $13.45 \pm 0.03$ & $0.27 \pm 0.04$ & $0.21 \pm 0.04$ & $12.67 \pm 0.03$ & $0.19 \pm 0.04$ & $1.79 \pm 0.03$ & $0.78 \pm 0.04$ & \checkmark & \checkmark \\
49 & J012416.76+633011.7 & 126.59 & $0.86$ & 13.10 & 0.66 & 0.49 & $11.53 \pm 0.02$ & $0.31 \pm 0.03$ & $0.16 \pm 0.02$ & $10.80 \pm 0.02$ & $0.17 \pm 0.03$ & $1.64 \pm 0.03$ & $0.73 \pm 0.03$ &  & \checkmark \\
50 & J012540.54+623025.6 & 126.87 & $-0.10$ & 13.42 & 0.48 & 0.52 & $12.05 \pm 0.02$ & $0.22 \pm 0.02$ & $0.19 \pm 0.02$ & $11.58 \pm 0.02$ & $0.24 \pm 0.03$ & $1.37 \pm 0.03$ & $0.48 \pm 0.03$ & \checkmark & \checkmark \\
51 & J012609.29+651618.0 & 126.55 & $2.64$ & 14.79 & 0.87 & 0.62 & $12.79 \pm 0.03$ & $0.45 \pm 0.04$ & $0.36 \pm 0.03$ & $11.46 \pm 0.02$ & $0.34 \pm 0.03$ & $2.46 \pm 0.03$ & $1.33 \pm 0.04$ & \checkmark & \checkmark \\
52 & J012634.69+641850.9 & 126.73 & $1.70$ & 12.77 & 0.64 & 0.55 & $11.07 \pm 0.02$ & $0.30 \pm 0.04$ & $0.23 \pm 0.04$ & $10.56 \pm 0.02$ & $0.29 \pm 0.03$ & $1.56 \pm 0.03$ & $0.51 \pm 0.03$ &  & \checkmark \\
53 & J012703.28+634333.3 & 126.86 & $1.13$ & 14.05 & 0.81 & 0.81 & $11.60 \pm 0.02$ & $0.37 \pm 0.04$ & $0.40 \pm 0.04$ & $10.86 \pm 0.02$ & $0.28 \pm 0.03$ & $2.38 \pm 0.03$ & $0.74 \pm 0.03$ & \checkmark & \checkmark \\
54 & J012745.08+625154.3 & 127.06 & $0.28$ & 13.46 & 0.45 & 0.58 & $12.04 \pm 0.02$ & $0.24 \pm 0.03$ & $0.21 \pm 0.04$ & $11.25 \pm 0.02$ & $0.27 \pm 0.03$ & $1.76 \pm 0.03$ & $0.80 \pm 0.03$ &  & \checkmark \\
55 & J012751.32+655104.2 & 126.65 & $3.24$ & 14.51 & 0.71 & 0.75 & $12.77 \pm 0.02$ & $0.37 \pm 0.04$ & $0.28 \pm 0.04$ & $11.67 \pm 0.02$ & $0.27 \pm 0.03$ & $2.14 \pm 0.03$ & $1.10 \pm 0.03$ & \checkmark & \checkmark \\
56* & J013000.21+631044.6 & 127.27 & $0.63$ & 13.76 & 1.02 & 1.09 & $11.33 \pm 0.02$ & $0.56 \pm 0.03$ & $0.53 \pm 0.04$ & $10.32 \pm 0.02$ & $0.50 \pm 0.03$ & $2.42 \pm 0.03$ & $1.01 \pm 0.03$ &  & \checkmark \\
57 & J013213.90+623717.2 & 127.60 & $0.12$ & 13.32 & 0.54 & 0.43 & $11.96 \pm 0.02$ & $0.22 \pm 0.04$ & $0.17 \pm 0.03$ & $11.37 \pm 0.02$ & $0.19 \pm 0.03$ & $1.42 \pm 0.03$ & $0.59 \pm 0.03$ &  & \checkmark \\
58 & J013245.71+645233.3 & 127.30 & $2.36$ & 15.42 & 0.76 & 1.02 & $13.31 \pm 0.03$ & $0.47 \pm 0.04$ & $0.40 \pm 0.05$ & $11.68 \pm 0.02$ & $0.32 \pm 0.03$ & $2.99 \pm 0.03$ & $1.63 \pm 0.04$ & \checkmark & \checkmark \\
59 & J013422.61+624459.7 & 127.82 & $0.29$ & 12.92 & 0.39 & 0.61 & $11.90 \pm 0.02$ & $0.14 \pm 0.04$ & $0.26 \pm 0.03$ & $11.57 \pm 0.02$ & $0.25 \pm 0.03$ & $0.96 \pm 0.03$ & $0.33 \pm 0.03$ &  & \checkmark \\
60 & J013739.40+613258.8 & 128.41 & $-0.83$ & 14.31 & 0.63 & 0.50 & $12.73 \pm 0.03$ & $0.31 \pm 0.04$ & $0.16 \pm 0.04$ & $12.26 \pm 0.02$ & $0.06 \pm 0.03$ & $1.42 \pm 0.03$ & $0.47 \pm 0.04$ &  & \checkmark \\
61 & J013819.58+635306.0 & 128.06 & $1.48$ & 12.98 & 0.61 & 0.91 & $11.50 \pm 0.03$ & $0.31 \pm 0.04$ & $0.34 \pm 0.03$ & $10.34 \pm 0.02$ & $0.29 \pm 0.03$ & $2.03 \pm 0.03$ & $1.16 \pm 0.03$ &  & \checkmark \\
62 & J013825.56+635008.5 & 128.08 & $1.43$ & 13.96 & 0.53 & 0.47 & $12.70 \pm 0.03$ & $0.24 \pm 0.04$ & $0.11 \pm 0.04$ & $12.01 \pm 0.02$ & $0.16 \pm 0.03$ & $1.42 \pm 0.03$ & $0.69 \pm 0.04$ &  & \checkmark \\
63 & J013834.27+634841.3 & 128.10 & $1.41$ & 15.28 & 0.65 & 0.56 & $14.16 \pm 0.03$ & $0.17 \pm 0.05$ & $0.15 \pm 0.07$ & $12.80 \pm 0.03$ & $0.26 \pm 0.04$ & $1.82 \pm 0.04$ & $1.36 \pm 0.04$ &  & \checkmark \\
64 & J013920.80+654338.7 & 127.83 & $3.31$ & 13.64 & 0.61 & 0.47 & $12.20 \pm 0.02$ & $0.24 \pm 0.04$ & $0.18 \pm 0.04$ & $11.75 \pm 0.02$ & $0.11 \pm 0.03$ & $1.28 \pm 0.03$ & $0.45 \pm 0.03$ &  & \checkmark \\
65 & J014218.76+624733.5 & 128.71 & $0.49$ & 14.56 & 0.60 & 0.50 & $12.90 \pm 0.03$ & $0.30 \pm 0.04$ & $0.23 \pm 0.04$ & $12.08 \pm 0.02$ & $0.17 \pm 0.03$ & $1.88 \pm 0.03$ & $0.82 \pm 0.03$ & \checkmark & \checkmark \\
66 & J014238.73+633753.1 & 128.58 & $1.32$ & 13.18 & 0.40 & 0.37 & $12.03 \pm 0.02$ & $0.22 \pm 0.03$ & $0.12 \pm 0.03$ & $11.68 \pm 0.02$ & $0.08 \pm 0.03$ & $1.10 \pm 0.03$ & $0.35 \pm 0.03$ &  & \checkmark \\
67 & J014244.86+623056.2 & 128.81 & $0.23$ & 13.22 & 0.38 & 0.40 & $12.06 \pm 0.02$ & $0.22 \pm 0.03$ & $0.14 \pm 0.03$ & $11.52 \pm 0.02$ & $0.15 \pm 0.03$ & $1.33 \pm 0.03$ & $0.54 \pm 0.03$ &  & \checkmark \\
68 & J014322.19+640118.5 & 128.58 & $1.72$ & 13.29 & 0.37 & 0.48 & $12.30 \pm 0.02$ & $0.16 \pm 0.04$ & $0.16 \pm 0.04$ & $11.77 \pm 0.03$ & $0.24 \pm 0.05$ & $1.16 \pm 0.04$ & $0.53 \pm 0.04$ &  & \checkmark \\
69 & J014401.85+640124.5 & 128.65 & $1.74$ & 15.49 & 0.54 & 0.53 & $14.07 \pm 0.03$ & $0.33 \pm 0.05$ & $0.18 \pm 0.05$ & $$ & $$ & $$ & $$ &  & \checkmark \\
70 & J014458.15+633244.0 & 128.85 & $1.29$ & 13.99 & 0.42 & 0.43 & $12.89 \pm 0.02$ & $0.18 \pm 0.04$ & $0.16 \pm 0.04$ & $12.34 \pm 0.02$ & $0.16 \pm 0.03$ & $1.23 \pm 0.03$ & $0.56 \pm 0.03$ & \checkmark & \checkmark \\
71 & J014458.27+625245.8 & 128.99 & $0.64$ & 13.35 & 0.59 & 0.61 & $11.71 \pm 0.02$ & $0.34 \pm 0.04$ & $0.28 \pm 0.04$ & $10.66 \pm 0.02$ & $0.26 \pm 0.03$ & $2.09 \pm 0.03$ & $1.04 \pm 0.03$ &  & \checkmark \\
72 & J014519.02+633559.1 & 128.88 & $1.35$ & 13.13 & 0.42 & 0.55 & $12.27 \pm 0.02$ & $0.21 \pm 0.04$ & $0.22 \pm 0.04$ & $11.27 \pm 0.02$ & $0.29 \pm 0.03$ & $1.44 \pm 0.03$ & $1.00 \pm 0.03$ &  & \checkmark \\
73 & J014539.64+611259.1 & 129.41 & $-0.97$ & 12.45 & 0.76 & 0.74 & $10.92 \pm 0.02$ & $0.19 \pm 0.03$ & $0.24 \pm 0.03$ & $10.23 \pm 0.02$ & $0.29 \pm 0.03$ & $1.45 \pm 0.03$ & $0.69 \pm 0.03$ &  & \checkmark \\
74 & J014602.11+611502.2 & 129.45 & $-0.93$ & 13.75 & 0.55 & 0.71 & $12.39 \pm 0.03$ & $0.26 \pm 0.05$ & $0.25 \pm 0.04$ & $11.68 \pm 0.03$ & $0.33 \pm 0.04$ & $1.51 \pm 0.04$ & $0.71 \pm 0.04$ &  & \checkmark \\
75 & J014620.49+644802.5 & 128.74 & $2.55$ & 14.41 & 0.48 & 0.43 & $13.25 \pm 0.03$ & $0.23 \pm 0.04$ & $0.17 \pm 0.04$ & $12.66 \pm 0.03$ & $0.21 \pm 0.04$ & $1.27 \pm 0.03$ & $0.59 \pm 0.04$ & \checkmark & \checkmark \\
76 & J014624.42+611037.3 & 129.51 & $-0.99$ & 13.36 & 0.45 & 0.34 & $12.22 \pm 0.02$ & $0.17 \pm 0.04$ & $0.14 \pm 0.04$ & $11.93 \pm 0.03$ & $0.02 \pm 0.04$ & $0.99 \pm 0.03$ & $0.30 \pm 0.03$ &  & \checkmark \\
77 & J014807.07+631613.2 & 129.25 & $1.10$ & 13.54 & 0.60 & 0.69 & $11.92 \pm 0.02$ & $0.25 \pm 0.03$ & $0.38 \pm 0.03$ & $10.77 \pm 0.04$ & $0.32 \pm 0.05$ & $2.18 \pm 0.04$ & $1.15 \pm 0.04$ &  & \checkmark \\
78 & J014843.39+642854.4 & 129.05 & $2.29$ & 13.74 & 0.45 & 0.49 & $12.66 \pm 0.02$ & $0.24 \pm 0.04$ & $0.15 \pm 0.04$ & $12.04 \pm 0.02$ & $0.14 \pm 0.03$ & $1.25 \pm 0.03$ & $0.62 \pm 0.03$ &  & \checkmark \\
79 & J014905.20+624912.3 & 129.46 & $0.68$ & 13.60 & 0.81 & 0.68 & $11.49 \pm 0.02$ & $0.34 \pm 0.04$ & $0.48 \pm 0.04$ & $9.97 \pm 0.02$ & $0.26 \pm 0.03$ & $2.82 \pm 0.03$ & $1.51 \pm 0.03$ & \checkmark & \checkmark \\
80 & J015001.17+642219.7 & 129.21 & $2.22$ & 15.76 & 0.56 & 0.45 & $14.14 \pm 0.02$ & $0.29 \pm 0.05$ & $0.14 \pm 0.06$ & $13.39 \pm 0.03$ & $0.12 \pm 0.04$ & $1.81 \pm 0.03$ & $0.75 \pm 0.04$ &  & \checkmark \\
81 & J015022.92+652743.2 & 129.01 & $3.29$ & 13.79 & 0.51 & 0.42 & $12.55 \pm 0.02$ & $0.26 \pm 0.03$ & $0.28 \pm 0.03$ & $12.16 \pm 0.02$ & $0.22 \pm 0.03$ & $1.12 \pm 0.03$ & $0.38 \pm 0.03$ &  & \checkmark \\
82 & J015030.72+634132.5 & 129.42 & $1.57$ & 15.70 & 0.60 & 0.50 & $13.84 \pm 0.03$ & $0.40 \pm 0.04$ & $0.24 \pm 0.04$ & $12.89 \pm 0.03$ & $0.20 \pm 0.04$ & $2.20 \pm 0.03$ & $0.94 \pm 0.04$ &  & \checkmark \\
83 & J015037.70+644447.0 & 129.19 & $2.60$ & 14.64 & 0.40 & 0.39 & $13.52 \pm 0.03$ & $0.21 \pm 0.04$ & $0.22 \pm 0.04$ & $13.36 \pm 0.03$ & $-0.07 \pm 0.05$ & $0.88 \pm 0.03$ & $0.17 \pm 0.04$ & \checkmark & \checkmark \\
84 & J015105.68+611602.6 & 130.04 & $-0.78$ & 13.71 & 0.50 & 0.37 & $12.50 \pm 0.03$ & $0.24 \pm 0.04$ & $0.18 \pm 0.03$ & $12.02 \pm 0.02$ & $0.09 \pm 0.03$ & $1.19 \pm 0.03$ & $0.48 \pm 0.03$ &  & \checkmark \\
85 & J015109.13+641421.8 & 129.36 & $2.12$ & 15.71 & 0.57 & 0.48 & $14.27 \pm 0.03$ & $0.33 \pm 0.05$ & $0.08 \pm 0.06$ & $13.80 \pm 0.03$ & $0.16 \pm 0.05$ & $1.33 \pm 0.04$ & $0.47 \pm 0.04$ &  & \checkmark \\
\hline
\multicolumn{16}{l}{*: IPHAS photometry for these classical Be stars is not included in DR2, 
because the corresponding fields were D-graded in \citet{Barentsen14}. Their photometry is to be considered accurate at the level of 
$\sim 10$~per cent.}\\
\end{tabular}
\end{minipage}
\end{table*}
\begin{table*}
\tiny
\begin{minipage}{180mm}
\contcaption{}
\centering
\begin{tabular}{@{}llrrrrrrrrrrrrcc@{}}
\hline
\# & Name & $\ell$ & $b$ & $r$ & $(r-i)$ & $(r-\Halpha)$ & $J$ & $(J-H)$ & $(H-K)$ & $W1$ & $(W1-W2)$ & $(i-W1)$ & $(J-W1)$ & LP & FF \\
 & Jhhmmss.ss+ddmmss.s & (deg) & (deg) & (mag) & (mag) & (mag) & (mag) & (mag) & (mag) & (mag) & (mag) & (mag) & (mag) &  &   \\
\hline
86 & J015213.08+624813.6 & 129.81 & $0.75$ & 13.04 & 0.31 & 0.35 & $12.11 \pm 0.02$ & $0.12 \pm 0.03$ & $0.11 \pm 0.03$ & $11.70 \pm 0.02$ & $0.09 \pm 0.03$ & $1.03 \pm 0.03$ & $0.40 \pm 0.03$ &  & \checkmark \\
87 & J015221.93+635739.6 & 129.56 & $1.88$ & 14.29 & 0.56 & 0.61 & $12.79 \pm 0.02$ & $0.27 \pm 0.04$ & $0.30 \pm 0.04$ & $11.87 \pm 0.02$ & $0.29 \pm 0.03$ & $1.86 \pm 0.03$ & $0.92 \pm 0.03$ &  & \checkmark \\
88 & J015246.27+630315.0 & 129.82 & $1.00$ & 14.35 & 0.51 & 0.40 & $12.90 \pm 0.02$ & $0.25 \pm 0.04$ & $0.18 \pm 0.04$ & $12.32 \pm 0.03$ & $0.07 \pm 0.04$ & $1.52 \pm 0.03$ & $0.59 \pm 0.03$ & \checkmark &  \\
89 & J015307.22+650110.4 & 129.39 & $2.92$ & 13.97 & 0.41 & 0.37 & $12.97 \pm 0.03$ & $0.20 \pm 0.04$ & $0.11 \pm 0.04$ & $12.47 \pm 0.03$ & $0.09 \pm 0.04$ & $1.08 \pm 0.03$ & $0.49 \pm 0.04$ &  & \checkmark \\
90 & J015314.56+620241.5 & 130.11 & $0.04$ & 12.89 & 0.39 & 0.31 & $11.88 \pm 0.02$ & $0.17 \pm 0.04$ & $0.07 \pm 0.03$ & $11.48 \pm 0.02$ & $0.06 \pm 0.03$ & $1.01 \pm 0.03$ & $0.40 \pm 0.03$ &  & \checkmark \\
91 & J015329.19+643128.1 & 129.54 & $2.45$ & 13.98 & 0.33 & 0.26 & $13.12 \pm 0.02$ & $0.11 \pm 0.03$ & $0.11 \pm 0.04$ & $12.78 \pm 0.03$ & $-0.01 \pm 0.04$ & $0.88 \pm 0.03$ & $0.35 \pm 0.03$ &  & \checkmark \\
92 & J015427.15+612204.7 & 130.41 & $-0.59$ & 14.38 & 1.08 & 0.84 & $11.55 \pm 0.02$ & $0.57 \pm 0.03$ & $0.51 \pm 0.02$ & $9.87 \pm 0.02$ & $0.38 \pm 0.03$ & $3.43 \pm 0.03$ & $1.68 \pm 0.03$ & \checkmark & \checkmark \\
93 & J015510.44+632108.5 & 130.01 & $1.36$ & 15.85 & 0.50 & 0.41 & $14.46 \pm 0.03$ & $0.15 \pm 0.05$ & $0.20 \pm 0.07$ & $13.92 \pm 0.04$ & $-0.03 \pm 0.06$ & $1.42 \pm 0.04$ & $0.54 \pm 0.04$ &  & \checkmark \\
94 & J015520.62+611752.8 & 130.53 & $-0.63$ & 13.46 & 0.50 & 0.43 & $12.24 \pm 0.02$ & $0.17 \pm 0.03$ & $0.14 \pm 0.02$ & $11.62 \pm 0.02$ & $0.16 \pm 0.03$ & $1.34 \pm 0.03$ & $0.62 \pm 0.03$ &  & \checkmark \\
95 & J015526.12+625056.4 & 130.16 & $0.88$ & 14.34 & 0.55 & 0.61 & $13.11 \pm 0.02$ & $0.25 \pm 0.03$ & $0.28 \pm 0.03$ & $12.12 \pm 0.03$ & $0.31 \pm 0.03$ & $1.67 \pm 0.03$ & $0.99 \pm 0.03$ &  & \checkmark \\
96 & J015601.25+633944.1 & 130.02 & $1.68$ & 14.58 & 0.56 & 0.46 & $13.14 \pm 0.02$ & $0.18 \pm 0.03$ & $0.22 \pm 0.03$ & $12.63 \pm 0.03$ & $0.02 \pm 0.04$ & $1.39 \pm 0.03$ & $0.52 \pm 0.03$ &  & \checkmark \\
97 & J015613.26+635623.7 & 129.97 & $1.96$ & 14.08 & 0.46 & 0.64 & $12.82 \pm 0.03$ & $0.31 \pm 0.04$ & $0.33 \pm 0.04$ & $$ & $$ & $$ & $$ & \checkmark & \checkmark \\
98 & J015627.82+612939.2 & 130.61 & $-0.40$ & 12.70 & 0.44 & 0.50 & $11.49 \pm 0.02$ & $0.23 \pm 0.03$ & $0.17 \pm 0.03$ & $11.26 \pm 0.02$ & $0.06 \pm 0.03$ & $1.00 \pm 0.03$ & $0.23 \pm 0.03$ &  & \checkmark \\
99 & J015630.87+630307.5 & 130.23 & $1.11$ & 13.28 & 0.27 & 0.51 & $12.52 \pm 0.02$ & $0.10 \pm 0.03$ & $0.08 \pm 0.03$ & $12.21 \pm 0.03$ & $0.15 \pm 0.04$ & $0.79 \pm 0.03$ & $0.31 \pm 0.03$ &  & \checkmark \\
100 & J015644.71+653640.6 & 129.61 & $3.59$ & 15.01 & 0.54 & 0.40 & $13.70 \pm 0.02$ & $0.14 \pm 0.03$ & $0.11 \pm 0.04$ & $13.39 \pm 0.03$ & $-0.02 \pm 0.04$ & $1.07 \pm 0.03$ & $0.31 \pm 0.03$ &  & \checkmark \\
101 & J015645.75+635259.8 & 130.05 & $1.92$ & 13.11 & 0.58 & 0.53 & $11.67 \pm 0.02$ & $0.27 \pm 0.02$ & $0.22 \pm 0.02$ & $10.88 \pm 0.02$ & $0.24 \pm 0.03$ & $1.66 \pm 0.03$ & $0.79 \pm 0.03$ &  & \checkmark \\
102 & J015741.35+605313.7 & 130.91 & $-0.95$ & 14.76 & 0.71 & 0.51 & $13.06 \pm 0.02$ & $0.31 \pm 0.03$ & $0.17 \pm 0.03$ & $12.32 \pm 0.03$ & $0.12 \pm 0.04$ & $1.72 \pm 0.03$ & $0.74 \pm 0.03$ &  & \checkmark \\
103 & J015804.46+653020.6 & 129.77 & $3.52$ & 13.14 & 0.39 & 0.36 & $12.15 \pm 0.02$ & $0.12 \pm 0.03$ & $0.17 \pm 0.03$ & $11.46 \pm 0.02$ & $0.22 \pm 0.03$ & $1.28 \pm 0.03$ & $0.69 \pm 0.03$ &  & \checkmark \\
104 & J015809.04+615813.6 & 130.68 & $0.11$ & 14.59 & 0.72 & 0.60 & $12.92 \pm 0.02$ & $0.29 \pm 0.03$ & $0.24 \pm 0.03$ & $12.35 \pm 0.02$ & $0.12 \pm 0.04$ & $1.52 \pm 0.03$ & $0.58 \pm 0.03$ &  & \checkmark \\
105 & J015918.33+654955.8 & 129.81 & $3.87$ & 15.19 & 0.55 & 0.54 & $13.80 \pm 0.02$ & $0.24 \pm 0.03$ & $0.16 \pm 0.04$ & $13.23 \pm 0.03$ & $0.19 \pm 0.04$ & $1.42 \pm 0.03$ & $0.58 \pm 0.03$ & \checkmark & \checkmark \\
106 & J015919.69+645053.4 & 130.07 & $2.92$ & 12.78 & 0.34 & 0.53 & $11.88 \pm 0.02$ & $0.24 \pm 0.03$ & $0.16 \pm 0.02$ & $10.89 \pm 0.02$ & $0.25 \pm 0.03$ & $1.56 \pm 0.03$ & $0.99 \pm 0.03$ &  & \checkmark \\
107 & J015922.56+635829.2 & 130.30 & $2.08$ & 15.15 & 0.56 & 0.82 & $13.37 \pm 0.02$ & $0.45 \pm 0.02$ & $0.36 \pm 0.03$ & $12.08 \pm 0.02$ & $0.41 \pm 0.03$ & $2.51 \pm 0.03$ & $1.29 \pm 0.03$ & \checkmark & \checkmark \\
108 & J015938.99+643615.4 & 130.17 & $2.69$ & 13.26 & 0.39 & 0.53 & $12.10 \pm 0.02$ & $0.27 \pm 0.03$ & $0.22 \pm 0.03$ & $11.43 \pm 0.02$ & $0.28 \pm 0.03$ & $1.45 \pm 0.03$ & $0.67 \pm 0.03$ &  & \checkmark \\
109 & J015945.33+630314.9 & 130.58 & $1.20$ & 14.63 & 0.53 & 0.39 & $13.04 \pm 0.02$ & $0.25 \pm 0.04$ & $0.12 \pm 0.04$ & $12.33 \pm 0.02$ & $0.08 \pm 0.04$ & $1.78 \pm 0.03$ & $0.71 \pm 0.03$ &  & \checkmark \\
110 & J020037.84+652133.9 & 130.07 & $3.45$ & 13.08 & 0.40 & 0.34 & $12.11 \pm 0.02$ & $0.21 \pm 0.03$ & $0.14 \pm 0.03$ & $11.79 \pm 0.02$ & $0.08 \pm 0.03$ & $0.89 \pm 0.03$ & $0.32 \pm 0.03$ &  & \checkmark \\
111 & J020049.43+635944.0 & 130.45 & $2.14$ & 13.91 & 0.84 & 0.57 & $12.39 \pm 0.02$ & $0.28 \pm 0.03$ & $0.31 \pm 0.03$ & $11.37 \pm 0.02$ & $0.33 \pm 0.03$ & $1.70 \pm 0.03$ & $1.03 \pm 0.03$ &  & \checkmark \\
112 & J020105.33+613403.0 & 131.12 & $-0.19$ & 14.46 & 0.67 & 0.68 & $12.78 \pm 0.02$ & $0.37 \pm 0.03$ & $0.31 \pm 0.03$ & $11.82 \pm 0.03$ & $0.30 \pm 0.04$ & $1.98 \pm 0.04$ & $0.96 \pm 0.04$ &  & \checkmark \\
113 & J020121.79+630117.3 & 130.77 & $1.22$ & 13.32 & 0.45 & 0.35 & $12.26 \pm 0.02$ & $0.14 \pm 0.04$ & $0.13 \pm 0.04$ & $12.05 \pm 0.02$ & $-0.06 \pm 0.03$ & $0.82 \pm 0.03$ & $0.21 \pm 0.03$ &  & \checkmark \\
114 & J020136.00+613207.6 & 131.19 & $-0.21$ & 13.14 & 0.47 & 0.39 & $11.93 \pm 0.02$ & $0.19 \pm 0.04$ & $0.15 \pm 0.04$ & $11.46 \pm 0.02$ & $0.05 \pm 0.03$ & $1.22 \pm 0.03$ & $0.47 \pm 0.03$ &  & \checkmark \\
115 & J020203.16+630213.4 & 130.84 & $1.25$ & 13.41 & 0.77 & 0.73 & $11.62 \pm 0.02$ & $0.40 \pm 0.04$ & $0.36 \pm 0.04$ & $10.43 \pm 0.02$ & $0.34 \pm 0.03$ & $2.21 \pm 0.03$ & $1.19 \pm 0.03$ &  & \checkmark \\
116 & J020252.26+620926.0 & 131.17 & $0.43$ & 15.21 & 0.60 & 0.57 & $12.91 \pm 0.03$ & $0.45 \pm 0.05$ & $0.28 \pm 0.05$ & $11.77 \pm 0.02$ & $0.09 \pm 0.03$ & $2.84 \pm 0.03$ & $1.14 \pm 0.04$ &  & \checkmark \\
117 & J020326.01+635943.1 & 130.72 & $2.22$ & 14.40 & 0.54 & 0.49 & $13.00 \pm 0.03$ & $0.22 \pm 0.04$ & $0.22 \pm 0.04$ & $11.98 \pm 0.02$ & $0.22 \pm 0.03$ & $1.88 \pm 0.03$ & $1.02 \pm 0.03$ &  & \checkmark \\
118 & J020328.03+624333.8 & 131.08 & $1.00$ & 13.73 & 0.41 & 0.42 & $12.55 \pm 0.03$ & $0.15 \pm 0.04$ & $0.14 \pm 0.04$ & $12.05 \pm 0.03$ & $0.12 \pm 0.04$ & $1.26 \pm 0.03$ & $0.50 \pm 0.04$ &  & \checkmark \\
119 & J020407.85+643122.2 & 130.65 & $2.75$ & 13.94 & 0.36 & 0.39 & $12.98 \pm 0.02$ & $0.19 \pm 0.03$ & $0.12 \pm 0.04$ & $12.47 \pm 0.03$ & $0.11 \pm 0.04$ & $1.10 \pm 0.03$ & $0.50 \pm 0.03$ &  & \checkmark \\
120 & J020504.17+630216.1 & 131.17 & $1.35$ & 15.07 & 0.57 & 0.41 & $13.68 \pm 0.03$ & $0.36 \pm 0.05$ & $0.13 \pm 0.04$ & $13.21 \pm 0.03$ & $0.02 \pm 0.04$ & $1.29 \pm 0.03$ & $0.47 \pm 0.04$ &  & \checkmark \\
121 & J020547.47+641051.7 & 130.92 & $2.47$ & 12.68 & 0.40 & 0.41 & $11.80 \pm 0.02$ & $0.17 \pm 0.03$ & $0.12 \pm 0.03$ & $11.52 \pm 0.02$ & $0.08 \pm 0.03$ & $0.76 \pm 0.03$ & $0.28 \pm 0.03$ &  & \checkmark \\
122 & J020618.67+644945.1 & 130.79 & $3.11$ & 14.88 & 0.42 & 0.57 & $13.85 \pm 0.03$ & $0.20 \pm 0.05$ & $0.15 \pm 0.06$ & $13.29 \pm 0.03$ & $0.07 \pm 0.05$ & $1.16 \pm 0.04$ & $0.55 \pm 0.04$ &  & \checkmark \\
123 & J020707.67+612422.7 & 131.86 & $-0.15$ & 13.28 & 0.39 & 0.37 & $12.27 \pm 0.02$ & $0.09 \pm 0.03$ & $0.16 \pm 0.03$ & $11.65 \pm 0.02$ & $0.19 \pm 0.03$ & $1.24 \pm 0.03$ & $0.62 \pm 0.03$ &  & \checkmark \\
124 & J020717.23+645046.2 & 130.88 & $3.15$ & 12.57 & 0.35 & 0.54 & $11.63 \pm 0.02$ & $0.18 \pm 0.04$ & $0.16 \pm 0.04$ & $11.00 \pm 0.02$ & $0.24 \pm 0.03$ & $1.22 \pm 0.03$ & $0.63 \pm 0.03$ &  & \checkmark \\
125 & J020731.12+634520.4 & 131.22 & $2.12$ & 14.54 & 0.67 & 0.45 & $12.78 \pm 0.02$ & $0.36 \pm 0.04$ & $0.18 \pm 0.04$ & $12.33 \pm 0.03$ & $0.02 \pm 0.04$ & $1.55 \pm 0.03$ & $0.45 \pm 0.03$ &  & \checkmark \\
126 & J020734.24+623601.1 & 131.56 & $1.01$ & 14.43 & 0.56 & 0.46 & $12.98 \pm 0.02$ & $0.27 \pm 0.04$ & $0.18 \pm 0.04$ & $12.36 \pm 0.02$ & $0.12 \pm 0.04$ & $1.50 \pm 0.03$ & $0.62 \pm 0.03$ & \checkmark & \checkmark \\
127 & J020753.51+644148.9 & 130.99 & $3.03$ & 15.91 & 0.33 & 0.34 & $14.89 \pm 0.04$ & $0.19 \pm 0.07$ & $0.08 \pm 0.12$ & $14.36 \pm 0.03$ & $-0.42 \pm 0.13$ & $1.22 \pm 0.04$ & $0.53 \pm 0.05$ &  & \checkmark \\
128 & J020817.76+614220.1 & 131.91 & $0.18$ & 13.43 & 0.55 & 0.61 & $11.82 \pm 0.02$ & $0.32 \pm 0.03$ & $0.26 \pm 0.03$ & $10.81 \pm 0.02$ & $0.22 \pm 0.03$ & $2.07 \pm 0.03$ & $1.01 \pm 0.03$ &  & \checkmark \\
129 & J020826.27+625745.9 & 131.55 & $1.39$ & 14.37 & 0.64 & 0.53 & $12.64 \pm 0.03$ & $0.28 \pm 0.04$ & $0.24 \pm 0.04$ & $11.81 \pm 0.03$ & $0.15 \pm 0.03$ & $1.92 \pm 0.03$ & $0.83 \pm 0.04$ &  & \checkmark \\
130 & J020837.59+652028.2 & 130.87 & $3.67$ & 14.74 & 0.53 & 0.54 & $13.34 \pm 0.02$ & $0.25 \pm 0.04$ & $0.17 \pm 0.04$ & $12.59 \pm 0.03$ & $0.18 \pm 0.04$ & $1.62 \pm 0.03$ & $0.75 \pm 0.03$ &  & \checkmark \\
131 & J020855.24+631501.1 & 131.52 & $1.68$ & 12.72 & 0.54 & 0.61 & $11.60 \pm 0.03$ & $0.16 \pm 0.04$ & $0.16 \pm 0.04$ & $10.88 \pm 0.02$ & $0.14 \pm 0.03$ & $1.30 \pm 0.03$ & $0.72 \pm 0.04$ &  & \checkmark \\
132 & J020859.72+635536.1 & 131.33 & $2.33$ & 14.94 & 0.41 & 0.35 & $13.68 \pm 0.03$ & $0.19 \pm 0.05$ & $0.18 \pm 0.05$ & $13.15 \pm 0.03$ & $0.02 \pm 0.05$ & $1.38 \pm 0.03$ & $0.54 \pm 0.04$ &  & \checkmark \\
133 & J020917.87+613045.2 & 132.08 & $0.03$ & 13.89 & 0.83 & 0.57 & $11.91 \pm 0.02$ & $0.34 \pm 0.04$ & $0.23 \pm 0.03$ & $11.11 \pm 0.02$ & $0.24 \pm 0.03$ & $1.96 \pm 0.03$ & $0.80 \pm 0.03$ &  & \checkmark \\
134 & J021000.05+640838.5 & 131.37 & $2.57$ & 12.81 & 0.35 & 0.35 & $11.84 \pm 0.02$ & $0.17 \pm 0.03$ & $0.13 \pm 0.03$ & $11.28 \pm 0.02$ & $0.10 \pm 0.03$ & $1.19 \pm 0.03$ & $0.56 \pm 0.03$ &  & \checkmark \\
135 & J021005.63+631100.3 & 131.67 & $1.65$ & 15.98 & 0.51 & 0.39 & $14.64 \pm 0.02$ & $0.27 \pm 0.05$ & $0.08 \pm 0.08$ & $14.23 \pm 0.03$ & $-0.08 \pm 0.07$ & $1.24 \pm 0.04$ & $0.41 \pm 0.04$ &  & \checkmark \\
136 & J021011.09+652230.2 & 131.02 & $3.75$ & 13.77 & 0.61 & 0.53 & $12.37 \pm 0.02$ & $0.27 \pm 0.03$ & $0.14 \pm 0.03$ & $11.72 \pm 0.03$ & $0.20 \pm 0.04$ & $1.43 \pm 0.03$ & $0.65 \pm 0.03$ &  & \checkmark \\
137 & J021036.32+611444.0 & 132.31 & $-0.18$ & 13.09 & 0.50 & 0.42 & $11.66 \pm 0.02$ & $0.26 \pm 0.03$ & $0.17 \pm 0.02$ & $11.09 \pm 0.02$ & $0.13 \pm 0.03$ & $1.51 \pm 0.03$ & $0.57 \pm 0.03$ &  & \checkmark \\
138 & J021057.06+624700.8 & 131.88 & $1.30$ & 14.02 & 0.47 & 0.31 & $12.73 \pm 0.02$ & $0.29 \pm 0.02$ & $0.08 \pm 0.03$ & $12.12 \pm 0.03$ & $0.11 \pm 0.03$ & $1.44 \pm 0.03$ & $0.62 \pm 0.03$ &  & \checkmark \\
139 & J021121.67+624707.5 & 131.92 & $1.32$ & 15.57 & 0.57 & 0.52 & $13.93 \pm 0.03$ & $0.35 \pm 0.05$ & $0.25 \pm 0.05$ & $$ & $$ & $$ & $$ & \checkmark & \checkmark \\
140 & J021128.86+634604.0 & 131.64 & $2.26$ & 13.93 & 0.48 & 0.42 & $12.54 \pm 0.02$ & $0.22 \pm 0.03$ & $0.15 \pm 0.03$ & $11.98 \pm 0.02$ & $0.11 \pm 0.03$ & $1.46 \pm 0.03$ & $0.56 \pm 0.03$ &  & \checkmark \\
141 & J021159.14+615639.9 & 132.25 & $0.54$ & 15.00 & 0.84 & 0.69 & $12.79 \pm 0.02$ & $0.44 \pm 0.03$ & $0.27 \pm 0.03$ & $11.97 \pm 0.02$ & $0.18 \pm 0.03$ & $2.19 \pm 0.03$ & $0.82 \pm 0.03$ &  & \checkmark \\
142 & J021202.03+613230.4 & 132.38 & $0.16$ & 15.41 & 0.83 & 0.59 & $13.27 \pm 0.02$ & $0.39 \pm 0.03$ & $0.36 \pm 0.03$ & $12.07 \pm 0.02$ & $0.32 \pm 0.03$ & $2.50 \pm 0.03$ & $1.20 \pm 0.03$ &  & \checkmark \\
143 & J021210.42+623242.3 & 132.09 & $1.12$ & 12.20 & 0.34 & 0.45 & $11.33 \pm 0.02$ & $0.13 \pm 0.03$ & $0.17 \pm 0.03$ & $10.72 \pm 0.02$ & $0.17 \pm 0.03$ & $1.14 \pm 0.03$ & $0.61 \pm 0.03$ &  & \checkmark \\
144 & J021320.12+613003.1 & 132.54 & $0.17$ & 12.89 & 0.67 & 0.57 & $11.26 \pm 0.02$ & $0.28 \pm 0.02$ & $0.23 \pm 0.02$ & $10.45 \pm 0.02$ & $0.23 \pm 0.03$ & $1.77 \pm 0.03$ & $0.82 \pm 0.03$ &  & \checkmark \\
145 & J021325.98+622043.1 & 132.29 & $0.97$ & 14.51 & 0.72 & 0.59 & $12.56 \pm 0.02$ & $0.49 \pm 0.03$ & $0.37 \pm 0.03$ & $$ & $$ & $$ & $$ &  & \checkmark \\
146 & J021336.53+601829.1 & 132.94 & $-0.96$ & 13.75 & 0.63 & 0.38 & $12.18 \pm 0.02$ & $0.26 \pm 0.03$ & $0.15 \pm 0.03$ & $11.64 \pm 0.02$ & $0.07 \pm 0.03$ & $1.49 \pm 0.03$ & $0.54 \pm 0.03$ &  & \checkmark \\
147 & J021352.00+642520.3 & 131.68 & $2.96$ & 15.03 & 0.47 & 0.39 & $13.96 \pm 0.03$ & $0.20 \pm 0.05$ & $0.19 \pm 0.05$ & $13.24 \pm 0.03$ & $0.11 \pm 0.04$ & $1.32 \pm 0.03$ & $0.72 \pm 0.04$ &  & \checkmark \\
148 & J021532.96+623236.9 & 132.46 & $1.24$ & 13.20 & 0.64 & 0.65 & $11.45 \pm 0.02$ & $0.32 \pm 0.03$ & $0.25 \pm 0.02$ & $10.21 \pm 0.02$ & $0.29 \pm 0.03$ & $2.35 \pm 0.03$ & $1.24 \pm 0.03$ &  & \checkmark \\
149 & J021647.40+642812.9 & 131.97 & $3.11$ & 14.59 & 0.52 & 0.46 & $13.28 \pm 0.02$ & $0.20 \pm 0.04$ & $0.23 \pm 0.04$ & $12.72 \pm 0.02$ & $0.18 \pm 0.04$ & $1.35 \pm 0.03$ & $0.57 \pm 0.03$ &  & \checkmark \\
150 & J021744.41+644335.2 & 131.98 & $3.38$ & 13.71 & 0.56 & 0.63 & $12.22 \pm 0.02$ & $0.29 \pm 0.03$ & $0.27 \pm 0.03$ & $11.89 \pm 0.02$ & $0.12 \pm 0.03$ & $1.25 \pm 0.03$ & $0.33 \pm 0.03$ &  & \checkmark \\
151 & J021747.84+643419.9 & 132.04 & $3.24$ & 14.66 & 0.57 & 0.62 & $13.14 \pm 0.03$ & $0.28 \pm 0.04$ & $0.27 \pm 0.03$ & $$ & $$ & $$ & $$ &  & \checkmark \\
152 & J022009.76+643605.9 & 132.27 & $3.35$ & 15.65 & 0.55 & 0.43 & $14.27 \pm 0.03$ & $0.29 \pm 0.05$ & $0.17 \pm 0.07$ & $$ & $$ & $$ & $$ &  & \checkmark \\
153 & J022033.45+625717.4 & 132.86 & $1.81$ & 15.71 & 0.71 & 0.85 & $13.90 \pm 0.02$ & $0.40 \pm 0.03$ & $0.45 \pm 0.04$ & $12.55 \pm 0.03$ & $0.38 \pm 0.03$ & $2.44 \pm 0.03$ & $1.35 \pm 0.03$ & \checkmark & \checkmark \\
154 & J022045.25+631642.8 & 132.78 & $2.13$ & 15.02 & 0.52 & 0.49 & $13.85 \pm 0.03$ & $0.22 \pm 0.05$ & $0.22 \pm 0.05$ & $13.19 \pm 0.03$ & $0.14 \pm 0.04$ & $1.32 \pm 0.03$ & $0.67 \pm 0.04$ &  & \checkmark \\
155 & J022053.65+642835.6 & 132.38 & $3.26$ & 15.91 & 0.61 & 0.49 & $14.28 \pm 0.03$ & $0.31 \pm 0.05$ & $0.20 \pm 0.06$ & $13.47 \pm 0.03$ & $0.08 \pm 0.05$ & $1.83 \pm 0.03$ & $0.81 \pm 0.04$ &  & \checkmark \\
156 & J022100.28+635435.2 & 132.59 & $2.73$ & 13.09 & 0.40 & 0.39 & $12.11 \pm 0.02$ & $0.13 \pm 0.04$ & $0.13 \pm 0.04$ & $11.66 \pm 0.03$ & $0.11 \pm 0.04$ & $1.03 \pm 0.03$ & $0.45 \pm 0.03$ &  & \checkmark \\
157 & J022107.83+625754.6 & 132.92 & $1.85$ & 14.94 & 0.68 & 0.57 & $13.23 \pm 0.03$ & $0.31 \pm 0.05$ & $0.36 \pm 0.05$ & $12.09 \pm 0.02$ & $0.26 \pm 0.03$ & $2.17 \pm 0.03$ & $1.14 \pm 0.04$ &  & \checkmark \\
158 & J022222.78+623841.9 & 133.17 & $1.59$ & 15.17 & 0.58 & 0.43 & $13.54 \pm 0.02$ & $0.33 \pm 0.04$ & $0.20 \pm 0.04$ & $12.81 \pm 0.03$ & $0.13 \pm 0.04$ & $1.77 \pm 0.03$ & $0.72 \pm 0.03$ &  & \checkmark \\
159 & J022227.73+623530.6 & 133.19 & $1.55$ & 14.32 & 0.41 & 0.71 & $13.12 \pm 0.03$ & $0.23 \pm 0.04$ & $0.16 \pm 0.04$ & $12.33 \pm 0.02$ & $0.34 \pm 0.03$ & $1.58 \pm 0.03$ & $0.79 \pm 0.03$ &  & \checkmark \\
160 & J022337.05+601602.8 & 134.13 & $-0.59$ & 14.00 & 0.58 & 0.49 & $12.54 \pm 0.02$ & $0.25 \pm 0.02$ & $0.16 \pm 0.03$ & $11.34 \pm 0.02$ & $0.28 \pm 0.03$ & $2.08 \pm 0.03$ & $1.20 \pm 0.03$ & \checkmark & \checkmark \\
161 & J022343.45+603545.6 & 134.02 & $-0.27$ & 12.94 & 0.66 & 0.44 & $11.24 \pm 0.03$ & $0.26 \pm 0.05$ & $0.19 \pm 0.05$ & $10.63 \pm 0.03$ & $0.11 \pm 0.04$ & $1.65 \pm 0.03$ & $0.62 \pm 0.04$ &  & \checkmark \\
162 & J022420.68+624842.5 & 133.32 & $1.83$ & 13.43 & 0.60 & 0.48 & $11.74 \pm 0.02$ & $0.32 \pm 0.03$ & $0.18 \pm 0.02$ & $11.01 \pm 0.02$ & $0.19 \pm 0.03$ & $1.82 \pm 0.03$ & $0.73 \pm 0.03$ &  & \checkmark \\
163 & J022502.69+644947.6 & 132.68 & $3.74$ & 13.26 & 0.53 & 0.44 & $11.80 \pm 0.02$ & $0.26 \pm 0.03$ & $0.19 \pm 0.03$ & $11.14 \pm 0.02$ & $0.12 \pm 0.03$ & $1.59 \pm 0.03$ & $0.66 \pm 0.03$ &  & \checkmark \\
164 & J022636.00+601401.8 & 134.49 & $-0.49$ & 14.50 & 0.96 & 0.99 & $12.13 \pm 0.02$ & $0.56 \pm 0.02$ & $0.49 \pm 0.02$ & $10.47 \pm 0.02$ & $0.42 \pm 0.03$ & $3.07 \pm 0.03$ & $1.66 \pm 0.03$ & \checkmark & \checkmark \\
165 & J022821.67+641216.0 & 133.24 & $3.29$ & 15.84 & 0.61 & 0.49 & $14.16 \pm 0.02$ & $0.22 \pm 0.04$ & $0.27 \pm 0.05$ & $13.43 \pm 0.03$ & $0.12 \pm 0.05$ & $1.79 \pm 0.04$ & $0.72 \pm 0.04$ &  & \checkmark \\
166 & J022823.86+631834.8 & 133.57 & $2.46$ & 12.78 & 0.55 & 0.63 & $11.45 \pm 0.02$ & $0.28 \pm 0.02$ & $0.30 \pm 0.02$ & $11.01 \pm 0.02$ & $0.23 \pm 0.03$ & $1.21 \pm 0.03$ & $0.44 \pm 0.03$ &  & \checkmark \\
167 & J022953.81+630742.2 & 133.79 & $2.35$ & 14.25 & 0.67 & 0.50 & $12.85 \pm 0.02$ & $0.22 \pm 0.03$ & $0.15 \pm 0.03$ & $11.15 \pm 0.02$ & $0.25 \pm 0.03$ & $2.43 \pm 0.03$ & $1.70 \pm 0.03$ & \checkmark & \checkmark \\
168 & J023003.21+643829.4 & 133.25 & $3.76$ & 15.31 & 0.56 & 0.53 & $13.84 \pm 0.02$ & $0.17 \pm 0.04$ & $0.24 \pm 0.05$ & $13.28 \pm 0.03$ & $0.10 \pm 0.04$ & $1.47 \pm 0.03$ & $0.56 \pm 0.03$ &  & \checkmark \\
169* & J023031.38+594127.1 & 135.14 & $-0.81$ & 14.48 & 0.57 & 0.43 & $12.96 \pm 0.02$ & $0.31 \pm 0.04$ & $0.17 \pm 0.04$ & $12.32 \pm 0.02$ & $0.12 \pm 0.03$ & $1.60 \pm 0.03$ & $0.65 \pm 0.03$ & \checkmark & \checkmark \\
170 & J023035.11+610005.9 & 134.66 & $0.41$ & 14.56 & 0.63 & 0.47 & $12.61 \pm 0.02$ & $0.42 \pm 0.03$ & $0.39 \pm 0.03$ & $11.68 \pm 0.02$ & $0.30 \pm 0.03$ & $2.25 \pm 0.03$ & $0.93 \pm 0.03$ &  & \checkmark \\
171 & J023150.04+604952.4 & 134.86 & $0.31$ & 13.85 & 0.41 & 0.35 & $12.75 \pm 0.02$ & $0.19 \pm 0.03$ & $0.12 \pm 0.03$ & $12.28 \pm 0.03$ & $0.10 \pm 0.04$ & $1.15 \pm 0.03$ & $0.47 \pm 0.03$ &  & \checkmark \\
172 & J023202.89+641033.2 & 133.62 & $3.41$ & 16.23 & 0.67 & 0.44 & $14.47 \pm 0.03$ & $0.25 \pm 0.05$ & $0.32 \pm 0.06$ & $13.78 \pm 0.03$ & $0.03 \pm 0.06$ & $1.78 \pm 0.03$ & $0.68 \pm 0.04$ &  & \checkmark \\
\hline
\multicolumn{16}{l}{*: IPHAS photometry for these classical Be stars is not included in DR2, 
because the corresponding fields were D-graded in \citet{Barentsen14}. Their photometry is to be considered accurate at the level of 
$\sim 10$~per cent.}\\
\end{tabular}
\end{minipage}
\end{table*}
\begin{table*}
\tiny
\begin{minipage}{180mm}
\contcaption{}
\centering
\begin{tabular}{@{}llrrrrrrrrrrrrcc@{}}
\hline
\# & Name & $\ell$ & $b$ & $r$ & $(r-i)$ & $(r-\Halpha)$ & $J$ & $(J-H)$ & $(H-K)$ & $W1$ & $(W1-W2)$ & $(i-W1)$ & $(J-W1)$ & LP & FF \\
 & Jhhmmss.ss+ddmmss.s & (deg) & (deg) & (mag) & (mag) & (mag) & (mag) & (mag) & (mag) & (mag) & (mag) & (mag) & (mag) &  &   \\
\hline
173 & J023235.10+640522.7 & 133.71 & $3.35$ & 13.53 & 0.76 & 0.76 & $11.62 \pm 0.02$ & $0.33 \pm 0.02$ & $0.27 \pm 0.02$ & $10.94 \pm 0.02$ & $0.31 \pm 0.03$ & $1.82 \pm 0.03$ & $0.69 \pm 0.03$ &  & \checkmark \\
174** & J023404.70+605914.4 & 135.06 & $0.55$ & 12.88 & 0.60 & 0.69 & $11.30 \pm 0.02$ & $0.42 \pm 0.03$ & $0.51 \pm 0.03$ & $8.73 \pm 0.02$ & $1.64 \pm 0.03$ & $3.55 \pm 0.03$ & $2.57 \pm 0.03$ & \checkmark & \checkmark \\
175 & J023411.97+595634.2 & 135.47 & $-0.40$ & 13.13 & 0.59 & 0.56 & $11.65 \pm 0.02$ & $0.28 \pm 0.03$ & $0.23 \pm 0.03$ & $11.27 \pm 0.02$ & $0.05 \pm 0.03$ & $1.27 \pm 0.03$ & $0.38 \pm 0.03$ &  & \checkmark \\
176 & J023431.08+601616.5 & 135.38 & $-0.08$ & 13.58 & 1.13 & 0.82 & $10.77 \pm 0.02$ & $0.55 \pm 0.03$ & $0.48 \pm 0.04$ & $9.34 \pm 0.02$ & $0.39 \pm 0.03$ & $3.12 \pm 0.03$ & $1.43 \pm 0.03$ & \checkmark & \checkmark \\
177 & J023439.79+641813.4 & 133.83 & $3.64$ & 17.02 & 0.77 & 0.49 & $15.11 \pm 0.04$ & $0.28 \pm 0.09$ & $0.26 \pm 0.11$ & $14.39 \pm 0.04$ & $-0.15 \pm 0.09$ & $1.86 \pm 0.04$ & $0.72 \pm 0.06$ &  & \checkmark \\
178 & J023536.82+625251.7 & 134.49 & $2.37$ & 16.68 & 0.59 & 0.46 & $14.96 \pm 0.06$ & $0.31 \pm 0.11$ & $0.23 \pm 0.13$ & $$ & $$ & $$ & $$ &  & \checkmark \\
179 & J023629.19+634245.8 & 134.25 & $3.17$ & 15.61 & 0.64 & 0.60 & $13.92 \pm 0.03$ & $0.32 \pm 0.05$ & $0.37 \pm 0.05$ & $13.53 \pm 0.03$ & $0.11 \pm 0.04$ & $1.44 \pm 0.03$ & $0.39 \pm 0.04$ &  & \checkmark \\
180 & J023642.66+614714.8 & 135.03 & $1.41$ & 15.38 & 0.54 & 0.41 & $13.95 \pm 0.05$ & $0.33 \pm 0.07$ & $0.14 \pm 0.06$ & $13.31 \pm 0.07$ & $0.18 \pm 0.10$ & $1.53 \pm 0.07$ & $0.64 \pm 0.09$ & \checkmark & \checkmark \\
181 & J023744.52+605352.8 & 135.50 & $0.65$ & 16.79 & 0.76 & 0.52 & $14.76 \pm 0.04$ & $0.27 \pm 0.07$ & $0.31 \pm 0.09$ & $14.06 \pm 0.03$ & $0.20 \pm 0.06$ & $1.97 \pm 0.04$ & $0.70 \pm 0.05$ & \checkmark &  \\
182 & J023753.78+620410.0 & 135.05 & $1.73$ & 13.03 & 0.58 & 0.47 & $11.58 \pm 0.02$ & $0.27 \pm 0.03$ & $0.26 \pm 0.03$ & $11.97 \pm 0.02$ & $-0.02 \pm 0.03$ & $0.49 \pm 0.03$ & $-0.39 \pm 0.03$ &  & \checkmark \\
183 & J023758.11+634635.6 & 134.38 & $3.30$ & 13.30 & 0.36 & 0.36 & $12.34 \pm 0.02$ & $0.18 \pm 0.03$ & $0.07 \pm 0.03$ & $11.89 \pm 0.02$ & $0.06 \pm 0.03$ & $1.05 \pm 0.03$ & $0.45 \pm 0.03$ &  & \checkmark \\
184 & J023809.91+620224.6 & 135.09 & $1.71$ & 13.21 & 0.57 & 0.54 & $11.71 \pm 0.02$ & $0.30 \pm 0.02$ & $0.19 \pm 0.02$ & $10.96 \pm 0.02$ & $0.17 \pm 0.03$ & $1.67 \pm 0.03$ & $0.75 \pm 0.03$ &  & \checkmark \\
185 & J023841.80+640826.3 & 134.30 & $3.66$ & 14.05 & 0.64 & 0.44 & $12.38 \pm 0.02$ & $0.30 \pm 0.03$ & $0.17 \pm 0.03$ & $11.63 \pm 0.02$ & $0.16 \pm 0.03$ & $1.79 \pm 0.03$ & $0.75 \pm 0.03$ &  & \checkmark \\
186 & J023948.17+604505.1 & 135.79 & $0.61$ & 13.26 & 0.37 & 0.42 & $12.28 \pm 0.02$ & $0.21 \pm 0.03$ & $0.12 \pm 0.03$ & $11.93 \pm 0.03$ & $0.07 \pm 0.04$ & $0.96 \pm 0.03$ & $0.35 \pm 0.03$ &  & \checkmark \\
187 & J024054.97+630009.8 & 134.99 & $2.72$ & 15.69 & 0.54 & 0.49 & $14.36 \pm 0.03$ & $0.29 \pm 0.05$ & $0.21 \pm 0.07$ & $13.73 \pm 0.03$ & $0.15 \pm 0.05$ & $1.41 \pm 0.04$ & $0.63 \pm 0.04$ & \checkmark & \checkmark \\
188 & J024146.73+602532.2 & 136.14 & $0.42$ & 14.05 & 0.63 & 0.71 & $12.31 \pm 0.02$ & $0.30 \pm 0.03$ & $0.25 \pm 0.03$ & $11.53 \pm 0.02$ & $0.29 \pm 0.03$ & $1.89 \pm 0.03$ & $0.78 \pm 0.03$ & \checkmark & \checkmark \\
189 & J024159.21+600106.0 & 136.34 & $0.06$ & 14.55 & 0.65 & 0.58 & $12.80 \pm 0.02$ & $0.29 \pm 0.02$ & $0.21 \pm 0.03$ & $12.07 \pm 0.02$ & $0.17 \pm 0.03$ & $1.83 \pm 0.03$ & $0.73 \pm 0.03$ & \checkmark & \checkmark \\
190 & J024221.54+593716.4 & 136.54 & $-0.29$ & 13.38 & 0.61 & 0.46 & $11.67 \pm 0.02$ & $0.23 \pm 0.03$ & $0.26 \pm 0.03$ & $10.74 \pm 0.02$ & $0.40 \pm 0.03$ & $2.03 \pm 0.03$ & $0.93 \pm 0.03$ &  & \checkmark \\
191 & J024252.56+611953.9 & 135.89 & $1.30$ & 15.74 & 0.64 & 0.73 & $13.95 \pm 0.03$ & $0.36 \pm 0.05$ & $0.37 \pm 0.05$ & $12.84 \pm 0.03$ & $0.33 \pm 0.04$ & $2.27 \pm 0.03$ & $1.12 \pm 0.04$ & \checkmark & \checkmark \\
192 & J024305.60+631614.7 & 135.10 & $3.07$ & 14.55 & 0.55 & 0.61 & $13.27 \pm 0.02$ & $0.27 \pm 0.03$ & $0.31 \pm 0.04$ & $12.24 \pm 0.02$ & $0.27 \pm 0.03$ & $1.77 \pm 0.03$ & $1.03 \pm 0.03$ &  & \checkmark \\
193 & J024317.68+603205.5 & 136.27 & $0.59$ & 13.67 & 0.66 & 0.68 & $11.98 \pm 0.03$ & $0.32 \pm 0.04$ & $0.26 \pm 0.04$ & $10.73 \pm 0.02$ & $0.26 \pm 0.03$ & $2.28 \pm 0.03$ & $1.25 \pm 0.04$ & \checkmark & \checkmark \\
194 & J024332.05+632150.1 & 135.11 & $3.17$ & 14.16 & 0.46 & 0.46 & $12.98 \pm 0.02$ & $0.21 \pm 0.03$ & $0.14 \pm 0.03$ & $12.37 \pm 0.02$ & $0.18 \pm 0.03$ & $1.34 \pm 0.03$ & $0.62 \pm 0.03$ &  & \checkmark \\
195 & J024405.37+621448.7 & 135.64 & $2.19$ & 15.22 & 0.49 & 0.48 & $13.80 \pm 0.02$ & $0.49 \pm 0.03$ & $0.49 \pm 0.04$ & $12.10 \pm 0.02$ & $0.56 \pm 0.03$ & $2.63 \pm 0.03$ & $1.69 \pm 0.03$ & \checkmark & \checkmark \\
196 & J024454.00+635608.0 & 135.01 & $3.76$ & 15.75 & 0.58 & 0.51 & $14.25 \pm 0.02$ & $0.23 \pm 0.05$ & $0.22 \pm 0.06$ & $13.56 \pm 0.03$ & $0.20 \pm 0.04$ & $1.61 \pm 0.03$ & $0.69 \pm 0.04$ &  & \checkmark \\
197 & J024504.86+612502.1 & 136.09 & $1.48$ & 15.35 & 0.56 & 0.47 & $13.81 \pm 0.02$ & $0.35 \pm 0.04$ & $0.21 \pm 0.04$ & $13.01 \pm 0.03$ & $0.14 \pm 0.04$ & $1.78 \pm 0.03$ & $0.81 \pm 0.03$ & \checkmark & \checkmark \\
198 & J024506.09+611409.1 & 136.17 & $1.32$ & 15.97 & 0.65 & 0.65 & $14.12 \pm 0.02$ & $0.35 \pm 0.03$ & $0.31 \pm 0.04$ & $13.07 \pm 0.03$ & $0.27 \pm 0.05$ & $2.25 \pm 0.04$ & $1.05 \pm 0.04$ & \checkmark & \checkmark \\
199 & J024509.54+612523.5 & 136.10 & $1.49$ & 15.11 & 0.53 & 0.34 & $13.37 \pm 0.02$ & $0.41 \pm 0.04$ & $0.38 \pm 0.04$ & $$ & $$ & $$ & $$ &  & \checkmark \\
200 & J024519.10+633755.1 & 135.18 & $3.50$ & 14.04 & 0.55 & 0.62 & $12.74 \pm 0.02$ & $0.30 \pm 0.03$ & $0.26 \pm 0.03$ & $11.84 \pm 0.02$ & $0.26 \pm 0.03$ & $1.66 \pm 0.03$ & $0.90 \pm 0.03$ &  & \checkmark \\
201 & J024521.28+625416.2 & 135.49 & $2.84$ & 13.59 & 0.44 & 0.46 & $12.48 \pm 0.02$ & $0.19 \pm 0.03$ & $0.16 \pm 0.03$ & $11.82 \pm 0.02$ & $0.24 \pm 0.03$ & $1.33 \pm 0.03$ & $0.66 \pm 0.03$ &  & \checkmark \\
202 & J024540.61+592151.2 & 137.03 & $-0.34$ & 13.28 & 0.47 & 0.55 & $11.99 \pm 0.02$ & $0.23 \pm 0.03$ & $0.20 \pm 0.03$ & $11.59 \pm 0.02$ & $0.13 \pm 0.03$ & $1.22 \pm 0.03$ & $0.40 \pm 0.03$ &  & \checkmark \\
203 & J024553.87+635414.7 & 135.12 & $3.77$ & 14.21 & 0.68 & 0.53 & $12.48 \pm 0.02$ & $0.34 \pm 0.03$ & $0.30 \pm 0.03$ & $11.30 \pm 0.03$ & $0.29 \pm 0.03$ & $2.23 \pm 0.03$ & $1.18 \pm 0.03$ &  & \checkmark \\
204 & J024618.12+613514.7 & 136.15 & $1.70$ & 15.70 & 0.50 & 0.47 & $14.33 \pm 0.03$ & $0.27 \pm 0.04$ & $0.13 \pm 0.05$ & $13.95 \pm 0.03$ & $0.16 \pm 0.06$ & $1.25 \pm 0.04$ & $0.38 \pm 0.04$ & \checkmark & \checkmark \\
205 & J024735.56+615530.9 & 136.15 & $2.07$ & 13.44 & 0.39 & 0.44 & $12.29 \pm 0.02$ & $0.23 \pm 0.03$ & $0.12 \pm 0.03$ & $11.72 \pm 0.03$ & $0.22 \pm 0.04$ & $1.33 \pm 0.03$ & $0.57 \pm 0.03$ &  & \checkmark \\
206 & J024753.07+613405.8 & 136.33 & $1.76$ & 14.37 & 0.68 & 0.74 & $12.61 \pm 0.02$ & $0.40 \pm 0.03$ & $0.27 \pm 0.03$ & $11.38 \pm 0.02$ & $0.39 \pm 0.03$ & $2.31 \pm 0.03$ & $1.24 \pm 0.03$ &  & \checkmark \\
207 & J024758.73+630156.8 & 135.71 & $3.09$ & 14.66 & 0.54 & 0.74 & $13.24 \pm 0.02$ & $0.36 \pm 0.03$ & $0.30 \pm 0.03$ & $12.13 \pm 0.03$ & $0.34 \pm 0.03$ & $1.99 \pm 0.03$ & $1.11 \pm 0.03$ &  & \checkmark \\
208 & J024823.01+614728.1 & 136.29 & $1.99$ & 12.91 & 0.38 & 0.48 & $11.72 \pm 0.02$ & $0.23 \pm 0.03$ & $0.13 \pm 0.03$ & $11.10 \pm 0.02$ & $0.16 \pm 0.03$ & $1.43 \pm 0.03$ & $0.63 \pm 0.03$ &  & \checkmark \\
209 & J024823.69+614107.1 & 136.34 & $1.90$ & 14.09 & 0.40 & 0.61 & $12.75 \pm 0.02$ & $0.29 \pm 0.03$ & $0.24 \pm 0.03$ & $11.86 \pm 0.02$ & $0.26 \pm 0.03$ & $1.82 \pm 0.03$ & $0.89 \pm 0.03$ & \checkmark & \checkmark \\
210 & J024838.04+630153.0 & 135.77 & $3.12$ & 15.21 & 0.45 & 0.65 & $13.89 \pm 0.03$ & $0.23 \pm 0.05$ & $0.18 \pm 0.06$ & $13.31 \pm 0.03$ & $0.22 \pm 0.04$ & $1.45 \pm 0.03$ & $0.58 \pm 0.04$ &  & \checkmark \\
211 & J024913.93+631042.4 & 135.77 & $3.28$ & 13.16 & 0.39 & 0.54 & $12.04 \pm 0.03$ & $0.25 \pm 0.05$ & $0.27 \pm 0.05$ & $$ & $$ & $$ & $$ &  & \checkmark \\
212 & J024928.47+595248.4 & 137.24 & $0.33$ & 13.37 & 0.83 & 0.70 & $11.25 \pm 0.02$ & $0.47 \pm 0.03$ & $0.41 \pm 0.03$ & $9.94 \pm 0.02$ & $0.38 \pm 0.03$ & $2.61 \pm 0.03$ & $1.31 \pm 0.03$ &  & \checkmark \\
213 & J024940.66+621424.8 & 136.23 & $2.46$ & 13.24 & 0.33 & 0.45 & $12.36 \pm 0.02$ & $0.13 \pm 0.04$ & $0.12 \pm 0.04$ & $11.85 \pm 0.02$ & $0.15 \pm 0.03$ & $1.05 \pm 0.03$ & $0.51 \pm 0.03$ &  & \checkmark \\
214 & J025016.65+624435.6 & 136.07 & $2.94$ & 14.48 & 0.42 & 0.37 & $13.21 \pm 0.02$ & $0.24 \pm 0.04$ & $0.11 \pm 0.04$ & $12.67 \pm 0.03$ & $0.09 \pm 0.04$ & $1.39 \pm 0.03$ & $0.55 \pm 0.03$ & \checkmark & \checkmark \\
215 & J025059.12+615648.7 & 136.50 & $2.26$ & 15.28 & 0.45 & 0.58 & $14.16 \pm 0.03$ & $0.21 \pm 0.04$ & $0.22 \pm 0.05$ & $13.59 \pm 0.03$ & $0.26 \pm 0.05$ & $1.24 \pm 0.04$ & $0.56 \pm 0.04$ & \checkmark & \checkmark \\
216 & J025102.21+615733.8 & 136.50 & $2.28$ & 14.07 & 0.51 & 0.81 & $12.70 \pm 0.02$ & $0.30 \pm 0.04$ & $0.37 \pm 0.04$ & $11.47 \pm 0.02$ & $0.30 \pm 0.03$ & $2.08 \pm 0.03$ & $1.22 \pm 0.03$ & \checkmark & \checkmark \\
217* & J025136.03+601557.5 & 137.31 & $0.79$ & 15.71 & 0.44 & 0.45 & $14.16 \pm 0.03$ & $0.50 \pm 0.05$ & $0.37 \pm 0.05$ & $12.12 \pm 0.03$ & $0.48 \pm 0.03$ & $3.15 \pm 0.03$ & $2.04 \pm 0.04$ &  & \checkmark \\
218 & J025200.23+621145.2 & 136.49 & $2.54$ & 13.85 & 0.40 & 0.45 & $12.65 \pm 0.03$ & $0.22 \pm 0.04$ & $0.16 \pm 0.04$ & $11.96 \pm 0.02$ & $0.19 \pm 0.03$ & $1.49 \pm 0.03$ & $0.70 \pm 0.03$ &  & \checkmark \\
219* & J025204.14+583423.4 & 138.12 & $-0.70$ & 12.96 & 0.35 & 0.46 & $11.84 \pm 0.02$ & $0.17 \pm 0.03$ & $0.17 \pm 0.03$ & $11.48 \pm 0.02$ & $0.17 \pm 0.03$ & $1.13 \pm 0.03$ & $0.35 \pm 0.03$ &  & \checkmark \\
220* & J025233.24+615902.2 & 136.64 & $2.38$ & 14.82 & 0.43 & 0.36 & $13.60 \pm 0.03$ & $0.14 \pm 0.05$ & $0.22 \pm 0.06$ & $13.05 \pm 0.03$ & $0.09 \pm 0.04$ & $1.33 \pm 0.03$ & $0.55 \pm 0.04$ & \checkmark & \checkmark \\
221 & J025324.51+614622.9 & 136.83 & $2.24$ & 15.46 & 0.59 & 0.58 & $13.79 \pm 0.03$ & $0.29 \pm 0.05$ & $0.42 \pm 0.06$ & $13.31 \pm 0.03$ & $0.12 \pm 0.04$ & $1.56 \pm 0.03$ & $0.48 \pm 0.04$ &  & \checkmark \\
222 & J025448.83+605832.1 & 137.34 & $1.60$ & 16.17 & 0.80 & 0.52 & $13.90 \pm 0.03$ & $0.41 \pm 0.05$ & $0.32 \pm 0.05$ & $12.88 \pm 0.03$ & $0.18 \pm 0.05$ & $2.49 \pm 0.04$ & $1.02 \pm 0.05$ & \checkmark & \checkmark \\
223 & J025502.38+605001.9 & 137.43 & $1.49$ & 14.52 & 0.54 & 0.42 & $12.96 \pm 0.02$ & $0.33 \pm 0.04$ & $0.22 \pm 0.04$ & $$ & $$ & $$ & $$ & \checkmark &  \\
224 & J025610.39+580629.6 & 138.81 & $-0.87$ & 13.81 & 0.62 & 0.25 & $12.30 \pm 0.02$ & $0.20 \pm 0.03$ & $0.15 \pm 0.03$ & $11.35 \pm 0.02$ & $0.29 \pm 0.03$ & $1.84 \pm 0.03$ & $0.95 \pm 0.03$ & \checkmark & \checkmark \\
225 & J025631.47+593648.4 & 138.16 & $0.49$ & 15.01 & 0.70 & 0.74 & $13.26 \pm 0.02$ & $0.36 \pm 0.03$ & $0.36 \pm 0.03$ & $12.12 \pm 0.02$ & $0.32 \pm 0.03$ & $2.19 \pm 0.03$ & $1.14 \pm 0.03$ &  & \checkmark \\
226 & J025700.47+575742.8 & 138.98 & $-0.94$ & 14.26 & 0.71 & 0.56 & $12.60 \pm 0.02$ & $0.24 \pm 0.03$ & $0.15 \pm 0.03$ & $11.80 \pm 0.02$ & $0.21 \pm 0.03$ & $1.74 \pm 0.03$ & $0.80 \pm 0.03$ & \checkmark & \checkmark \\
227 & J025704.89+584311.6 & 138.63 & $-0.27$ & 16.25 & 0.81 & 0.51 & $14.13 \pm 0.02$ & $0.43 \pm 0.03$ & $0.21 \pm 0.04$ & $13.26 \pm 0.03$ & $0.17 \pm 0.04$ & $2.19 \pm 0.03$ & $0.87 \pm 0.04$ & \checkmark & \checkmark \\
228 & J025737.78+624703.4 & 136.80 & $3.36$ & 14.77 & 0.51 & 0.39 & $13.53 \pm 0.02$ & $0.23 \pm 0.04$ & $0.15 \pm 0.05$ & $12.92 \pm 0.03$ & $0.18 \pm 0.04$ & $1.34 \pm 0.03$ & $0.61 \pm 0.03$ &  & \checkmark \\
229 & J025904.88+621459.5 & 137.20 & $2.97$ & 15.59 & 0.39 & 0.34 & $14.48 \pm 0.03$ & $0.21 \pm 0.07$ & $0.17 \pm 0.08$ & $14.18 \pm 0.03$ & $-0.19 \pm 0.08$ & $1.02 \pm 0.04$ & $0.30 \pm 0.05$ &  & \checkmark \\
230 & J025935.04+603207.2 & 138.06 & $1.48$ & 13.88 & 0.43 & 0.49 & $12.52 \pm 0.02$ & $0.27 \pm 0.04$ & $0.21 \pm 0.04$ & $11.75 \pm 0.02$ & $0.16 \pm 0.03$ & $1.71 \pm 0.03$ & $0.77 \pm 0.03$ &  & \checkmark \\
231 & J025959.45+582929.5 & 139.07 & $-0.29$ & 13.31 & 0.63 & 0.49 & $11.69 \pm 0.02$ & $0.31 \pm 0.03$ & $0.13 \pm 0.03$ & $11.18 \pm 0.02$ & $0.07 \pm 0.03$ & $1.50 \pm 0.03$ & $0.51 \pm 0.03$ &  & \checkmark \\
232 & J030010.27+613223.2 & 137.65 & $2.40$ & 13.73 & 0.54 & 0.42 & $12.49 \pm 0.02$ & $0.11 \pm 0.04$ & $0.34 \pm 0.04$ & $11.91 \pm 0.02$ & $0.23 \pm 0.03$ & $1.28 \pm 0.03$ & $0.58 \pm 0.03$ &  & \checkmark \\
233 & J030056.68+615940.7 & 137.51 & $2.85$ & 13.04 & 0.35 & 0.31 & $12.17 \pm 0.02$ & $0.13 \pm 0.03$ & $0.10 \pm 0.03$ & $11.85 \pm 0.03$ & $-0.05 \pm 0.04$ & $0.84 \pm 0.03$ & $0.33 \pm 0.03$ &  & \checkmark \\
234 & J030120.82+602444.9 & 138.31 & $1.48$ & 14.38 & 0.55 & 0.48 & $12.83 \pm 0.02$ & $0.23 \pm 0.04$ & $0.25 \pm 0.03$ & $12.19 \pm 0.03$ & $0.14 \pm 0.04$ & $1.64 \pm 0.03$ & $0.64 \pm 0.03$ &  & \checkmark \\
235 & J030124.02+610221.3 & 138.02 & $2.03$ & 14.85 & 0.69 & 0.64 & $13.02 \pm 0.02$ & $0.32 \pm 0.04$ & $0.35 \pm 0.04$ & $11.92 \pm 0.02$ & $0.28 \pm 0.03$ & $2.24 \pm 0.03$ & $1.10 \pm 0.03$ &  & \checkmark \\
236 & J030133.63+593518.5 & 138.73 & $0.77$ & 14.07 & 0.67 & 0.73 & $12.42 \pm 0.02$ & $0.40 \pm 0.03$ & $0.33 \pm 0.02$ & $11.26 \pm 0.02$ & $0.34 \pm 0.03$ & $2.15 \pm 0.03$ & $1.17 \pm 0.03$ &  & \checkmark \\
237 & J030317.45+583402.0 & 139.42 & $-0.02$ & 14.57 & 0.80 & 0.61 & $12.51 \pm 0.02$ & $0.35 \pm 0.03$ & $0.22 \pm 0.03$ & $11.63 \pm 0.02$ & $0.19 \pm 0.03$ & $2.14 \pm 0.03$ & $0.87 \pm 0.03$ &  & \checkmark \\
238 & J030331.33+585234.6 & 139.29 & $0.27$ & 14.64 & 0.77 & 0.51 & $12.60 \pm 0.03$ & $0.37 \pm 0.05$ & $0.16 \pm 0.04$ & $11.79 \pm 0.03$ & $0.11 \pm 0.04$ & $2.09 \pm 0.04$ & $0.81 \pm 0.04$ &  & \checkmark \\
239 & J030332.31+623856.7 & 137.46 & $3.56$ & 14.32 & 0.55 & 0.38 & $12.80 \pm 0.02$ & $0.27 \pm 0.04$ & $0.17 \pm 0.04$ & $12.17 \pm 0.02$ & $0.10 \pm 0.03$ & $1.60 \pm 0.03$ & $0.63 \pm 0.03$ &  & \checkmark \\
240 & J030422.01+574820.4 & 139.91 & $-0.62$ & 14.40 & 0.65 & 0.54 & $12.91 \pm 0.02$ & $0.26 \pm 0.03$ & $0.17 \pm 0.03$ & $12.00 \pm 0.02$ & $0.25 \pm 0.03$ & $1.75 \pm 0.03$ & $0.92 \pm 0.03$ &  & \checkmark \\
241 & J030423.32+622900.9 & 137.63 & $3.47$ & 13.70 & 0.60 & 0.59 & $12.10 \pm 0.02$ & $0.31 \pm 0.03$ & $0.22 \pm 0.03$ & $11.38 \pm 0.02$ & $0.22 \pm 0.03$ & $1.72 \pm 0.03$ & $0.71 \pm 0.03$ &  & \checkmark \\
242 & J030451.26+594758.8 & 138.99 & $1.15$ & 15.09 & 0.76 & 0.81 & $13.06 \pm 0.02$ & $0.44 \pm 0.05$ & $0.38 \pm 0.05$ & $11.72 \pm 0.02$ & $0.32 \pm 0.03$ & $2.61 \pm 0.03$ & $1.34 \pm 0.03$ &  & \checkmark \\
243 & J030501.73+585146.8 & 139.47 & $0.35$ & 13.46 & 0.81 & 0.59 & $11.00 \pm 0.01$ & $0.35 \pm 0.03$ & $0.32 \pm 0.03$ & $10.75 \pm 0.02$ & $0.04 \pm 0.03$ & $1.89 \pm 0.03$ & $0.24 \pm 0.03$ &  & \checkmark \\
244 & J030539.90+610725.6 & 138.43 & $2.36$ & 14.47 & 0.81 & 0.73 & $12.21 \pm 0.02$ & $0.54 \pm 0.03$ & $0.32 \pm 0.03$ & $10.78 \pm 0.02$ & $0.28 \pm 0.03$ & $2.88 \pm 0.03$ & $1.43 \pm 0.03$ &  & \checkmark \\
245 & J030625.00+614359.0 & 138.20 & $2.93$ & 15.87 & 0.71 & 0.58 & $13.90 \pm 0.03$ & $0.40 \pm 0.04$ & $0.24 \pm 0.05$ & $$ & $$ & $$ & $$ &  & \checkmark \\
246 & J031141.73+614847.9 & 138.70 & $3.31$ & 15.50 & 0.54 & 0.38 & $14.00 \pm 0.03$ & $0.24 \pm 0.04$ & $0.18 \pm 0.05$ & $13.38 \pm 0.04$ & $0.03 \pm 0.06$ & $1.57 \pm 0.04$ & $0.61 \pm 0.05$ &  & \checkmark \\
247 & J031208.92+605534.5 & 139.21 & $2.58$ & 15.11 & 0.78 & 0.67 & $12.90 \pm 0.02$ & $0.45 \pm 0.03$ & $0.41 \pm 0.03$ & $11.58 \pm 0.02$ & $0.33 \pm 0.03$ & $2.76 \pm 0.03$ & $1.33 \pm 0.03$ & \checkmark & \checkmark \\
248 & J031333.27+603307.1 & 139.55 & $2.35$ & 15.74 & 0.89 & 0.59 & $13.69 \pm 0.03$ & $0.44 \pm 0.05$ & $0.36 \pm 0.05$ & $12.34 \pm 0.02$ & $0.27 \pm 0.03$ & $2.51 \pm 0.03$ & $1.35 \pm 0.04$ &  & \checkmark \\
\hline
\multicolumn{16}{l}{*: IPHAS photometry for these classical Be stars is not included in DR2, 
because the corresponding fields were D-graded in \citet{Barentsen14}. Their photometry is to be considered accurate at the level of 
$\sim 10$~per cent.}\\
\multicolumn{16}{l}{**: This object is no more classified as a classical Be star but rather like a Herbig~Be star.}
\end{tabular}
\end{minipage}
\end{table*}
\section{Stellar parameters}
\begin{table*}
\tiny
\begin{minipage}{180mm}
\caption[]{Stellar parameters of the 247 classical Be stars. The columns list: ID number; spectral types and S/N measured at $\lambda\, 4500$\,\AA; $(r-i)$ colours corrected to
zero $\Halpha$ emission; intrinsic $(r-i)$ colours; colour excess; photometry $EW(\Halpha)$ (typical uncertainties are in the range of 20~per cent); disc fraction; circumstellar colour excess;
interstellar reddenings in the $(r-i)$ and $(B-V)$ colours; interstellar reddenings from \citetalias{Raddi13}; total Galactic reddenings from \citetalias{SFD98} and \citetalias{RF09};
maximum reddenings from the \citet{Sale14} extinction-distance maps. The reddenings from \citetalias{SFD98} are multiplied by a correction factor of
 0.86, as suggested by \citep[][]{Schlafly10}. The objects labelled with a ``*'' carry a larger photometric uncertainty, because their photometry does not meet the quality controls of DR2 
\citep[cf. Table\,\ref{t:photometry} and][]{Barentsen14}.}
\centering
\begin{tabular}{@{}llrlrccrcrcrrrrrr@{}}
\hline
\# & \multicolumn{4}{c}{Spectral typing} & $(r-i)_{c}$ & $(r-i)_{\circ}$ & $E(r-i)$ & $EW(\Halpha)$ & $f_{D}$ & $E^{cs}(r-i)$ & $E^{is}(r-i)$ & $E^{is}(B-V)$ & $E^{is}(B-V)_{\rm{R13}}$ & SFD98\,$\times 0.86$ & RF09 & Sale14  \label{t:parameters}\\
  & R13 & S/N & This work & S/N & (mag) & (mag) & (mag) & (mag) & & (mag) & (mag) & (mag) & (mag) & (mag) & (mag) &  (mag)\\
\hline
1 & B5III & 38 &   &  & 0.84 & $-0.08$ & $0.92 \pm 0.01$ & $-25$ & 0.08 & 0.11 & $0.81 \pm 0.04$ & $1.25 \pm 0.06$ & $1.36 \pm 0.08$ & 1.30 & 0.45 & 1.30 \\
2 &  &  & Mid-B & 29 & 0.60 & $-0.08$ & $0.68 \pm 0.05$ & $-16$ & 0.05 & 0.07 & $0.61 \pm 0.06$ & $0.94 \pm 0.09$ &  & 1.37 & 0.35 & 1.13 \\
3 &  &  & Late-B & 16 & 0.48 & $-0.04$ & $0.52 \pm 0.04$ & $-15$ & 0.05 & 0.09 & $0.43 \pm 0.05$ & $0.66 \pm 0.08$ &  & 0.81 & 0.27 & 0.79 \\
4 & B7V & 66 & B6 & 56 & 0.41 & $-0.06$ & $0.47 \pm 0.02$ & $-11$ & 0.04 & 0.07 & $0.40 \pm 0.04$ & $0.62 \pm 0.06$ & $0.64 \pm 0.07$ & 1.36 & 0.34 & 1.23 \\
5 &  &  & B6 & 44 & 0.48 & $-0.07$ & $0.55 \pm 0.01$ & $-25$ & 0.08 & 0.12 & $0.43 \pm 0.04$ & $0.66 \pm 0.06$ &  & 1.63 & 0.35 & 1.14 \\
6 & B3III & 31 & Early-B & 38 & 1.06 & $-0.12$ & $1.18 \pm 0.02$ & $-51$ & 0.17 & 0.19 & $0.99 \pm 0.05$ & $1.52 \pm 0.08$ & $1.54 \pm 0.08$ & 1.57 & 0.74 & 1.59 \\
7 & A0V & 26 & Late-B & 26 & 0.73 & $0.00$ & $0.73 \pm 0.02$ & $-28$ & 0.09 & 0.19 & $0.54 \pm 0.05$ & $0.83 \pm 0.08$ & $0.98 \pm 0.09$ & 1.53 & 0.63 & 1.55 \\
8 & B2V & 34 &  &  & 0.71 & $-0.13$ & $0.84 \pm 0.02$ & $-52$ & 0.17 & 0.18 & $0.66 \pm 0.04$ & $1.02 \pm 0.06$ & $1.07 \pm 0.08$ & 1.33 & 0.34 & 1.17 \\
9 &  &  & Early-B & 29 & 0.68 & $-0.13$ & $0.81 \pm 0.05$ & $-17$ & 0.06 & 0.07 & $0.74 \pm 0.05$ & $1.14 \pm 0.08$ &  & 1.32 & 0.53 & 1.15 \\
10 & B3V & 24 & Early-B & 25 & 0.96 & $-0.12$ & $1.08 \pm 0.02$ & $-26$ & 0.09 & 0.11 & $0.97 \pm 0.04$ & $1.49 \pm 0.06$ & $1.34 \pm 0.10$ & 1.30 & 0.48 & 1.29 \\
11 &  &  & Mid-B & 48 & 0.64 & $-0.08$ & $0.71 \pm 0.05$ & $-23$ & 0.08 & 0.11 & $0.60 \pm 0.06$ & $0.92 \pm 0.09$ &  & 1.30 & 0.41 & 1.23 \\
12 & B7V & 32 & Late-B & 35 & 0.60 & $-0.06$ & $0.66 \pm 0.02$ & $-14$ & 0.05 & 0.08 & $0.58 \pm 0.04$ & $0.89 \pm 0.06$ & $0.81 \pm 0.07$ & 1.30 & 0.42 & 1.11 \\
13 &  &  & B7 & 34 & 0.46 & $-0.06$ & $0.52 \pm 0.02$ & $-16$ & 0.05 & 0.08 & $0.44 \pm 0.04$ & $0.68 \pm 0.06$ &  & 1.20 & 0.38 & 1.12 \\
14 &  &  & B6 & 48 & 0.58 & $-0.07$ & $0.65 \pm 0.01$ & $-12$ & 0.04 & 0.06 & $0.59 \pm 0.03$ & $0.91 \pm 0.05$ &  & 1.59 & 0.60 & 1.25 \\
15 &  &  & Late-B & 26 & 0.87 & $-0.04$ & $0.91 \pm 0.04$ & $-19$ & 0.06 & 0.10 & $0.81 \pm 0.06$ & $1.25 \pm 0.09$ &  & 1.43 & 0.50 & 1.19 \\
16 &  &  & B5 & 32 & 0.41 & $-0.08$ & $0.49 \pm 0.01$ & $-29$ & 0.10 & 0.13 & $0.36 \pm 0.04$ & $0.55 \pm 0.06$ &  & 1.13 & 0.28 & 1.14 \\
17 & B3V & 29 & Mid-B & 31 & 0.66 & $-0.12$ & $0.78 \pm 0.02$ & $-30$ & 0.10 & 0.12 & $0.66 \pm 0.04$ & $1.02 \pm 0.06$ & $1.07 \pm 0.07$ & 1.10 & 0.30 & 1.07 \\
18 & B5V & 33 & B5 & 66 & 0.56 & $-0.08$ & $0.64 \pm 0.01$ & $-19$ & 0.06 & 0.08 & $0.56 \pm 0.04$ & $0.86 \pm 0.06$ & $0.91 \pm 0.08$ & 1.38 & 0.62 & 1.27 \\
19 & B7IV & 40 & Mid-B & 40 & 0.71 & $-0.06$ & $0.77 \pm 0.02$ & $-34$ & 0.11 & 0.17 & $0.60 \pm 0.04$ & $0.92 \pm 0.06$ & $1.05 \pm 0.08$ & 1.43 & 0.49 & 1.14 \\
20 &  &  & Late-B & 15 & 0.59 & $-0.04$ & $0.63 \pm 0.04$ & $-17$ & 0.06 & 0.10 & $0.53 \pm 0.05$ & $0.82 \pm 0.08$ &  & 1.07 & 0.34 & 1.11 \\
21 & B2-3V & 54 & Mid-B & 23 & 0.67 & $-0.12$ & $0.79 \pm 0.02$ & $-51$ & 0.17 & 0.18 & $0.61 \pm 0.05$ & $0.94 \pm 0.08$ & $0.88 \pm 0.08$ & 0.92 & 0.38 & 0.88 \\
22 & B5V & 49 & Late-B & 34 & 0.56 & $-0.08$ & $0.64 \pm 0.01$ & $-17$ & 0.06 & 0.08 & $0.56 \pm 0.03$ & $0.86 \pm 0.05$ & $0.83 \pm 0.09$ & 0.99 & 0.23 & 1.06 \\
23 & B5V & 81 & Late-B & 36 & 0.42 & $-0.08$ & $0.50 \pm 0.01$ & $-11$ & 0.04 & 0.06 & $0.44 \pm 0.03$ & $0.68 \pm 0.05$ & $0.64 \pm 0.08$ & 0.88 & 0.18 & 0.88 \\
24 &  &  & B5 & 42 & 0.45 & $-0.08$ & $0.53 \pm 0.01$ & $-13$ & 0.04 & 0.06 & $0.47 \pm 0.03$ & $0.72 \pm 0.05$ &  & 1.30 & 0.41 & 1.31 \\
25 &  &  & Late-B & 23 & 0.51 & $-0.04$ & $0.55 \pm 0.04$ & $-30$ & 0.10 & 0.16 & $0.39 \pm 0.06$ & $0.60 \pm 0.09$ &  & 0.91 & 0.20 & 0.84 \\
26 &  &  & B3 & 26 & 0.47 & $-0.12$ & $0.59 \pm 0.02$ & $-54$ & 0.18 & 0.20 & $0.39 \pm 0.05$ & $0.60 \pm 0.08$ &  & 1.10 & 0.25 & 1.18 \\
27 &  &  & B8 & 45 & 0.49 & $-0.04$ & $0.53 \pm 0.02$ & $-28$ & 0.09 & 0.15 & $0.38 \pm 0.05$ & $0.58 \pm 0.08$ &  & 1.45 & 0.37 & 1.02 \\
28 & B4V & 46 & continuum+Balmer & 8 & 0.78 & $-0.09$ & $0.87 \pm 0.02$ & $-30$ & 0.10 & 0.13 & $0.74 \pm 0.04$ & $1.14 \pm 0.06$ & $1.09 \pm 0.08$ & 1.18 & 0.34 & 1.09 \\
29 &  &  & Late-B & 37 & 0.53 & $-0.04$ & $0.57 \pm 0.04$ & $-13$ & 0.04 & 0.07 & $0.50 \pm 0.05$ & $0.77 \pm 0.08$ &  & 1.65 & 0.48 & 1.30 \\
30 &  &  & Late-B & 28 & 0.62 & $-0.04$ & $0.66 \pm 0.04$ & $-27$ & 0.09 & 0.15 & $0.51 \pm 0.06$ & $0.78 \pm 0.09$ &  & 1.66 & 0.57 & 1.42 \\
31 &  &  & Late-B & 39 & 0.58 & $-0.04$ & $0.62 \pm 0.04$ & $-19$ & 0.06 & 0.10 & $0.52 \pm 0.06$ & $0.80 \pm 0.09$ &  & 1.20 & 0.41 & 1.17 \\
32 &  &  & Late-B & 41 & 0.65 & $-0.04$ & $0.69 \pm 0.04$ & $-45$ & 0.15 & 0.23 & $0.46 \pm 0.06$ & $0.71 \pm 0.09$ &  & 1.65 & 0.48 & 1.25 \\
33 & B5V & 58 & continuum+Balmer & 10 & 0.76 & $-0.08$ & $0.83 \pm 0.01$ & $-24$ & 0.08 & 0.11 & $0.72 \pm 0.04$ & $1.11 \pm 0.06$ & $1.06 \pm 0.07$ & 1.43 & 0.47 & 1.26 \\
34 &  &  & B6 & 53 & 0.38 & $-0.07$ & $0.45 \pm 0.01$ & $-20$ & 0.07 & 0.10 & $0.35 \pm 0.04$ & $0.54 \pm 0.06$ &  & 0.72 & 0.19 & 0.77 \\
35 & B3III & 53 & continuum+Balmer & 12 & 0.87 & $-0.12$ & $0.99 \pm 0.02$ & $-30$ & 0.10 & 0.12 & $0.87 \pm 0.04$ & $1.34 \pm 0.06$ & $1.22 \pm 0.08$ & 1.64 & 0.52 & 1.20 \\
36 &  &  & Late-B & 30 & 0.34 & $-0.04$ & $0.38 \pm 0.04$ & $-3$ & 0.01 & 0.02 & $0.36 \pm 0.05$ & $0.55 \pm 0.08$ &  & 0.99 & 0.24 & 0.84 \\
37 &  &  & B9 & 38 & 0.69 & $-0.02$ & $0.71 \pm 0.02$ & $-31$ & 0.10 & 0.19 & $0.52 \pm 0.05$ & $0.80 \pm 0.08$ &  & 1.68 & 0.56 & 1.45 \\
38 &  &  & B1 & 39 & 0.56 & $-0.15$ & $0.71 \pm 0.02$ & $-4$ & 0.01 & 0.01 & $0.70 \pm 0.02$ & $1.08 \pm 0.03$ &  & 1.51 & 0.71 & 1.43 \\
39 & B3IV & 35 & continuum+Balmer & 16 & 1.03 & $-0.12$ & $1.15 \pm 0.02$ & $-99$ & 0.33 & 0.32 & $0.83 \pm 0.06$ & $1.28 \pm 0.09$ & $1.36 \pm 0.08$ & 1.26 & 0.61 & 1.12 \\
40 &  &  & Late-B & 15 & 0.77 & $-0.04$ & $0.81 \pm 0.04$ & $-19$ & 0.06 & 0.10 & $0.71 \pm 0.06$ & $1.09 \pm 0.09$ &  & 1.61 & 0.54 & 1.38 \\
41 &  &  & Mid-B & 42 & 0.57 & $-0.08$ & $0.65 \pm 0.05$ & $-48$ & 0.16 & 0.20 & $0.45 \pm 0.06$ & $0.69 \pm 0.09$ &  & 1.23 & 0.43 & 1.09 \\
42 & B4V & 49 & Mid-B & 14 & 0.70 & $-0.09$ & $0.79 \pm 0.02$ & $-47$ & 0.16 & 0.19 & $0.60 \pm 0.05$ & $0.92 \pm 0.08$ & $0.99 \pm 0.08$ & 1.20 & 0.41 & 1.10 \\
43 & B3V & 41 & continuum+Balmer & 26 & 0.95 & $-0.12$ & $1.07 \pm 0.02$ & $-75$ & 0.25 & 0.26 & $0.81 \pm 0.06$ & $1.25 \pm 0.09$ & $1.28 \pm 0.08$ & 1.63 & 0.71 & 1.23 \\
44 &  &  & Late-B & 18 & 0.68 & $-0.04$ & $0.72 \pm 0.04$ & $-25$ & 0.08 & 0.14 & $0.58 \pm 0.06$ & $0.89 \pm 0.09$ &  & 1.28 & 0.41 & 1.15 \\
45 &  &  & Late-B & 13 & 0.82 & $-0.04$ & $0.86 \pm 0.04$ & $-26$ & 0.09 & 0.15 & $0.71 \pm 0.06$ & $1.09 \pm 0.09$ &  & 1.69 & 0.65 & 1.45 \\
46 & B5V & 27 & Late-B & 14 & 0.84 & $-0.08$ & $0.92 \pm 0.01$ & $-44$ & 0.15 & 0.19 & $0.73 \pm 0.04$ & $1.12 \pm 0.06$ & $1.33 \pm 0.09$ & 2.06 & 0.51 & 1.23 \\
47 &  &  & B7 & 36 & 0.48 & $-0.06$ & $0.54 \pm 0.02$ & $-20$ & 0.07 & 0.11 & $0.43 \pm 0.04$ & $0.66 \pm 0.06$ &  & 1.20 & 0.38 & 0.98 \\
48 & B6IV & 27 & Late-B & 16 & 0.64 & $-0.07$ & $0.71 \pm 0.01$ & $-27$ & 0.09 & 0.13 & $0.58 \pm 0.04$ & $0.89 \pm 0.06$ & $1.10 \pm 0.07$ & 1.02 & 0.62 & 0.90 \\
49 &  &  & B3 & 47 & 0.67 & $-0.12$ & $0.79 \pm 0.02$ & $-17$ & 0.06 & 0.08 & $0.71 \pm 0.03$ & $1.09 \pm 0.05$ &  & 1.60 & 0.46 & 1.28 \\
50 & B5V & 79 & B5 & 39 & 0.48 & $-0.08$ & $0.56 \pm 0.01$ & $-26$ & 0.09 & 0.12 & $0.44 \pm 0.04$ & $0.68 \pm 0.06$ & $0.82 \pm 0.07$ & 1.02 & 0.37 & 1.27 \\
51 & B4III & 39 & continuum+Balmer & 11 & 0.89 & $-0.09$ & $0.98 \pm 0.02$ & $-26$ & 0.09 & 0.12 & $0.86 \pm 0.04$ & $1.32 \pm 0.06$ & $1.29 \pm 0.07$ & 1.20 & 0.60 & 1.27 \\
52 &  &  & B3 & 44 & 0.65 & $-0.12$ & $0.77 \pm 0.02$ & $-24$ & 0.08 & 0.10 & $0.67 \pm 0.04$ & $1.03 \pm 0.06$ &  & 1.20 & 0.51 & 1.24 \\
53 & B3V & 48 & Mid-B & 30 & 0.84 & $-0.12$ & $0.96 \pm 0.02$ & $-53$ & 0.18 & 0.20 & $0.76 \pm 0.05$ & $1.17 \pm 0.08$ & $1.31 \pm 0.07$ & 1.69 & 0.50 & 1.33 \\
54 &  &  & B3 & 32 & 0.46 & $-0.12$ & $0.58 \pm 0.02$ & $-34$ & 0.11 & 0.13 & $0.45 \pm 0.04$ & $0.69 \pm 0.06$ &  & 1.02 & 0.34 & 1.22 \\
55 & B7V & 37 & continuum+Balmer & 15 & 0.74 & $-0.06$ & $0.80 \pm 0.02$ & $-48$ & 0.16 & 0.23 & $0.57 \pm 0.05$ & $0.88 \pm 0.08$ & $0.99 \pm 0.09$ & 1.20 & 0.42 & 1.15 \\
56* &  &  & Early-B & 24 & 1.10 & $-0.13$ & $1.23 \pm 0.15$ & $-92$ & 0.31 & 0.29 & $0.94 \pm 0.16$ & $1.45 \pm 0.25$ &  & 1.37 & 0.30 & 1.17 \\
57 &  &  & B4 & 37 & 0.54 & $-0.09$ & $0.63 \pm 0.02$ & $-14$ & 0.05 & 0.07 & $0.56 \pm 0.03$ & $0.86 \pm 0.05$ &  & 1.20 & 0.57 & 1.08 \\
58 & B3V & 29 & continuum+Balmer & 18 & 0.83 & $-0.12$ & $0.95 \pm 0.02$ & $-91$ & 0.30 & 0.30 & $0.65 \pm 0.06$ & $1.00 \pm 0.09$ & $1.00 \pm 0.09$ & 1.18 & 0.37 & 0.96 \\
59 &  &  & B6 & 49 & 0.41 & $-0.07$ & $0.48 \pm 0.01$ & $-39$ & 0.13 & 0.18 & $0.30 \pm 0.05$ & $0.46 \pm 0.08$ &  & 1.14 & 0.30 & 1.05 \\
60 &  &  & Late-B & 21 & 0.64 & $-0.04$ & $0.68 \pm 0.04$ & $-19$ & 0.06 & 0.10 & $0.58 \pm 0.06$ & $0.89 \pm 0.09$ &  & 1.22 & 0.52 & 1.27 \\
61 &  &  & B3 & 33 & 0.66 & $-0.12$ & $0.78 \pm 0.02$ & $-76$ & 0.25 & 0.26 & $0.52 \pm 0.06$ & $0.80 \pm 0.09$ &  & 0.96 & 0.31 & 0.90 \\
62 &  &  & Mid-B & 23 & 0.54 & $-0.08$ & $0.62 \pm 0.05$ & $-18$ & 0.06 & 0.08 & $0.54 \pm 0.06$ & $0.83 \pm 0.09$ &  & 0.96 & 0.40 & 0.90 \\
63 &  &  & Mid-B & 19 & 0.66 & $-0.08$ & $0.74 \pm 0.05$ & $-25$ & 0.08 & 0.11 & $0.63 \pm 0.06$ & $0.97 \pm 0.09$ &  & 0.97 & 0.40 & 0.90 \\
64 &  &  & B6 & 46 & 0.62 & $-0.07$ & $0.69 \pm 0.01$ & $-16$ & 0.05 & 0.08 & $0.61 \pm 0.03$ & $0.94 \pm 0.05$ &  & 1.05 & 0.48 & 1.23 \\
65 & B5III & 42 & Late-B & 36 & 0.61 & $-0.08$ & $0.69 \pm 0.01$ & $-20$ & 0.07 & 0.10 & $0.59 \pm 0.04$ & $0.91 \pm 0.06$ & $1.05 \pm 0.07$ & 1.12 & 0.34 & 0.90 \\
66 &  &  & Late-B & 34 & 0.41 & $-0.04$ & $0.45 \pm 0.04$ & $-11$ & 0.04 & 0.07 & $0.38 \pm 0.05$ & $0.58 \pm 0.08$ &  & 1.24 & 0.43 & 1.15 \\
67 &  &  & B6 & 31 & 0.38 & $-0.07$ & $0.45 \pm 0.01$ & $-16$ & 0.05 & 0.08 & $0.37 \pm 0.03$ & $0.57 \pm 0.05$ &  & 1.39 & 0.38 & 1.04 \\
68 &  &  & B7 & 34 & 0.37 & $-0.06$ & $0.43 \pm 0.02$ & $-25$ & 0.08 & 0.13 & $0.30 \pm 0.04$ & $0.46 \pm 0.06$ &  & 1.02 & 0.02 & 0.86 \\
69 &  &  & Late-B & 20 & 0.55 & $-0.04$ & $0.59 \pm 0.04$ & $-25$ & 0.08 & 0.14 & $0.45 \pm 0.06$ & $0.69 \pm 0.09$ &  & 1.02 & 0.29 & 0.86 \\
70 & B7IV & 66 & Late-B & 28 & 0.43 & $-0.06$ & $0.49 \pm 0.02$ & $-18$ & 0.06 & 0.10 & $0.39 \pm 0.04$ & $0.60 \pm 0.06$ & $0.77 \pm 0.07$ & 1.22 & 0.42 & 1.07 \\
71 &  &  & Early-B & 24 & 0.61 & $-0.13$ & $0.74 \pm 0.05$ & $-33$ & 0.11 & 0.12 & $0.62 \pm 0.06$ & $0.95 \pm 0.09$ &  & 1.37 & 0.34 & 0.81 \\
72 &  &  & Mid-B & 36 & 0.43 & $-0.08$ & $0.51 \pm 0.05$ & $-31$ & 0.10 & 0.13 & $0.38 \pm 0.06$ & $0.58 \pm 0.09$ &  & 1.17 & 0.41 & 1.05 \\
73 &  &  & B2 & 53 & 0.79 & $-0.13$ & $0.92 \pm 0.02$ & $-45$ & 0.15 & 0.16 & $0.76 \pm 0.04$ & $1.17 \pm 0.06$ &  & 1.80 & 0.38 & 1.36 \\
74 &  &  & B4 & 40 & 0.57 & $-0.09$ & $0.66 \pm 0.02$ & $-48$ & 0.16 & 0.19 & $0.47 \pm 0.05$ & $0.72 \pm 0.08$ &  & 1.97 & 0.05 & 1.36 \\
75 & B7V & 44 & B5 & 32 & 0.49 & $-0.06$ & $0.55 \pm 0.02$ & $-15$ & 0.05 & 0.08 & $0.47 \pm 0.04$ & $0.72 \pm 0.06$ & $0.80 \pm 0.08$ & 0.96 & 0.40 & 0.96 \\
76 &  &  & B3 & 43 & 0.45 & $-0.12$ & $0.57 \pm 0.02$ & $-7$ & 0.02 & 0.03 & $0.54 \pm 0.04$ & $0.83 \pm 0.06$ &  & 1.95 & -0.08 & 1.24 \\
77 &  &  & Mid-B & 29 & 0.62 & $-0.08$ & $0.70 \pm 0.05$ & $-43$ & 0.14 & 0.18 & $0.52 \pm 0.06$ & $0.80 \pm 0.09$ &  & 0.97 & 0.23 & 0.87 \\
78 &  &  & Late-B & 26 & 0.45 & $-0.04$ & $0.49 \pm 0.04$ & $-24$ & 0.08 & 0.14 & $0.35 \pm 0.06$ & $0.54 \pm 0.09$ &  & 1.10 & 0.42 & 1.03 \\
79 & B3IV & 46 & Mid-B & 31 & 0.83 & $-0.12$ & $0.95 \pm 0.02$ & $-35$ & 0.12 & 0.14 & $0.81 \pm 0.05$ & $1.25 \pm 0.08$ & $1.24 \pm 0.07$ & 1.50 & 0.36 & 1.05 \\
80 &  &  & Late-B & 17 & 0.56 & $-0.04$ & $0.60 \pm 0.04$ & $-15$ & 0.05 & 0.09 & $0.51 \pm 0.05$ & $0.78 \pm 0.08$ &  & 1.15 & 0.29 & 0.95 \\
81 &  &  & B6 & 35 & 0.51 & $-0.07$ & $0.58 \pm 0.01$ & $-14$ & 0.05 & 0.08 & $0.50 \pm 0.03$ & $0.77 \pm 0.05$ &  & 0.88 & 0.42 & 0.89 \\
82 &  &  & B9 & 24 & 0.61 & $-0.02$ & $0.63 \pm 0.02$ & $-20$ & 0.07 & 0.14 & $0.49 \pm 0.06$ & $0.75 \pm 0.09$ &  & 1.47 & 0.43 & 0.99 \\
83 & B4V & 44 & Mid-B & 33 & 0.41 & $-0.09$ & $0.49 \pm 0.02$ & $-14$ & 0.05 & 0.07 & $0.42 \pm 0.03$ & $0.65 \pm 0.05$ & $0.77 \pm 0.07$ & 0.83 & 0.17 & 0.65 \\
84 &  &  & B7 & 82 & 0.50 & $-0.06$ & $0.56 \pm 0.02$ & $-8$ & 0.03 & 0.05 & $0.51 \pm 0.04$ & $0.78 \pm 0.06$ &  & 1.45 & 0.42 & 1.19 \\
85 &  &  & Mid-B & 22 & 0.58 & $-0.08$ & $0.66 \pm 0.05$ & $-18$ & 0.06 & 0.08 & $0.58 \pm 0.06$ & $0.89 \pm 0.09$ &  & 1.12 & 0.35 & 1.01 \\
\hline
\end{tabular}
\end{minipage}
\end{table*}

\begin{table*}
\tiny
\begin{minipage}{180mm}
\contcaption{}
\centering
\begin{tabular}{@{}llrlrccrcrcrrrrrr@{}}
\hline
\# & \multicolumn{4}{c}{Spectral typing} & $(r-i)_{c}$ & $(r-i)_{\circ}$ & $E(r-i)$ & $EW(\Halpha)$ & $f_{D}$ & $E^{cs}(r-i)$ & $E^{is}(r-i)$ & $E^{is}(B-V)$ & $E^{is}(B-V)_{\rm{R13}}$ & SFD98\,$\times 0.86$ & RF09 & Sale14 \\
  & R13 & S/N & This work & S/N & (mag) & (mag) & (mag) & (mag) & & (mag) & (mag) & (mag) & (mag) & (mag) & (mag) & (mag) )\\
\hline
86 &  &  & B7 & 67 & 0.31 & $-0.06$ & $0.37 \pm 0.02$ & $-12$ & 0.04 & 0.07 & $0.30 \pm 0.04$ & $0.46 \pm 0.06$ &  & 1.07 & 0.32 & 0.85 \\
87 &  &  & Mid-B & 17 & 0.58 & $-0.08$ & $0.66 \pm 0.05$ & $-34$ & 0.11 & 0.15 & $0.51 \pm 0.06$ & $0.78 \pm 0.09$ &  & 1.06 & 0.37 & 0.83 \\
88 & B8-9III & 66 &  &  & 0.51 & $-0.03$ & $0.54 \pm 0.02$ & $-11$ & 0.04 & 0.08 & $0.46 \pm 0.05$ & $0.71 \pm 0.08$ & $0.86 \pm 0.07$ & 0.88 & 0.25 & 0.78 \\
89 &  &  & B5 & 36 & 0.41 & $-0.08$ & $0.49 \pm 0.01$ & $-11$ & 0.04 & 0.06 & $0.43 \pm 0.03$ & $0.66 \pm 0.05$ &  & 0.83 & 0.25 & 0.76 \\
90 &  &  & B6 & 34 & 0.39 & $-0.07$ & $0.46 \pm 0.01$ & $-6$ & 0.02 & 0.03 & $0.43 \pm 0.03$ & $0.66 \pm 0.05$ &  & 0.93 & 0.33 & 0.99 \\
91 &  &  & B7 & 26 & 0.33 & $-0.06$ & $0.39 \pm 0.02$ & $-3$ & 0.01 & 0.02 & $0.36 \pm 0.03$ & $0.55 \pm 0.05$ &  & 0.78 & 0.31 & 0.70 \\
92 & B4V & 31 & continuum+Balmer & 17 & 1.12 & $-0.09$ & $1.21 \pm 0.02$ & $-48$ & 0.16 & 0.19 & $1.02 \pm 0.05$ & $1.57 \pm 0.08$ & $1.43 \pm 0.10$ & 1.77 & 0.63 & 1.35 \\
93 &  &  & Late-B & 17 & 0.50 & $-0.04$ & $0.54 \pm 0.04$ & $-12$ & 0.04 & 0.07 & $0.47 \pm 0.05$ & $0.72 \pm 0.08$ &  & 1.11 & 0.28 & 0.80 \\
94 &  &  & B3 & 52 & 0.50 & $-0.12$ & $0.62 \pm 0.02$ & $-15$ & 0.05 & 0.06 & $0.56 \pm 0.03$ & $0.86 \pm 0.05$ &  & 1.43 & 0.52 & 1.24 \\
95 &  &  & B3 & 32 & 0.56 & $-0.12$ & $0.68 \pm 0.02$ & $-34$ & 0.11 & 0.13 & $0.55 \pm 0.04$ & $0.85 \pm 0.06$ &  & 0.91 & 0.30 & 0.68 \\
96 &  &  & Late-B & 15 & 0.57 & $-0.04$ & $0.61 \pm 0.04$ & $-17$ & 0.06 & 0.10 & $0.51 \pm 0.05$ & $0.78 \pm 0.08$ &  & 1.36 & 0.36 & 1.01 \\
97 & B3V & 46 & B3 & 36 & 0.48 & $-0.12$ & $0.60 \pm 0.02$ & $-41$ & 0.14 & 0.16 & $0.44 \pm 0.05$ & $0.68 \pm 0.08$ & $0.64 \pm 0.08$ & 1.01 & 0.34 & 0.83 \\
98 &  &  & B1 & 72 & 0.44 & $-0.15$ & $0.59 \pm 0.02$ & $-25$ & 0.08 & 0.09 & $0.50 \pm 0.04$ & $0.77 \pm 0.06$ &  & 1.19 & 0.47 & 1.14 \\
99 &  &  & B4-5 & 75 & 0.28 & $-0.08$ & $0.36 \pm 0.02$ & $-32$ & 0.11 & 0.14 & $0.22 \pm 0.04$ & $0.34 \pm 0.06$ &  & 0.83 & 0.15 & 0.69 \\
100 &  &  & Mid-B & 22 & 0.55 & $-0.08$ & $0.63 \pm 0.05$ & $-10$ & 0.03 & 0.04 & $0.59 \pm 0.06$ & $0.91 \pm 0.09$ &  & 0.93 & 0.24 & 0.85 \\
101 &  &  & B3 & 40 & 0.59 & $-0.12$ & $0.71 \pm 0.02$ & $-23$ & 0.08 & 0.10 & $0.61 \pm 0.04$ & $0.94 \pm 0.06$ &  & 1.40 & 0.25 & 0.88 \\
102 &  &  & Late-B & 15 & 0.72 & $-0.04$ & $0.76 \pm 0.04$ & $-17$ & 0.06 & 0.10 & $0.66 \pm 0.05$ & $1.02 \pm 0.08$ &  & 2.10 & 0.79 & 1.72 \\
103 &  &  & B4 & 70 & 0.39 & $-0.09$ & $0.48 \pm 0.02$ & $-11$ & 0.04 & 0.05 & $0.43 \pm 0.04$ & $0.66 \pm 0.06$ &  & 0.75 & 0.16 & 0.71 \\
104 &  &  & Mid-B & 23 & 0.73 & $-0.08$ & $0.81 \pm 0.05$ & $-28$ & 0.09 & 0.12 & $0.70 \pm 0.06$ & $1.08 \pm 0.09$ &  & 1.12 & 0.30 & 1.17 \\
105 & B6IV & 59 & Late-B & 20 & 0.56 & $-0.07$ & $0.63 \pm 0.01$ & $-26$ & 0.09 & 0.13 & $0.50 \pm 0.04$ & $0.77 \pm 0.06$ & $0.97 \pm 0.08$ & 0.80 & 0.31 & 1.04 \\
106 &  &  & B4 & 81 & 0.34 & $-0.09$ & $0.43 \pm 0.02$ & $-31$ & 0.10 & 0.13 & $0.30 \pm 0.04$ & $0.46 \pm 0.06$ &  & 0.78 & 0.30 & 0.67 \\
107 & B2-3V & 46 & continuum+Balmer & 14 & 0.59 & $-0.12$ & $0.72 \pm 0.02$ & $-64$ & 0.21 & 0.22 & $0.50 \pm 0.05$ & $0.77 \pm 0.08$ & $0.84 \pm 0.08$ & 0.94 & 0.18 & 0.78 \\
108 &  &  & B5 & 38 & 0.40 & $-0.08$ & $0.48 \pm 0.01$ & $-30$ & 0.10 & 0.13 & $0.34 \pm 0.04$ & $0.52 \pm 0.06$ &  & 0.82 & 0.27 & 0.65 \\
109 &  &  & Late-B & 21 & 0.53 & $-0.04$ & $0.57 \pm 0.04$ & $-10$ & 0.03 & 0.06 & $0.51 \pm 0.06$ & $0.78 \pm 0.09$ &  & 1.12 & 0.32 & 0.89 \\
110 &  &  & Mid-B & 34 & 0.40 & $-0.08$ & $0.48 \pm 0.05$ & $-8$ & 0.03 & 0.04 & $0.44 \pm 0.06$ & $0.68 \pm 0.09$ &  & 0.71 & 0.17 & 0.70 \\
111 &  &  & B3 & 38 & 0.85 & $-0.12$ & $0.97 \pm 0.02$ & $-20$ & 0.07 & 0.09 & $0.88 \pm 0.04$ & $1.35 \pm 0.06$ &  & 0.90 & 0.36 & 0.84 \\
112 &  &  & Mid-B & 31 & 0.69 & $-0.08$ & $0.77 \pm 0.05$ & $-39$ & 0.13 & 0.17 & $0.60 \pm 0.06$ & $0.92 \pm 0.09$ &  & 1.09 & 0.34 & 0.93 \\
113 &  &  & Mid-B & 27 & 0.45 & $-0.08$ & $0.53 \pm 0.05$ & $-8$ & 0.03 & 0.04 & $0.49 \pm 0.06$ & $0.75 \pm 0.09$ &  & 1.01 & 0.48 & 0.89 \\
114 &  &  & B7 & 66 & 0.47 & $-0.06$ & $0.53 \pm 0.02$ & $-11$ & 0.04 & 0.07 & $0.46 \pm 0.04$ & $0.71 \pm 0.06$ &  & 1.13 & 0.41 & 0.98 \\
115 &  &  & Early-B & 20 & 0.80 & $-0.13$ & $0.93 \pm 0.05$ & $-43$ & 0.14 & 0.15 & $0.78 \pm 0.06$ & $1.20 \pm 0.09$ &  & 1.03 & 0.25 & 0.78 \\
116 &  &  & Mid-B & 20 & 0.61 & $-0.08$ & $0.69 \pm 0.05$ & $-27$ & 0.09 & 0.12 & $0.57 \pm 0.06$ & $0.88 \pm 0.09$ &  & 1.09 & 0.32 & 1.10 \\
117 &  &  & Mid-B & 31 & 0.55 & $-0.08$ & $0.62 \pm 0.05$ & $-21$ & 0.07 & 0.10 & $0.52 \pm 0.06$ & $0.80 \pm 0.09$ &  & 0.91 & 0.35 & 0.80 \\
118 &  &  & B7 & 52 & 0.42 & $-0.06$ & $0.48 \pm 0.02$ & $-17$ & 0.06 & 0.10 & $0.38 \pm 0.04$ & $0.58 \pm 0.06$ &  & 0.79 & 0.22 & 0.62 \\
119 &  &  & B5 & 32 & 0.36 & $-0.08$ & $0.44 \pm 0.01$ & $-15$ & 0.05 & 0.07 & $0.37 \pm 0.03$ & $0.57 \pm 0.05$ &  & 0.77 & 0.16 & 0.75 \\
120 &  &  & Late-B & 24 & 0.57 & $-0.04$ & $0.61 \pm 0.04$ & $-11$ & 0.04 & 0.07 & $0.54 \pm 0.05$ & $0.83 \pm 0.08$ &  & 0.98 & 0.23 & 0.87 \\
121 &  &  & B1 & 62 & 0.40 & $-0.15$ & $0.55 \pm 0.02$ & $-16$ & 0.05 & 0.06 & $0.49 \pm 0.03$ & $0.75 \pm 0.05$ &  & 0.89 & 0.14 & 0.75 \\
122 &  &  & Mid-B & 36 & 0.43 & $-0.08$ & $0.51 \pm 0.05$ & $-34$ & 0.11 & 0.15 & $0.36 \pm 0.06$ & $0.55 \pm 0.09$ &  & 0.78 & 0.27 & 0.73 \\
123 &  &  & Mid-B & 43 & 0.39 & $-0.08$ & $0.47 \pm 0.05$ & $-11$ & 0.04 & 0.06 & $0.41 \pm 0.06$ & $0.63 \pm 0.09$ &  & 1.14 & 0.32 & 1.23 \\
124 &  &  & B3 & 48 & 0.36 & $-0.12$ & $0.48 \pm 0.02$ & $-33$ & 0.11 & 0.13 & $0.35 \pm 0.04$ & $0.54 \pm 0.06$ &  & 0.77 & 0.38 & 0.74 \\
125 &  &  & Late-B & 23 & 0.67 & $-0.04$ & $0.71 \pm 0.04$ & $-12$ & 0.04 & 0.07 & $0.64 \pm 0.05$ & $0.98 \pm 0.08$ &  & 0.88 & 0.29 & 0.91 \\
126 & B6V & 49 & Late-B & 27 & 0.57 & $-0.07$ & $0.64 \pm 0.01$ & $-16$ & 0.05 & 0.08 & $0.56 \pm 0.03$ & $0.86 \pm 0.05$ & $0.89 \pm 0.08$ & 1.02 & 0.32 & 0.91 \\
127 &  &  & Late-B & 19 & 0.33 & $-0.04$ & $0.37 \pm 0.04$ & $-11$ & 0.04 & 0.07 & $0.30 \pm 0.05$ & $0.46 \pm 0.08$ &  & 0.73 & 0.17 & 0.74 \\
128 &  &  & B3 & 36 & 0.57 & $-0.12$ & $0.69 \pm 0.02$ & $-35$ & 0.12 & 0.14 & $0.55 \pm 0.05$ & $0.85 \pm 0.08$ &  & 1.23 & 0.58 & 1.21 \\
129 &  &  & A0 & 28 & 0.65 & $0.00$ & $0.65 \pm 0.01$ & $-22$ & 0.07 & 0.15 & $0.50 \pm 0.06$ & $0.77 \pm 0.09$ &  & 1.08 & 0.34 & 0.86 \\
130 &  &  & Late-B & 19 & 0.54 & $-0.04$ & $0.58 \pm 0.04$ & $-27$ & 0.09 & 0.15 & $0.43 \pm 0.06$ & $0.66 \pm 0.09$ &  & 0.87 & 0.49 & 0.77 \\
131 &  &  & B3 & 41 & 0.55 & $-0.12$ & $0.67 \pm 0.02$ & $-35$ & 0.12 & 0.14 & $0.53 \pm 0.05$ & $0.82 \pm 0.08$ &  & 0.96 & 0.22 & 0.74 \\
132 &  &  & Late-B & 34 & 0.41 & $-0.04$ & $0.45 \pm 0.04$ & $-9$ & 0.03 & 0.06 & $0.39 \pm 0.06$ & $0.60 \pm 0.09$ &  & 0.79 & 0.35 & 0.72 \\
133 &  &  & B9 & 35 & 0.84 & $-0.02$ & $0.86 \pm 0.02$ & $-21$ & 0.07 & 0.14 & $0.72 \pm 0.05$ & $1.11 \pm 0.08$ &  & 1.34 & 0.54 & 1.43 \\
134 &  &  & B8 & 53 & 0.35 & $-0.04$ & $0.39 \pm 0.02$ & $-11$ & 0.04 & 0.07 & $0.32 \pm 0.04$ & $0.49 \pm 0.06$ &  & 0.81 & 0.22 & 0.74 \\
135 &  &  & Late-B & 13 & 0.51 & $-0.04$ & $0.55 \pm 0.04$ & $-10$ & 0.03 & 0.06 & $0.49 \pm 0.06$ & $0.75 \pm 0.09$ &  & 1.01 & 0.13 & 0.96 \\
136 &  &  & Mid-B & 27 & 0.62 & $-0.08$ & $0.70 \pm 0.05$ & $-23$ & 0.08 & 0.11 & $0.59 \pm 0.06$ & $0.91 \pm 0.09$ &  & 1.06 & 0.40 & 1.10 \\
137 &  &  & B8 & 48 & 0.50 & $-0.04$ & $0.54 \pm 0.02$ & $-14$ & 0.05 & 0.09 & $0.45 \pm 0.04$ & $0.69 \pm 0.06$ &  & 1.23 & 0.58 & 1.23 \\
138 &  &  & B6 & 37 & 0.47 & $-0.07$ & $0.54 \pm 0.01$ & $-4$ & 0.01 & 0.02 & $0.52 \pm 0.02$ & $0.80 \pm 0.03$ &  & 1.54 & 0.40 & 1.13 \\
139 & B5V & 22 & Mid-B & 25 & 0.58 & $-0.08$ & $0.66 \pm 0.02$ & $-23$ & 0.08 & 0.11 & $0.55 \pm 0.04$ & $0.85 \pm 0.06$ & $0.87 \pm 0.15$ & 1.38 & 0.54 & 0.99 \\
140 &  &  & B6 & 38 & 0.48 & $-0.07$ & $0.55 \pm 0.01$ & $-14$ & 0.05 & 0.08 & $0.47 \pm 0.03$ & $0.72 \pm 0.05$ &  & 0.87 & 0.37 & 0.73 \\
141 &  &  & Late-B & 18 & 0.87 & $-0.04$ & $0.91 \pm 0.04$ & $-35$ & 0.12 & 0.19 & $0.72 \pm 0.06$ & $1.11 \pm 0.09$ &  & 1.28 & 0.69 & 1.37 \\
142 &  &  & Mid-B & 13 & 0.85 & $-0.08$ & $0.93 \pm 0.05$ & $-24$ & 0.08 & 0.11 & $0.82 \pm 0.06$ & $1.26 \pm 0.09$ &  & 1.27 & 0.56 & 1.22 \\
143 &  &  & B5 & 65 & 0.34 & $-0.08$ & $0.42 \pm 0.01$ & $-23$ & 0.08 & 0.11 & $0.31 \pm 0.04$ & $0.48 \pm 0.06$ &  & 0.88 & 0.33 & 0.77 \\
144 &  &  & Mid-B & 42 & 0.68 & $-0.08$ & $0.76 \pm 0.05$ & $-26$ & 0.09 & 0.12 & $0.64 \pm 0.06$ & $0.98 \pm 0.09$ &  & 1.29 & 0.51 & 1.22 \\
145 &  &  & Early-B & 25 & 0.73 & $-0.13$ & $0.86 \pm 0.05$ & $-26$ & 0.09 & 0.10 & $0.76 \pm 0.06$ & $1.17 \pm 0.09$ &  & 1.11 & 0.46 & 1.03 \\
146 &  &  & Late-B & 28 & 0.63 & $-0.04$ & $0.67 \pm 0.04$ & $-6$ & 0.02 & 0.04 & $0.63 \pm 0.06$ & $0.97 \pm 0.09$ &  & 1.81 & 0.74 & 1.46 \\
147 &  &  & B7 & 39 & 0.47 & $-0.06$ & $0.53 \pm 0.02$ & $-12$ & 0.04 & 0.07 & $0.46 \pm 0.04$ & $0.71 \pm 0.06$ &  & 0.86 & 0.29 & 0.83 \\
148 &  &  & Early-B & 32 & 0.65 & $-0.13$ & $0.78 \pm 0.05$ & $-37$ & 0.12 & 0.13 & $0.65 \pm 0.06$ & $1.00 \pm 0.09$ &  & 0.88 & 0.34 & 0.93 \\
149 &  &  & Late-B & 21 & 0.52 & $-0.04$ & $0.56 \pm 0.04$ & $-17$ & 0.06 & 0.10 & $0.46 \pm 0.05$ & $0.71 \pm 0.08$ &  & 0.91 & 0.38 & 1.00 \\
150 &  &  & B4 & 30 & 0.58 & $-0.09$ & $0.67 \pm 0.02$ & $-36$ & 0.12 & 0.15 & $0.52 \pm 0.05$ & $0.80 \pm 0.08$ &  & 0.97 & 0.46 & 1.10 \\
151 &  &  & Mid-B & 22 & 0.59 & $-0.08$ & $0.67 \pm 0.05$ & $-35$ & 0.12 & 0.16 & $0.51 \pm 0.07$ & $0.78 \pm 0.11$ &  & 0.91 & 0.53 & 0.81 \\
152 &  &  & Late-B & 22 & 0.56 & $-0.04$ & $0.60 \pm 0.04$ & $-14$ & 0.05 & 0.09 & $0.51 \pm 0.05$ & $0.78 \pm 0.08$ &  & 0.83 & 0.41 & 0.79 \\
153 & B4V & 25 & Mid-B & 20 & 0.76 & $-0.09$ & $0.84 \pm 0.02$ & $-62$ & 0.21 & 0.24 & $0.60 \pm 0.06$ & $0.92 \pm 0.09$ & $0.82 \pm 0.12$ & 0.80 & 0.28 & 0.72 \\
154 &  &  & Late-B & 28 & 0.52 & $-0.04$ & $0.56 \pm 0.04$ & $-21$ & 0.07 & 0.12 & $0.44 \pm 0.06$ & $0.68 \pm 0.09$ &  & 0.55 & 0.32 & 0.69 \\
155 &  &  & Late-B & 18 & 0.61 & $-0.04$ & $0.65 \pm 0.04$ & $-18$ & 0.06 & 0.10 & $0.55 \pm 0.05$ & $0.85 \pm 0.08$ &  & 0.88 & 0.41 & 0.95 \\
156 &  &  & Late-B & 35 & 0.40 & $-0.04$ & $0.44 \pm 0.04$ & $-13$ & 0.04 & 0.07 & $0.37 \pm 0.05$ & $0.57 \pm 0.08$ &  & 0.79 & 0.26 & 0.82 \\
157 &  &  & Mid-B & 23 & 0.69 & $-0.08$ & $0.77 \pm 0.05$ & $-25$ & 0.08 & 0.11 & $0.66 \pm 0.06$ & $1.02 \pm 0.09$ &  & 0.97 & 0.25 & 0.87 \\
158 &  &  & Late-B & 15 & 0.59 & $-0.04$ & $0.63 \pm 0.04$ & $-12$ & 0.04 & 0.07 & $0.56 \pm 0.05$ & $0.86 \pm 0.08$ &  & 0.70 & 0.41 & 0.85 \\
159 &  &  & B8 & 27 & 0.43 & $-0.04$ & $0.47 \pm 0.02$ & $-53$ & 0.18 & 0.27 & $0.20 \pm 0.06$ & $0.31 \pm 0.09$ &  & 0.71 & 0.44 & 0.93 \\
160 & B7IV & 57 & Late-B & 17 & 0.58 & $-0.06$ & $0.64 \pm 0.02$ & $-19$ & 0.06 & 0.10 & $0.54 \pm 0.04$ & $0.83 \pm 0.06$ & $0.99 \pm 0.07$ & 0.82 & 0.56 & 1.24 \\
161 &  &  & B7 & 37 & 0.66 & $-0.06$ & $0.72 \pm 0.02$ & $-11$ & 0.04 & 0.07 & $0.65 \pm 0.04$ & $1.00 \pm 0.06$ &  & 0.62 & 0.61 & 1.41 \\
162 &  &  & Late-B & 24 & 0.61 & $-0.04$ & $0.65 \pm 0.04$ & $-17$ & 0.06 & 0.10 & $0.55 \pm 0.05$ & $0.85 \pm 0.08$ &  & 0.59 & 0.51 & 0.91 \\
163 &  &  & B5 & 34 & 0.53 & $-0.08$ & $0.61 \pm 0.01$ & $-15$ & 0.05 & 0.07 & $0.54 \pm 0.03$ & $0.83 \pm 0.05$ &  & 0.95 & 0.43 & 0.93 \\
164 & B2V & 21 & continuum+Balmer & 7 & 1.03 & $-0.13$ & $1.16 \pm 0.02$ & $-76$ & 0.25 & 0.24 & $0.92 \pm 0.05$ & $1.42 \pm 0.08$ & $1.35 \pm 0.08$ & 0.81 & 0.58 & 1.34 \\
165 &  &  & Late-B & 11 & 0.62 & $-0.04$ & $0.66 \pm 0.04$ & $-18$ & 0.06 & 0.10 & $0.56 \pm 0.05$ & $0.86 \pm 0.08$ &  & 0.87 & 0.57 & 1.12 \\
166 &  &  & B3 & 47 & 0.57 & $-0.12$ & $0.69 \pm 0.02$ & $-37$ & 0.12 & 0.14 & $0.55 \pm 0.05$ & $0.85 \pm 0.08$ &  & 0.38 & 0.39 & 0.82 \\
167 & B2III & 47 & Mid-B & 34 & 0.67 & $-0.13$ & $0.80 \pm 0.02$ & $-17$ & 0.06 & 0.07 & $0.73 \pm 0.03$ & $1.12 \pm 0.05$ & $1.02 \pm 0.07$ & 0.28 & 0.28 & 0.86 \\
168 &  &  & Late-B & 23 & 0.57 & $-0.04$ & $0.61 \pm 0.04$ & $-24$ & 0.08 & 0.14 & $0.47 \pm 0.06$ & $0.72 \pm 0.09$ &  & 0.87 & 0.46 & 0.88 \\
169* & B9V & 38 & Late-B & 22 & 0.57 & $-0.02$ & $0.59 \pm 0.14$ & $-13$ & 0.04 & 0.08 & $0.51 \pm 0.15$ & $0.78 \pm 0.23$ & $0.81 \pm 0.07$ & 1.02 & 0.46 & 1.41 \\
170 &  &  & Early-B & 19 & 0.64 & $-0.13$ & $0.77 \pm 0.05$ & $-15$ & 0.05 & 0.06 & $0.71 \pm 0.05$ & $1.09 \pm 0.08$ &  & 0.83 & 0.51 & 1.39 \\
\hline
\end{tabular}
\end{minipage}
\end{table*}

\begin{table*}
\tiny
\begin{minipage}{180mm}
\contcaption{}
\centering
\begin{tabular}{@{}llrlrccrcrcrrrrrr@{}}
\hline
\# & \multicolumn{4}{c}{Spectral typing} & $(r-i)_{c}$ & $(r-i)_{\circ}$ & $E(r-i)$ & $EW(\Halpha)$ & $f_{D}$ & $E^{cs}(r-i)$ &$E^{is}(r-i)$ & $E^{is}(B-V)$ & $E^{is}(B-V)_{\rm{R13}}$ & SFD98\,$\times0.86$  & RF09 & Sale14 \\
  & R13 & S/N & This work & S/N & (mag) & (mag) & (mag) & (mag) & & (mag) & (mag) & (mag) & (mag) & (mag) & (mag) & (mag) \\
\hline
171 &  &  & B4 & 28 & 0.41 & $-0.09$ & $0.50 \pm 0.02$ & $-9$ & 0.03 & 0.04 & $0.46 \pm 0.04$ & $0.71 \pm 0.06$ &  & 0.52 & 0.47 & 1.05 \\
172 &  &  & Late-B & 25 & 0.67 & $-0.04$ & $0.71 \pm 0.04$ & $-11$ & 0.04 & 0.07 & $0.64 \pm 0.05$ & $0.98 \pm 0.08$ &  & 0.82 & 0.46 & 0.99 \\
173 &  &  & Mid-B & 25 & 0.79 & $-0.08$ & $0.87 \pm 0.05$ & $-48$ & 0.16 & 0.20 & $0.67 \pm 0.06$ & $1.03 \pm 0.09$ &  & 0.85 & 0.45 & 1.05 \\
175 &  &  & B3 & 30 & 0.60 & $-0.12$ & $0.72 \pm 0.02$ & $-27$ & 0.09 & 0.11 & $0.61 \pm 0.04$ & $0.94 \pm 0.06$ &  & 0.91 & 0.59 & 1.35 \\
176 & B3IV & 32 & continuum+Balmer & 18 & 1.17 & $-0.12$ & $1.29 \pm 0.02$ & $-43$ & 0.14 & 0.16 & $1.13 \pm 0.05$ & $1.74 \pm 0.08$ & $1.58 \pm 0.07$ & 1.70 & 0.74 & 1.57 \\
177 &  &  & Mid-B & 13 & 0.78 & $-0.08$ & $0.86 \pm 0.05$ & $-14$ & 0.05 & 0.07 & $0.79 \pm 0.06$ & $1.22 \pm 0.09$ &  & 0.95 & 0.44 & 1.12 \\
178 &  &  & Late-B & 12 & 0.60 & $-0.04$ & $0.64 \pm 0.04$ & $-16$ & 0.05 & 0.09 & $0.55 \pm 0.05$ & $0.85 \pm 0.08$ &  & 0.53 & 0.52 & 0.78 \\
179 &  &  & Mid-B & 17 & 0.66 & $-0.08$ & $0.73 \pm 0.05$ & $-30$ & 0.10 & 0.13 & $0.60 \pm 0.06$ & $0.92 \pm 0.09$ &  & 0.82 & 0.39 & 0.85 \\
180 & B5V & 50 & continuum+Balmer & 9 & 0.54 & $-0.08$ & $0.62 \pm 0.01$ & $-12$ & 0.04 & 0.06 & $0.56 \pm 0.03$ & $0.86 \pm 0.05$ & $0.78 \pm 0.07$ & 0.73 & 0.38 & 0.89 \\
181 & B8V & 22 &  &  & 0.77 & $-0.04$ & $0.81 \pm 0.02$ & $-17$ & 0.06 & 0.10 & $0.71 \pm 0.04$ & $1.09 \pm 0.06$ &  & 1.06 & 0.41 & 1.13 \\
182 &  &  & Early-B & 20 & 0.58 & $-0.13$ & $0.71 \pm 0.05$ & $-17$ & 0.06 & 0.07 & $0.64 \pm 0.05$ & $0.98 \pm 0.08$ &  & 0.46 & 0.38 & 0.75 \\
183 &  &  & B7 & 45 & 0.36 & $-0.06$ & $0.42 \pm 0.02$ & $-12$ & 0.04 & 0.07 & $0.35 \pm 0.04$ & $0.54 \pm 0.06$ &  & 0.68 & 0.27 & 0.67 \\
184 &  &  & B4 & 42 & 0.58 & $-0.09$ & $0.67 \pm 0.02$ & $-25$ & 0.08 & 0.10 & $0.57 \pm 0.05$ & $0.88 \pm 0.08$ &  & 0.49 & 0.38 & 0.75 \\
185 &  &  & Late-B & 32 & 0.64 & $-0.04$ & $0.68 \pm 0.04$ & $-12$ & 0.04 & 0.07 & $0.61 \pm 0.05$ & $0.94 \pm 0.08$ &  & 0.87 & 0.55 & 0.94 \\
186 &  &  & Late-B & 39 & 0.37 & $-0.04$ & $0.41 \pm 0.04$ & $-18$ & 0.06 & 0.10 & $0.31 \pm 0.05$ & $0.48 \pm 0.08$ &  & 0.73 & 0.32 & 0.90 \\
187 & B6V & 45 & continuum+Balmer & 24 & 0.55 & $-0.07$ & $0.62 \pm 0.02$ & $-20$ & 0.07 & 0.10 & $0.52 \pm 0.04$ & $0.80 \pm 0.06$ &  & 0.58 & 0.30 & 0.76 \\
188 & B7V & 41 & Late-B & 25 & 0.66 & $-0.06$ & $0.72 \pm 0.02$ & $-44$ & 0.15 & 0.22 & $0.50 \pm 0.05$ & $0.77 \pm 0.08$ & $1.04 \pm 0.07$ & 1.26 & 0.70 & 1.59 \\
189 & B7-8V & 39 & continuum+Balmer & 7 & 0.66 & $-0.05$ & $0.71 \pm 0.02$ & $-27$ & 0.09 & 0.15 & $0.56 \pm 0.04$ & $0.86 \pm 0.06$ &  & 1.17 & 0.51 & 1.28 \\
190 &  &  & Late-B & 28 & 0.61 & $-0.04$ & $0.65 \pm 0.04$ & $-15$ & 0.05 & 0.09 & $0.56 \pm 0.05$ & $0.86 \pm 0.08$ &  & 1.92 & 0.52 & 1.33 \\
191 & B3V & 33 & continuum+Balmer & 13 & 0.67 & $-0.12$ & $0.79 \pm 0.02$ & $-47$ & 0.16 & 0.18 & $0.61 \pm 0.05$ & $0.94 \pm 0.08$ & $0.92 \pm 0.08$ & 1.08 & 0.39 & 0.94 \\
192 &  &  & Early-B & 27 & 0.56 & $-0.13$ & $0.69 \pm 0.05$ & $-34$ & 0.11 & 0.12 & $0.57 \pm 0.06$ & $0.88 \pm 0.09$ &  & 0.65 & 0.33 & 0.63 \\
193 & B7V & 39 & Mid-B & 19 & 0.69 & $-0.06$ & $0.75 \pm 0.02$ & $-39$ & 0.13 & 0.19 & $0.56 \pm 0.05$ & $0.86 \pm 0.08$ & $0.94 \pm 0.07$ & 1.48 & 0.68 & 1.37 \\
194 &  &  & B3 & 39 & 0.46 & $-0.12$ & $0.58 \pm 0.02$ & $-19$ & 0.06 & 0.08 & $0.50 \pm 0.04$ & $0.77 \pm 0.06$ &  & 0.67 & 0.24 & 0.62 \\
195 & A0III & 42 & Late-B & 14 & 0.50 & $0.00$ & $0.50 \pm 0.01$ & $-21$ & 0.07 & 0.15 & $0.35 \pm 0.06$ & $0.54 \pm 0.09$ & $0.70 \pm 0.07$ & 0.62 & 0.26 & 0.67 \\
196 &  &  & Mid-B & 15 & 0.59 & $-0.08$ & $0.67 \pm 0.05$ & $-21$ & 0.07 & 0.10 & $0.57 \pm 0.06$ & $0.88 \pm 0.09$ &  & 0.83 & 0.36 & 0.79 \\
197 & B7IV & 40 & Late-B & 15 & 0.57 & $-0.06$ & $0.63 \pm 0.02$ & $-17$ & 0.06 & 0.10 & $0.53 \pm 0.04$ & $0.82 \pm 0.06$ & $0.87 \pm 0.08$ & 0.84 & 0.29 & 0.73 \\
198 & B3-4V & 27 & Late-B & 16 & 0.67 & $-0.10$ & $0.78 \pm 0.02$ & $-36$ & 0.12 & 0.14 & $0.64 \pm 0.05$ & $0.98 \pm 0.08$ & $0.98 \pm 0.08$ & 1.05 & 0.38 & 1.14 \\
199 &  &  & Early-B & 19 & 0.53 & $-0.13$ & $0.66 \pm 0.05$ & $-5$ & 0.02 & 0.02 & $0.64 \pm 0.05$ & $0.98 \pm 0.08$ &  & 0.84 & 0.25 & 0.86 \\
200 &  &  & Early-B & 32 & 0.56 & $-0.13$ & $0.69 \pm 0.05$ & $-35$ & 0.12 & 0.13 & $0.56 \pm 0.06$ & $0.86 \pm 0.09$ &  & 0.78 & 0.26 & 0.79 \\
201 &  &  & Mid-B & 39 & 0.44 & $-0.08$ & $0.52 \pm 0.05$ & $-20$ & 0.07 & 0.10 & $0.42 \pm 0.06$ & $0.65 \pm 0.09$ &  & 0.80 & 0.33 & 0.71 \\
202 &  &  & B3 & 53 & 0.48 & $-0.12$ & $0.60 \pm 0.02$ & $-30$ & 0.10 & 0.12 & $0.48 \pm 0.04$ & $0.74 \pm 0.06$ &  & 0.81 & 0.38 & 1.16 \\
203 &  &  & Mid-B & 23 & 0.69 & $-0.08$ & $0.77 \pm 0.05$ & $-21$ & 0.07 & 0.10 & $0.67 \pm 0.06$ & $1.03 \pm 0.09$ &  & 0.83 & 0.57 & 0.79 \\
204 & B3V & 50 & Mid-B & 16 & 0.51 & $-0.12$ & $0.63 \pm 0.02$ & $-20$ & 0.07 & 0.09 & $0.54 \pm 0.04$ & $0.83 \pm 0.06$ & $0.81 \pm 0.07$ & 0.96 & 0.45 & 0.89 \\
205 &  &  & B5 & 31 & 0.39 & $-0.08$ & $0.47 \pm 0.01$ & $-20$ & 0.07 & 0.10 & $0.37 \pm 0.04$ & $0.57 \pm 0.06$ &  & 2.12 & 0.35 & 0.89 \\
206 &  &  & B3 & 26 & 0.71 & $-0.12$ & $0.83 \pm 0.02$ & $-47$ & 0.16 & 0.18 & $0.65 \pm 0.05$ & $1.00 \pm 0.08$ &  & 0.87 & 0.19 & 0.82 \\
207 &  &  & Early-B & 43 & 0.57 & $-0.13$ & $0.70 \pm 0.05$ & $-52$ & 0.17 & 0.18 & $0.52 \pm 0.06$ & $0.80 \pm 0.09$ &  & 1.25 & 0.29 & 0.61 \\
208 &  &  & B5 & 38 & 0.38 & $-0.08$ & $0.46 \pm 0.01$ & $-25$ & 0.08 & 0.11 & $0.35 \pm 0.04$ & $0.54 \pm 0.06$ &  & 0.87 & 0.49 & 0.76 \\
209 & B3V & 43 & B4 & 36 & 0.42 & $-0.12$ & $0.54 \pm 0.02$ & $-40$ & 0.13 & 0.15 & $0.39 \pm 0.05$ & $0.60 \pm 0.08$ & $0.67 \pm 0.07$ & 0.77 & 0.32 & 0.72 \\
210 &  &  & Mid-B & 23 & 0.47 & $-0.08$ & $0.55 \pm 0.05$ & $-43$ & 0.14 & 0.18 & $0.37 \pm 0.06$ & $0.57 \pm 0.09$ &  & 0.99 & 0.23 & 0.61 \\
211 &  &  & Mid-B & 32 & 0.40 & $-0.08$ & $0.48 \pm 0.05$ & $-32$ & 0.11 & 0.15 & $0.33 \pm 0.06$ & $0.51 \pm 0.09$ &  & 0.83 & 0.48 & 0.73 \\
212 &  &  & Early-B & 29 & 0.85 & $-0.13$ & $0.98 \pm 0.05$ & $-37$ & 0.12 & 0.13 & $0.85 \pm 0.06$ & $1.31 \pm 0.09$ &  & 1.61 & 0.62 & 1.16 \\
213 &  &  & B5 & 40 & 0.34 & $-0.08$ & $0.42 \pm 0.01$ & $-23$ & 0.08 & 0.11 & $0.31 \pm 0.04$ & $0.48 \pm 0.06$ &  & 0.75 & 0.28 & 0.65 \\
214 & B8-9III & 61 & Late-B & 19 & 0.42 & $-0.03$ & $0.45 \pm 0.02$ & $-11$ & 0.04 & 0.08 & $0.37 \pm 0.05$ & $0.57 \pm 0.08$ & $0.71 \pm 0.07$ & 0.67 & 0.39 & 0.64 \\
215 & B3-4V & 35 & Mid-B & 20 & 0.46 & $-0.10$ & $0.56 \pm 0.02$ & $-34$ & 0.11 & 0.13 & $0.43 \pm 0.04$ & $0.66 \pm 0.06$ & $0.77 \pm 0.08$ & 1.07 & 0.30 & 0.95 \\
216 & B5IV & 44 & Mid-B & 26 & 0.55 & $-0.08$ & $0.63 \pm 0.01$ & $-65$ & 0.22 & 0.26 & $0.37 \pm 0.06$ & $0.57 \pm 0.09$ & $0.69 \pm 0.08$ & 1.07 & 0.35 & 0.95 \\
217* &  &  & Late-B & 24 & 0.44 & $-0.04$ & $0.48 \pm 0.15$ & $-19$ & 0.06 & 0.10 & $0.38 \pm 0.16$ & $0.58 \pm 0.25$ &  & 1.26 & 0.67 & 0.97 \\
218 &  &  & B6 & 35 & 0.40 & $-0.07$ & $0.47 \pm 0.01$ & $-21$ & 0.07 & 0.10 & $0.37 \pm 0.04$ & $0.57 \pm 0.06$ &  & 1.63 & 0.31 & 0.62 \\
219* &  &  & B7 & 45 & 0.35 & $-0.06$ & $0.41 \pm 0.14$ & $-24$ & 0.08 & 0.13 & $0.28 \pm 0.15$ & $0.43 \pm 0.23$ &  & 1.08 & 0.42 & 1.41 \\
220* & B7V & 51 & Late-B & 28 & 0.43 & $-0.06$ & $0.49 \pm 0.14$ & $-9$ & 0.03 & 0.05 & $0.44 \pm 0.15$ & $0.68 \pm 0.23$ & $0.72 \pm 0.08$ & 0.70 & 0.31 & 0.72 \\
221 &  &  & Mid-B & 21 & 0.60 & $-0.08$ & $0.68 \pm 0.05$ & $-30$ & 0.10 & 0.13 & $0.55 \pm 0.06$ & $0.85 \pm 0.09$ &  & 0.60 & 0.42 & 0.87 \\
222 & B6V & 25 & Late-B & 19 & 0.81 & $-0.07$ & $0.88 \pm 0.02$ & $-16$ & 0.05 & 0.08 & $0.80 \pm 0.03$ & $1.23 \pm 0.05$ & $1.20 \pm 0.10$ & 1.39 & 0.68 & 1.38 \\
223 & B7V & 55 &  &  & 0.54 & $-0.06$ & $0.60 \pm 0.02$ & $-13$ & 0.04 & 0.07 & $0.53 \pm 0.04$ & $0.82 \pm 0.06$ &  & 1.44 & 0.48 & 1.08 \\
224 & B5V & 45 & Mid-B & 24 & 0.60 & $-0.08$ & $0.68 \pm 0.14$ & $-12$ & 0.04 & 0.06 & $0.62 \pm 0.15$ & $0.95 \pm 0.23$ & $0.93 \pm 0.07$ & 1.33 & 0.72 & 1.39 \\
225 &  &  & Mid-B & 20 & 0.73 & $-0.08$ & $0.81 \pm 0.05$ & $-46$ & 0.15 & 0.19 & $0.62 \pm 0.06$ & $0.95 \pm 0.09$ &  & 0.65 & 0.44 & 0.99 \\
226 & B4V & 30 & Mid-B & 27 & 0.73 & $-0.09$ & $0.82 \pm 0.02$ & $-24$ & 0.08 & 0.10 & $0.72 \pm 0.05$ & $1.11 \pm 0.08$ & $1.06 \pm 0.08$ & 1.48 & 0.65 & 1.49 \\
227 & B5V & 33 & Mid-B & 15 & 0.81 & $-0.08$ & $0.89 \pm 0.02$ & $-14$ & 0.05 & 0.07 & $0.82 \pm 0.03$ & $1.26 \pm 0.05$ & $1.22 \pm 0.09$ & 1.14 & 0.55 & 1.16 \\
228 &  &  & Late-B & 20 & 0.52 & $-0.04$ & $0.56 \pm 0.04$ & $-10$ & 0.03 & 0.06 & $0.50 \pm 0.06$ & $0.77 \pm 0.09$ &  & 0.77 & 0.35 & 0.87 \\
229 &  &  & Late-B & 18 & 0.39 & $-0.04$ & $0.43 \pm 0.04$ & $-9$ & 0.03 & 0.06 & $0.37 \pm 0.06$ & $0.57 \pm 0.09$ &  & 0.65 & 0.38 & 0.71 \\
230 &  &  & Late-B & 26 & 0.43 & $-0.04$ & $0.47 \pm 0.04$ & $-24$ & 0.08 & 0.14 & $0.33 \pm 0.06$ & $0.51 \pm 0.09$ &  & 1.38 & 0.90 & 1.00 \\
231 &  &  & Late-B & 27 & 0.64 & $-0.04$ & $0.68 \pm 0.04$ & $-18$ & 0.06 & 0.10 & $0.58 \pm 0.05$ & $0.89 \pm 0.08$ &  & 1.29 & 0.49 & 1.27 \\
232 &  &  & Early-B & 44 & 0.54 & $-0.13$ & $0.67 \pm 0.05$ & $-12$ & 0.04 & 0.05 & $0.62 \pm 0.05$ & $0.95 \pm 0.08$ &  & 0.54 & 0.36 & 0.93 \\
233 &  &  & Mid-B & 48 & 0.35 & $-0.08$ & $0.43 \pm 0.05$ & $-7$ & 0.02 & 0.03 & $0.40 \pm 0.06$ & $0.62 \pm 0.09$ &  & 0.62 & 0.30 & 0.81 \\
234 &  &  & Mid-B & 16 & 0.55 & $-0.08$ & $0.63 \pm 0.05$ & $-19$ & 0.06 & 0.08 & $0.55 \pm 0.06$ & $0.85 \pm 0.09$ &  & 2.74 & 0.74 & 1.00 \\
235 &  &  & Mid-B & 30 & 0.70 & $-0.08$ & $0.78 \pm 0.05$ & $-33$ & 0.11 & 0.15 & $0.63 \pm 0.06$ & $0.97 \pm 0.09$ &  & 0.97 & 0.64 & 1.05 \\
236 &  &  & B3 & 36 & 0.69 & $-0.12$ & $0.81 \pm 0.02$ & $-47$ & 0.16 & 0.18 & $0.63 \pm 0.05$ & $0.97 \pm 0.08$ &  & 0.86 & 0.52 & 1.00 \\
237 &  &  & Late-B & 25 & 0.81 & $-0.04$ & $0.85 \pm 0.04$ & $-27$ & 0.09 & 0.15 & $0.70 \pm 0.06$ & $1.08 \pm 0.09$ &  & 1.14 & 0.59 & 1.24 \\
238 &  &  & Mid-B & 23 & 0.78 & $-0.08$ & $0.86 \pm 0.05$ & $-16$ & 0.05 & 0.07 & $0.79 \pm 0.06$ & $1.22 \pm 0.09$ &  & 1.32 & 0.67 & 1.37 \\
239 &  &  & Late-B & 31 & 0.55 & $-0.04$ & $0.59 \pm 0.04$ & $-8$ & 0.03 & 0.06 & $0.53 \pm 0.06$ & $0.82 \pm 0.09$ &  & 0.90 & 0.46 & 0.77 \\
240 &  &  & Late-B & 27 & 0.66 & $-0.04$ & $0.70 \pm 0.04$ & $-23$ & 0.08 & 0.14 & $0.56 \pm 0.06$ & $0.86 \pm 0.09$ &  & 1.08 & 0.52 & 1.50 \\
241 &  &  & Mid-B & 31 & 0.61 & $-0.08$ & $0.69 \pm 0.05$ & $-30$ & 0.10 & 0.13 & $0.56 \pm 0.06$ & $0.86 \pm 0.09$ &  & 0.91 & 0.42 & 0.87 \\
242 &  &  & Early-B & 15 & 0.80 & $-0.13$ & $0.93 \pm 0.05$ & $-55$ & 0.18 & 0.19 & $0.74 \pm 0.06$ & $1.14 \pm 0.09$ &  & 1.14 & 0.57 & 1.09 \\
243 &  &  & Mid-B & 30 & 0.83 & $-0.08$ & $0.91 \pm 0.05$ & $-24$ & 0.08 & 0.11 & $0.80 \pm 0.06$ & $1.23 \pm 0.09$ &  & 1.18 & 0.62 & 1.13 \\
244 &  &  & Mid-B & 23 & 0.84 & $-0.08$ & $0.92 \pm 0.05$ & $-42$ & 0.14 & 0.18 & $0.74 \pm 0.06$ & $1.14 \pm 0.09$ &  & 0.93 & 0.98 & 1.39 \\
245 &  &  & Late-B & 11 & 0.73 & $-0.04$ & $0.77 \pm 0.04$ & $-25$ & 0.08 & 0.14 & $0.63 \pm 0.06$ & $0.97 \pm 0.09$ &  & 0.81 & 0.31 & 0.98 \\
246 &  &  & Late-B & 24 & 0.55 & $-0.04$ & $0.59 \pm 0.04$ & $-9$ & 0.03 & 0.06 & $0.53 \pm 0.06$ & $0.82 \pm 0.09$ &  & 0.84 & 0.33 & 0.73 \\
247 & B4V & 35 & Early-B & 11 & 0.80 & $-0.09$ & $0.89 \pm 0.02$ & $-34$ & 0.11 & 0.14 & $0.74 \pm 0.04$ & $1.14 \pm 0.06$ & $1.11 \pm 0.11$ & 1.30 & 0.72 & 1.30 \\
248 &  &  & Late-B & 14 & 0.90 & $-0.04$ & $0.94 \pm 0.04$ & $-22$ & 0.07 & 0.12 & $0.82 \pm 0.06$ & $1.26 \pm 0.09$ &  & 1.35 & 0.66 & 1.23 \\
\hline
\end{tabular}
\end{minipage}
\end{table*}

\label{lastpage}
\end{document}